\documentclass[prd,nofootinbib,showpacs,superscriptaddress, preprint]{revtex4-1}
\pdfoutput=1
\usepackage{amsmath,amsfonts,amssymb,graphicx,natbib,bm}
\usepackage{color}
\usepackage{slashed}    
\usepackage[colorlinks,citecolor=blue]{hyperref}
\usepackage{color}
\usepackage{url}
\usepackage{cancel}
\usepackage{slashed}
\usepackage{multirow}
\usepackage{rotating}
\usepackage{braket}
\usepackage{float}
\usepackage{subfigure}

\newcommand{\be}{\begin{eqnarray*}}
\newcommand{\ee}{\end{eqnarray*}}

\newcommand{\bee}{\begin{eqnarray}}
\newcommand{\eee}{\end{eqnarray}}
\newcommand{\beeq}{\begin{equation}}
\newcommand{\eeq}{\end{equation}}

\usepackage{xspace}

\newcommand{\ba}{\begin{array}}
\newcommand{\ea}{\end{array}}
\newcommand{\bd}{\begin{displaymath}}
\newcommand{\ed}{\end{displaymath}}
\newcommand{\besub}{\begin{subequations}}
\newcommand{\eesub}{\end{subequations}}
\newcommand{\bea}{\begin{eqnarray}}
\newcommand{\eea}{\end{eqnarray}}

\def\q2 {q^2}

\def\bt{\begin{table}}
\def\et{\end{table}}

\begin{document}
%\preprint{IP/BBSR/2015-4}
\title{Non-thermal leptogenesis and UV freeze-in of dark matter: impact of inflationary reheating }

\author{Basabendu Barman}
\email{bb1988@iitg.ac.in}
\affiliation{Department of Physics, Indian Institute of Technology Guwahati, Assam 781039, India}

\author{Debasish Borah}
\email{dborah@iitg.ac.in}
\affiliation{Department of Physics, Indian Institute of Technology Guwahati, Assam 781039, India}

\author{Rishav Roshan}
\email{rishav.roshan@iitg.ac.in}
\affiliation{Department of Physics, Indian Institute of Technology Guwahati, Assam 781039, India}

\begin{abstract}
We study a minimal scenario to realize non-thermal leptogenesis and UV freeze-in of a Standard Model (SM) gauge singlet fermionic dark matter (DM) simultaneously, with inflaton field playing a non-trivial role in their yields. The renormalizable interactions are restricted to the SM fields, two right handed neutrinos (RHN) and inflaton coupling exclusively to the RHNs, while the DM couples to both the SM and the RHNs only via operators of dimension $d>4$. Considering two separate cases of $d=\{5,6\}$, we show that for $d=5$, inflaton decay into RHNs followed by their subsequent decay into SM particles lead to both reheating as well as DM production from the SM bath. This requires a cut-off scale as large as $\Lambda\sim 10^{17}~\rm GeV$ depending on the DM mass. On the other hand, for $d=6$, DM production happens directly from scattering of RHNs (for $\Lambda\gtrsim 10^{14}~\rm GeV$) that results in a very non-trivial evolution of the DM yield. In both these cases, it is possible to explain the observed baryon asymmetry through successful non-thermal leptogenesis via the decay of the RHNs, together with the PLANCK observed relic density of the DM via pure UV freeze-in mechanism. Taking into account both instantaneous as well as non-instantaneous reheating separately, we constrain the parameter space of this minimal scenario from relevant phenomenological requirements including sub-eV scale active neutrino masses and their mixing.
\end{abstract}
%%%%%%%%%%%%%%%%%%%%%%%%%%%

\maketitle

{
  \hypersetup{linkcolor=black}
  \tableofcontents
}

%%%%%%%%%%%%%%%%%%%%
\section{Introduction}
\label{sec:intro}
%%%%%%%%%%%%%%%%%%%
The observed matter-antimatter asymmetry of the universe has been a longstanding puzzle~\cite{Zyla:2020zbs,Aghanim:2018eyx}. While approximately $5\%$ of our present universe is dominated by visible matter, most of it is made up of baryons. Quantitatively, the asymmetry is  quoted as a ratio of excess of baryons over antibaryons to photons. According to the latest data from Planck satellite it is given as~\cite{Aghanim:2018eyx} 

\begin{equation}
\eta_B = \frac{n_{B}-n_{\bar{B}}}{n_{\gamma}} = 6.1 \times 10^{-10}.
\label{etaBobs}
\end{equation}

The estimate from the cosmic microwave background (CMB) measurements also matches very well with the ones obtained from big bang nucleosynthesis (BBN). Since the universe is expected to originate in a matter-antimatter symmetric manner and any pre-existing initial asymmetry is likely to be diluted due to accelerated expansion during inflation, there arises the requirement of a dynamical explanation for the observed asymmetry.  Such a dynamical process requires to satisfy certain conditions, given by Sakharov~\cite{Sakharov:1967dj} as (1) baryon number (B) violation, (2) C and CP violation and (3) departure from thermal equilibrium, all of which can not be satisfied in required amount in the standard model (SM) of particle physics and considering an expanding Friedmann-Lemaitre-Robertson-Walker (FLRW) universe. Out-of-equilibrium decay of a heavy particle leading to the generation of baryon asymmetry of the Universe (BAU) has been a well-known mechanism for baryogenesis~\cite{Weinberg:1979bt,Kolb:1979qa}. One interesting way to implement such a mechanism is leptogenesis~\cite{Fukugita:1986hr}, where a net leptonic asymmetry is generated first which gets converted into baryon asymmetry through $(B+L)$-violating EW sphaleron transitions~\cite{Kuzmin:1985mm}. For the lepton asymmetry to be converted into baryon asymmetry, it is important that the processes giving rise to the leptonic asymmetry freeze out before the onset of the sphaleron transitions to prevent wash-out of the asymmetry~\cite{Fong:2013wr}. An interesting feature of this scenario is that the required lepton asymmetry can be generated through CP violating out-of-equilibrium decays of the same heavy fields that take part in the seesaw mechanism~\cite{Mohapatra:1979ia,Yanagida:1979as,GellMann:1980vs,Glashow:1979nm,Schechter:1980gr,Buchmuller:2004nz,DuttaBanik:2020vfr} that explains the origin of tiny neutrino masses~\cite{Zyla:2020zbs}, another observed phenomenon the SM fails to address. In such generic seesaw models for light neutrino masses, the scale of leptogenesis remains very high. For example, in Type-I seesaw framework, the requirement of producing the correct lepton asymmetry pushes the scale of right handed neutrinos to a very high scale $M >10^9$ GeV, known as the Davidson-Ibarra bound~\cite{Davidson:2002qv} of high scale or vanilla leptogenesis\footnote{For right handed neutrino masses lower than the Davidson-Ibarra bound, say around TeV scale, it is still possible to generate correct lepton asymmetry by resorting to a resonant enhancement of the CP-asymmetry with a quasi-degenerate right handed neutrino spectrum~\cite{Pilaftsis:2003gt,Dev:2017wwc}, known as resonant leptogenesis.}. If the temperature of the universe can reach a value equal or above the mass of right handed neutrinos, they can be thermally produced from the bath particles. However, the reheat temperature of the universe after inflation $(T_{\rm RH})$ can, in principle, be as low as a few MeV, typical temperature around the BBN epoch. Therefore, a low reheat temperature will forbid the production of right handed neutrinos in thermal bath. However, they can still be produced non-thermally from the inflaton field, leading to the scenario of non-thermal leptogenesis~\cite{Lazarides:1991wu,Murayama:1992ua,Kolb:1996jt,Giudice:1999fb,Asaka:1999yd,Asaka:1999jb,Hamaguchi:2001gw,Jeannerot:2001qu,Fujii:2002jw,Giudice:2003jh,Pascoli:2003rq,Asaka:2002zu,Panotopoulos:2006wj,HahnWoernle:2008pq,Buchmuller:2013dja,Croon:2019dfw,Borah:2020wyc,Samanta:2020gdw}.

% Minkowski:1977sc, 

Similar to matter-antimatter asymmetry, observations of a mysterious non-luminous, non-baryonic form of matter, known as dark matter (DM) in the universe has been another puzzle. Cosmology based experiments like Planck~\cite{Aghanim:2018eyx}, through precise measurements of cosmic microwave background (CMB) anisotropies, suggest around $26\%$ of the present universe's energy density to be made up of DM.  In terms of density parameter $\Omega_{\rm DM}$ and $h = \text{Hubble Parameter}/(100 \;\text{km} ~\text{s}^{-1} 
\text{Mpc}^{-1})$, the present DM abundance is conventionally reported as \cite{Aghanim:2018eyx}: $\Omega_{\text{DM}} h^2 = 0.120\pm 0.001$ at 68\% CL. Astrophysics related observations have also been suggesting the same for a much longer period~\cite{Zwicky:1933gu,Rubin:1970zza,Clowe:2006eq}. While none of  the SM particles can be a viable DM candidate, several beyond the standard model (BSM) proposals have been studied out of which the weakly interacting massive particle (WIMP) paradigm is the most popular one. WIMP paradigm relies upon weak Type-Interactions between the DM and SM sectors which not only play crucial role in generating its relic abundance from thermal freeze-out~\cite{Srednicki:1988ce,Gondolo:1990dk,Kolb:1990vq} but can also lead to observable DM-nucleon scattering rates at direct detection experiments. However, null results at such direct detection experiments like XENON1T~\cite{Aprile:2018dbl} have led to growing interest in alternative DM formalisms in recent times. One such popular alternative is the feebly interacting massive particle (FIMP) scenario where DM interaction with the SM bath is so weak that it never attains thermal equilibrium at any epoch in the early universe~\cite{Hall:2009bx, Bernal:2017kxu}. The initial abundance of such a particle is considered to be negligible and it can start populating (or freeze-in) the universe due to decay or annihilation of other particles in the thermal bath. If there exists renormalizable DM-SM coupling in FIMP scenario, the non-thermal criterion enforces the couplings to be extremely tiny via the following condition
$\left|\dfrac{\Gamma}{{\bf H}}\right|_{T\simeq M_0}<1$~\cite{Arcadi:2013aba}, where $\Gamma$
is the decay width. For the case of scattering, one has to replace $\Gamma$ by the interaction rate $n_{\rm eq}\, \langle \sigma {\rm v}\rangle$, $n_{\rm eq}$ being the equilibrium number density of mother particle. These types of freeze-in scenarios are known as infra-red (IR)-freeze-in~\cite{Yaguna:2011qn,Chu:2011be,Blennow:2013jba,Merle:2015oja,Shakya:2015xnx,Hessler:2016kwm,Biswas:2016bfo,Konig:2016dzg,Biswas:2016iyh,Biswas:2016yjr,Duch:2017khv,Bernal:2017kxu,Biswas:2018aib,Heeba:2018wtf,Zakeri:2018hhe,Becker:2018rve,Heeba:2019jho,Lebedev:2019ton,Barman:2019lvm,Bhattacharya:2019tqq,Datta:2021elq,Bhattacharya:2021jli} where DM production is dominated by the lowest possible temperature at which it can occur i.e. $T\sim M_0$, since for $T<M_0$, the number density of mother particle becomes Boltzmann suppressed. Another possibility to realize feeble DM-SM interaction is to consider higher dimensional operators  (dimension $d > 4$) connecting DM and SM sectors. Unlike the IR freeze-in scenario, here the DM production is effective at high temperatures and relies upon the details of reheating after the end of inflation. Since DM-SM operators are of dimension five or above, typically, DM is produced via scattering specially at temperatures above the electroweak scale.  This particular scenario is known as the ultra-violet (UV) freeze-in~\cite{Hall:2009bx,Elahi:2014fsa,McDonald:2015ljz,Chen:2017kvz,Biswas:2019iqm,Bernal:2019mhf,Bernal:2020bfj,Bernal:2020qyu,Barman:2020plp,Barman:2020ifq,Barman:2020jrf}.

Motivated by these, here we study a minimal scenario where both non-thermal leptogenesis and UV freeze-in of DM can be realized simultaneously. While two singlet right handed neutrinos (RHN) as in Type-I seesaw framework\footnote{Even though three RHNs are typically introduced in Type-I seesaw, two are sufficient to generate observed neutrino oscillation data with vanishing lightest neutrino mass.} take care of leptogenesis, a $Z_2$ odd SM gauge singlet fermion is introduced as DM which interact with the SM along with the RHNs only via operators of dimension $d>4$, which is necessary for UV freeze-in. Considering the inflaton to be a singlet scalar, we assume its renormalizable coupling with the RHNs alone so that the inflaton decays into a pair of RHNs followed by the latter's out-of-equilibrium decay into the SM particles leading to reheating. We consider both dimension five and dimension six operators separately to study the UV freeze-in of DM while simultaneously solving the coupled Boltzmann equations for non-thermal leptogenesis and reheating. Depending upon inflaton coupling with the RHNs, we outline two cases: instantaneous reheating where the inflation energy density is converted into the heavy neutrinos instantaneously) as well as non-instantaneous (where the inflation energy density is converted into the heavy neutrinos over some finite time) reheating where corresponding evolution of DM and lepton asymmetry are determined with suitable choice of parameters and initial conditions. Without going into the details of the dynamics of inflation, in the present scenario we look into the post slow-roll era when the inflaton energy starts converting into the energy of the heavy neutrinos. See \cite{Garcia:2020eof} for similar studies separating the period of inflation from reheating and references therein for specific models of inflation which allow this possibility. This separation also justifies the fact that we have a cut-off scale in the theory which is lower than the Planck scale while inflationary field excursions can, in general, be super-Planckian. We typically probe the $T_\text{RH}\lesssim M_\varphi$ scenario where $M_\varphi$ is the inflaton mass and $T_\text{RH}$ is the reheating temperature. This is in contrast to most of the cases where generally $M_\varphi\gg T_\text{RH}$ is assumed. From the requirement of successful non-thermal leptogenesis leading to observed BAU, DM relic also the correct neutrino mass, we finally constrain the key parameters namely, the DM mass, reheat temperature, cut-off scale $\Lambda$ as well as the inflaton coupling with RHNs. Also, for the dim.6 case we find the DM abundance shows a very non-trivial evolution as it is produced only from the heavy right handed neutrinos which is absent when it is produced from the radiation as in the case of dim.5.

The paper is organised as follows. In Sec.~\ref{sec:model}, we briefly discuss our framework followed by discussion of non-thermal leptogenesis in Sec.~\ref{sec:leptogen}. In Sec.~\ref{sec:dm-uv-yld} we elaborate the UV freeze-in set-up for calculating the DM yield following the details of relevant coupled Boltzmann equations in Sec.~\ref{sec:numeric}. Sec.~\ref{sec:results} summarizes our results for both instantaneous and non-instantaneous scenarios. In Sec.~\ref{sec:uvcomplete} we briefly discuss possible UV completion of DM-SM operators and finally in Sec.~\ref{sec:concl} we conclude.

%%%%%%%%%%%%%%%%%%%
\section{The Framework}
\label{sec:model}
%%%%%%%%%%%%%%%%

We extend the Standard Model effective field theory (SMEFT) with two right handed neutrinos $N_R$ which is commonly known as the $\nu$SMEFT~\cite{delAguila:2008ir,Aparici:2009fh,Bhattacharya:2015vja}. Assuming only SM gauge symmetries, the renormalizable interaction is given by the SM-Lagrangian with a sterile Majorana neutrino contribution as

\bea\begin{aligned}
&\mathcal{L}_\text{renorm} = \mathcal{L}_\text{SM} + i\overline{N_R^c}\slashed{\partial}N_R-\Bigl(\frac{1}{2}M_N\overline{N_R^c}N_R+\text{h.c}\Bigr)+\Bigl(-y_N\overline{\ell_L}\widetilde{H}N_R+\text{h.c.}\Bigr).     
    \end{aligned}\label{eq:nusmeft-renorm}
\eea

This Lagrangian is then extended by adding the dimension five (dim.5) and dim.6 terms of the form

\bea\begin{aligned}
& \mathcal{L}_\text{eff} =  \mathcal{L}_\text{renorm}+\frac{1}{\Lambda}\sum_i \mathcal{C}_i\mathcal{O}_i^{d=5}+\frac{1}{\Lambda^2}\sum_i \mathcal{C'}_i\mathcal{O}_i^{d=6}
    \end{aligned}
    \label{eqn3}
\eea

where a basis of the operators $\mathcal{O}_i$ upto and including dim.6 will be elaborated in subsection~\ref{sec:d5nusmeft} and subsection~\ref{sec:d6nusmeft} assuming CP conservation. We further consider a SM gauge singlet fermion $\chi$ odd under an imposed $Z_2$ symmetry that makes it absolutely stable over the age of the universe and hence a potential dark matter (DM) candidate. Since it is a gauge singlet fermion (like the right handed neutrinos) and $Z_2$-odd, it does not have any tree-level interaction either with the SM fields or with the RHNs. One can then write generalised DM-SM non-renormalizable interactions as

\bea
\mathcal{L}\supset \frac{c_{ij}\mathcal{O}^{(d)j}_{\rm SM}\mathcal{O}^{(d^{'})i}_{\rm DM}}{\Lambda_{ij}^{d+d^{'}-4}},
\label{eq:nonren-op}
\eea

where $\mathcal{O}^{(d^{'})}_{\rm DM}$ is a dark sector operator of mass dimension $d^{'}$, $\mathcal{O}^{(d)}_{\rm SM}$ is the operator in the visible sector of mass dimension $d$. The parameter $\Lambda_{ij}$ is a dimensionful scale and $c_{ij}$ is the dimensionless Wilson coefficient. If $d+d^{'}>4$, then the interaction is associated with an effective non-renormalizable operator of the form presented in Eq.~\eqref{eq:nonren-op}. All DM-SM interactions are encoded by higher-dimensional operators, with a cut-off scale $\Lambda$, which is the mass scale of the heavy fields integrated out to obtain the low-energy Lagrangian: 

\bea
\mathcal{L}=\mathcal{L}_{\text{SM}}+\mathcal{L}_{\text{DM}}+\mathcal{L}_{d>4},
\eea

where $\mathcal{L}_{\text{SM(DM)}}$ is the renormalizable SM (DM) Lagrangian and $\mathcal{L}_{d>4}$ corresponds to the operators of dimension $d>4$. In the renormalizable level the Lagrangian for the DM field has the  form

\bea
\mathcal{L}_{\text{DM}} = i\overline{\chi^c}\slashed{\partial}\chi-\frac{1}{2}M_\chi\overline{\chi^c}\chi, 
\eea

as the DM is electroweak singlet with zero hypercharge. Note that, the cut-off scale in Eq.~\eqref{eqn3} and Eq.~\eqref{eq:nonren-op} are the same implying the low energy theory containing the right handed neutrinos and the DM is generated by integrating out all the heavy fields at a single scale $\Lambda$. For simplicity, we shall stick to scalar DM bilinears which tell that at the level of dim.5, the DM production takes place from the thermal bath through the gauge invariant interaction of the form~\cite{Barman:2020plp}

\bea
\mathcal{O}_5^{\text{DM-SM}}=\frac{1}{\Lambda}\overline{\chi}\chi H^\dagger H
\eea

where $H$ is the SM Higgs doublet. The only source for the heavy neutrinos, on the other hand, is the on-shell decay of the inflaton which is assumed to couple exclusively to them\footnote{Heavy right handed neutrinos can also be produced from the SM fields by virtue of their Yukawa couplings if the highest temperature of the SM bath exceeds the mass of heavy neutrinos. However, this is not the case in non-thermal leptogenesis pursued here.}. This also implies that the kinematical condition $M_\varphi>2M_N$ always holds where $M_\varphi$ is the inflaton mass and $M_N$ is the mass of the heavy neutrino. In our case $M_N\equiv M_1$ which is the lightest heavy neutrino that is responsible for generating the required lepton asymmetry. As we shall argue, in the non-thermal leptogenesis set-up we consider hierarchical right handed neutrinos $M_2\gg M_1$ which, in turn, also implies $2M_1<M_\varphi<2M_2$. Now, since we are considering an effective description, hence we consider $\Lambda$ to be the highest scale of the theory. For completeness, we write down the full list of operators upto and including dim.6 involving SM fields and right handed neutrino $N_R$~\cite{delAguila:2008ir,Bhattacharya:2015vja,Liao:2016qyd,Chala:2020vqp,Bischer:2020sop}, along with the operators involving the DM and the SM fields plus $N_R$. However, as we shall discuss, for simplicity, in the main analysis we will confine ourselves only to the scalar operators involving different fields.
%%%%%%%%%%%%%%%%%%%
\subsection{Dim.5 $\nu$SMEFT operators}\label{sec:d5nusmeft}
%%%%%%%%%%%%%%%%%%%

At dim.5 the SMEFT contains only the Weinberg operator, while the RHN extension allows additional term

\bea
\mathcal{O}_5 = i\Bigl(H^\dagger H\Bigr)\Bigl(\overline{N^c_R} N_R\Bigr)+\text{h.c}.
\label{eq:op-dim5}
\eea

%%%%%%%%%%%%%%%%%%%
\subsection{Dim.6 $\nu$SMEFT operators}\label{sec:d6nusmeft}
%%%%%%%%%%%%%%%%%%%

In dim.6 the tree-generated (TG) operators can be classified on the basis of number of Higgs fields associated:

\bea\begin{aligned}
&\mathcal{O}_6^{2h}\Bigl[\psi^2\varphi^2\Bigr] = \Bigl(H^\dagger iD^\mu H\Bigr)\Bigl(\overline{N_R^c}\Gamma_\mu N_R\Bigr),\\&\Bigl(\overline{N_R^c}\gamma_\mu e_R\Bigr)\Bigl(\widetilde{H}^\dagger iD^\mu H\Bigr)+\text{h.c}\\& \mathcal{O}_6^{3h}\Bigl[\psi^2\varphi^3\Bigr] = \Bigl(H^\dagger H\Bigr)\Bigl(\overline{\ell_L}\widetilde{H}N_R\Bigr)+\text{h.c}.
\label{eq:op-dim6-h}
    \end{aligned}
\eea

The dim.6 operators involving 1-Higgs {\it e.g.,} $\Bigl(\overline{L}\sigma^{\mu\nu}N_R\Bigr)\widetilde{H}B_{\mu\nu}$, $\Bigl(\overline{L}\sigma^{\mu\nu}N\Bigr)\tau^i\widetilde{H}W_{\mu\nu}^i$ are loop-generated (LG) operators and hence suppressed by $1/16\pi^2$~\cite{delAguila:2008ir}. There are also 4-fermion contact terms in dim.6 level as follows.

\bea\begin{aligned}
&\mathcal{O}_6^{LLRR} = \Bigl(\overline{\ell_L}\gamma_\mu \ell_L\Bigr)\Bigl(\overline{N_R^c}\Gamma^\mu N_R\Bigr),\Bigl(\overline{Q_L}\gamma_\mu Q_L\Bigr)\Bigl(\overline{N_R^c}\Gamma^\mu N_R\Bigr);\\&\mathcal{O}_6^{RRRR} = \Bigl(\overline{e_R}\gamma_\mu e_R\Bigr)\Bigl(\overline{N_R^c}\Gamma^\mu N_R\Bigr),\Bigl(\overline{u_R}\gamma_\mu u_R\Bigr)\Bigl(\overline{N_R^c}\Gamma^\mu N_R\Bigr),\\&\Bigl(\overline{d_R}\gamma_\mu d_R\Bigr)\Bigl(\overline{N_R^c}\Gamma^\mu N_R\Bigr), \Bigl(\overline{d_R}\gamma_\mu u_R\Bigr)\Bigl(\overline{N_R^c}\Gamma^\mu e\Bigr)+\text{h.c.};\\& \mathcal{O}_6^{LRLR} = \Bigl(\overline{\ell_L}N_R\Bigr)\epsilon \Bigl(\overline{\ell_L}e_R\Bigr)+\text{h.c.},\\&\Bigl(\overline{\ell_L}N_R\Bigr)\epsilon \Bigl(\overline{Q_L}d_R\Bigr)+\text{h.c.}, \Bigl(\overline{\ell_L}d_R\Bigr)\epsilon \Bigl(\overline{Q_L}N_R\Bigr)+\text{h.c.};\\& \mathcal{O}_6^{LRRL} =  \Bigl(\overline{Q_L}u_R\Bigr)\Bigl(\overline{N_R^c}\ell_L\Bigr)+\text{h.c}.  
\label{eq:op-dim6-4f}
     \end{aligned}
\eea

To this end we have listed only TG operators which conserve baryon number. Note that as $N_R$ is a Majorana field, hence $\Gamma^\mu\equiv \gamma^\mu\gamma^5$. All the SM left handed quark and lepton $SU\left(2\right)_L$ doublets are denoted by $Q_L,\ell_L$ respectively while the right handed singlet up-type quarks, down-type quarks and leptons are denoted respectively by $u_R,d_R,e_R$. As per~\cite{Liao:2016qyd} there are 52 dim.7 independent operators containing sterile or right handed neutrinos. All these operators carry two units of lepton number, five of which further carry one unit of baryon number, and are thus all non-Hermitian. In the Dirac case, all operators violating lepton number are absent, and we need to replace the Majorana kinetic and mass terms by their Dirac fermion counterparts. Also, in the Dirac case $\mathcal{O}_5$ operator should become $\overline{N}NH^\dagger H$.

% \begin{itemize}
%  \item 
%  
%  \item 
% \end{itemize}

%%%%%%%%%%%%%%%%%%%%%%%%%%%%
\subsection{$N_R$-DM operators}\label{sec:nudmops}
%%%%%%%%%%%%%%%%%%%%%%%%%%%%

The lowest dimensional operator that can be formed out of the DM and the sterile neutrino fields are dim.6 four fermion operators

\bea\begin{aligned}
& \mathcal{O}_{\chi N}^6 = \overline{N_R^c}\Gamma N_R\overline{\chi^c}\Gamma\chi,    
    \end{aligned}\label{eq:dm-rhn-op}
\eea

\noindent where $\Gamma\in\{1,\gamma^\mu\gamma^5,\gamma^5\}$. As one can understand, these operators are suppressed by $1/\Lambda^2$ and hence in the presence of dim.5 DM-SM interaction the production of DM from the heavy neutrinos is negligible and can can safely be ignored. One can certainly write DM-SM interactions following Eq.~\eqref{eq:dm-rhn-op} by replacing the right handed neutrino field with the SM fermion fields. But as we are considering only DM scalar bilinears, hence in a full SM basis such DM-SM operators shall not arise~\cite{Barman:2020plp}. Thus, the sole source of DM production in dim.6 is via the 4-fermion interaction involving heavy neutrinos. With this, now we have all the necessary ingredients to address baryogenesis via (non-thermal) leptogenesis from the decay of the heavy neutrinos together with DM genesis via freeze-in in the early universe.

% interaction in Eq.~\eqref{eq:dm-rhn-op} or in other words annihilation of the heavy neutrinos.

%%%%%%%%%%%%%%%%%%%%%%%%%%
\section{Non-thermal Leptogenesis}
\label{sec:leptogen}
%%%%%%%%%%%%%%%%%%%%%%%%%%

In the standard vanilla leptogenesis~\cite{Buchmuller:2004nz} scenario the right-handed neutrinos are produced thermally by scattering processes in the thermal bath or are assumed to be initially in thermal equilibrium. The lower bound on the mass of the lightest right-handed neutrino, in turn, puts a lower bound on the reheat temperature: $M_1\gtrsim 10^{10}~\rm GeV$~\cite{Davidson:2002qv, Buchmuller:2005eh}. Such a large reheat temperature can lead to gravitino overproduction in supersymmetric scenarios~\cite{Kawasaki:2004qu}. An elegant alternative to avoid such a catastrophe is to consider non-thermal production of the right handed neutrinos. In the present case we assume the inflaton $\varphi$ decays exclusively to a pair of RHNs~\cite{SENOGUZ20046,HahnWoernle:2008pq} via

\bea
\mathcal{L}\supset y_\varphi\varphi \overline{N_R^c} N_R
\eea

where $y_\phi$ is the corresponding coupling strength\footnote{Here we consider the inflaton-Higgs coupling to be absent.}. The decay width for such a process:

\bea
\Gamma_\varphi\simeq\frac{y_\varphi^2}{4\pi} M_\varphi
\eea

\noindent After the inflaton condensate has decayed away, the heavy neutrinos dominate the energy density of the universe. Subsequently, their out-of-equilibrium decay generates the required lepton asymmetry, and reheat the universe since their decay products, SM lepton and Higgs doublets, thermalise quickly. Thus, the decay of the heavy right handed neutrinos not only triggers leptogenesis but produce the thermal bath as well. This leads to interesting consequences contrary to the usual cases where the radiation is produced directly from inflaton decay. Assuming that the energy stored in in the inflaton condensate is {\it instantaneously} transformed into radiation one finds

\bea\begin{aligned}
& T_\text{RH} = \Biggl(\frac{90}{8\pi^3 g_\star}\Biggr)^{1/4}\sqrt{\Gamma_\varphi M_\text{pl}}     
    \end{aligned}\label{eq:inst-inf-decy}
\eea

\noindent which we shall use to parametrize the inflation-heavy neutrino coupling $y_\varphi$ in one of our scenarios, where the Hubble parameter at the end of reheating is given by~\cite{Kaneta:2019zgw}

\bea
H\left(T_\text{RH}\right) = \Biggl(\frac{8\pi^3 g_\star}{90}\Biggr)^{1/2}\frac{T_\text{RH}^2}{M_\text{pl}}.
\eea

It is important to note here the instantaneous decay of inflaton energy density to radiation energy density is an assumption which may or may not always hold. As a result, we shall treat two cases separately: (i) {\it instantaneous} inflaton decay where Eq.~\eqref{eq:inst-inf-decy} holds and the Yukawa coupling $y_\varphi$ is parametrized via the reheat temperature $T_\text{RH}$ and (ii) {\it non-instantaneous} inflaton decay where the Yukawa coupling $y_\varphi$ is a free parameter. In the latter case $T_\text{RH}$ is defined as the temperature where the inflaton energy density becomes equal to the radiation energy density and can be determined by solving the coupled Boltzmann equations as we shall elaborate \cite{Garcia:2020eof}. In the later case, however, it is not possible to establish an analytical relation between $y_\varphi$ and $T_\text{RH}$.

It is possible to define a reheating temperature for the reheating process initiated by the RHNs in the same spirit we define reheating temperature due to inflaton decay~\cite{HahnWoernle:2008pq}

\bea\begin{aligned}
& T_\text{RH}^N = \Biggl(\frac{90}{8\pi^3 g^\star}\Biggr)^\frac{1}{4}\sqrt{\Gamma_N M_\text{pl}}     
    \end{aligned}
\eea

% We also assume a mass hierarchy amongst the three right handed neutrinos: $M_{2,3}\gg M_1$ which implies that the lightest heavy neutrino $N_1$ with mass $M_1$ contributes dominantly to the asymmetry generation.

\noindent where $\Gamma_N$ is the decay width in the RHN rest frame. As advocated earlier, we are typically interested in the non-thermal leptogenesis scenario where we assume $T_\text{RH}<M_1$. For the choice of parameters which gives rise to the right DM abundance with observed baryon asymmetry (together with correct light neutrino mass), we find that $\Gamma_\varphi\ll\Gamma_N$ always holds as long as $T_\text{RH}<M_1$. 
%As an example, in dim.5 we see that for $T_\text{RH}\simeq 10^{10}~\rm GeV$ it is possible to have right DM abundance with $\Lambda\gtrsim 10^{16}~\rm GeV$ depending on the DM mass. For this particular choice of parameters we find $\Gamma_\varphi\simeq 6.5~\rm GeV$, while $\Gamma_N\simeq 10^7~\rm GeV$ where $M_\varphi=10^{13}~\rm GeV$ and $M_1=10^{12}~\rm GeV$ can produce the right baryon asymmetry. We ensure that this condition always holds in our numerical analysis}. 
This points out that the time span of the right handed neutrino dominated universe is extremely small, and hence the true reheat temperature of the universe is $T_\text{RH}$ due to inflaton decay and not $T_\text{RH}^N$. It is important to note here, as we are considering $M_\varphi\gtrsim 2M_1$, hence the RHNs can be considered to be non-relativistic (NR). Obviously this is an assumption that depends on the inflaton mass. If $M_\varphi\gg M_1$ the produced RHNs are relativistic and their energy density contributes to the radiation energy density. However that is not the case here because of the chosen mass hierarchy, and the heavy neutrinos are non-relativistic. Finally, note that once we have $T_\text{RH}<M_i$ the lepton asymmetry washout processes can be neglected as the thermal bath does not have sufficient energy to produce the RHNs. We suppose that the RHN masses are hierarchical, with $M_1\ll M_2$ similar to vanilla leptogenesis but $M_1>T_\text{RH}$ such that the lightest heavy neutrino $N_1$ with mass $M_1$ contributes dominantly to the asymmetry generation. This is precisely due to the fact that any source of lepton number violation at low scale washes out the asymmetries created due to similar violation at high energy scale. Additionally, in our case, we take $M_\varphi\gtrsim 2M_1$ and hence inflaton decay into the lightest RHN is more favoured kinematically. Thus, if $M_2>M_\varphi/2$, then the inflaton only decays exclusively into two $N_1$. 

%This ensures the ineffectiveness of the inverse decay whereas scattering processes involving RHNs will have additional Boltzmann suppression. More precisely, the washout factor is proportional to $e^{-z}$ with $z=M_1/T_\text{RH}$~\cite{SENOGUZ20046} and can be neglected for $z\gtrsim10$.%

% \textcolor{red}{Also, we assume $N_3$ is superheavy and practically decoupled. Why $N_3$ specifically? I thought we consider both $N_2, N_3$ decoupled from lepton asymmetry generation. Otherwise, we need to solve BE for $N_2$ as well.}

%%%%%%%%%%%%%%%%%%%%%
\subsection{Light neutrino mass and Casas-Ibarra parametrization}\label{sec:ci}
%%%%%%%%%%%%%%%%%%%%%

The extension of the SM particle spectrum with SM singlet RHNs allows us to write its Yukawa interaction with the SM lepton doublet and Higgs whereas at the same time, being a neutral SM singlet one cannot forbid its bare mass term which violates the lepton number (an essential ingredient for leptogenesis) by two units. The moment neutral component of the SM Higgs doublet acquires a vacuum expectation value (VEV) leading to the spontaneous breaking of the gauge symmetry, neutrinos which were massless in the SM obtains a Dirac mass which can be written as

\bea
m_D= \frac{y_{N}}{\sqrt{2}}v.
\eea

Here, $v=246$ GeV denotes the VEV of the SM Higgs field. The presence of this Dirac mass together with the bare mass term of the heavy RHNs can explain the non-zero light neutrino masses through \textit{Type-I seesaw}~\cite{GellMann:1980vs, Mohapatra:1979ia}. Here, the light neutrino masses can be expressed as,

\bea
m_{\nu}\simeq m_DM^{-1}_N(m_D)^T,
\label{NM}
\eea

with $M_N$ being the mass of the heavy RHN. The mass eigenvalues and mixing are then obtained by diagonalising the light neutrino mass matrix as

\bea
m_{\nu}= U^*(m_{\nu}^d)U^{\dagger}
\eea

with $m_{\nu}^d={\rm diag}(m_1,m_2,m_3)$ consisting of mass eigenvalues and $U$ being the Pontecorvo-Maki-Nakagawa-Sakata (PMNS) matrix~\cite{Zyla:2020zbs} (the charged lepton mass matrix is considered to be diagonal). These discussions suggest the importance of the Yukawa coupling in generating the neutrino masses. The same Yukawa couplings also play a non-trivial role in determining the lepton asymmetry of the universe as they dictate the decay width of RHNs into leptons. In order to obtain a complex structure of the Yukawa coupling which is essential from the leptogenesis point of view we use the well-known Casas-Ibarra (CI) parametrization~\cite{Casas:2001sr}. Using the CI parametrization one can write Yukawa coupling as,

\bea
y_N = \frac{\sqrt{2}}{v}\,U\sqrt{m_{\nu}^d}\,\mathbb{R}^T\,\sqrt{M}
\label{CI}
\eea

where $\mathbb{R}$ is a complex orthogonal matrix $\mathbb{R}^T \mathbb{R} = I$, which we choose as

\begin{align}
\mathbb{R} =
\begin{pmatrix}
0 & \cos{z} & \sin{z}\\
0 & -\sin{z} & \cos{z}
\end{pmatrix}\label{eq:rot-mat}
\end{align} 

where $z=a+ib$ is a complex angle. In order to obtain $m_{\nu}^d={\rm diag}(m_1,m_2,m_3)$, we
consider the lightest neutrino mass eigenvalue to be vanishingly small with $m_1 = 0~ (m_1 < m_2 < m_3 )$ for normal hierarchy of the neutrino masses. Then $m_{\nu}^d$ is calculable using the best fit values of solar and atmospheric mass alignedtings obtained from the latest neutrino oscillation data~\cite{Zyla:2020zbs}. Now, the elements of Yukawa coupling matrix, $y_N$ for a specific value of $z$ can be obtained for different choices of the heavy neutrino masses. The complex angle $z$ can be chosen in a way that enhances the CP asymmetry while keeping the Yukawa couplings within perturbative limits. The CP asymmetry parameter is defined as

\bea\begin{aligned}
& \epsilon_i = \frac{\sum_j\Gamma\left(N_i\to\ell_j H\right)-\sum_j\Gamma\left(N_i\to\overline{\ell_j}~ \overline{H}\right)}{\sum_j\Gamma\left(N_i\to\ell_j H\right)+\sum_j\Gamma\left(N_i\to\overline{\ell_j}~ \overline{H}\right)}      
    \end{aligned}
\eea

Neglecting flavour effects we can write the final expression for the CP-asymmetry as~\cite{Buchmuller:2004nz,Davidson:2008bu}

% \bea\label{eq:cpasym}
% 	\epsilon_1&\simeq&\frac{3}{16\pi} \sum_{m \neq 1} \frac{\text{Im}[(y_N^\dagger y_N)^2_{1m}]}{(y_N^\dagger y_N)_{11}}\left(\frac{M_1}{M_m}\right)\,
% \eea

\bea\label{eq:cpasym}
	\epsilon_1&\simeq&\frac{3}{16\pi} \frac{\text{Im}[(y_N^\dagger y_N)^2_{12}]}{(y_N^\dagger y_N)_{11}}\left(\frac{M_1}{M_2}\right)\,
\eea

which occurs due to the decay of the lightest right handed neutrino $N_1$. Note that since we are working in two RHN set-up, hence no summation is involved on the right hand side of above equation. Although $N_2$ decay can also generate lepton asymmetry, in principle, we consider the asymmetry generated by $N_2$ decay or any pre-existing asymmetry to be negligible due to strong washout effects due to entropy injection. Finally, we would like to give an explicit structure for the complex Yukawa matrix $y_N$ for the set of parameters: $M_1=10^{12}~\rm GeV$, $M_2=10M_1$ with $a=0.3,b=0.04$ that provides the desired CP asymmetry satisfying light neutrino mass
%As an example, for $M_1=10^{12}~\rm GeV$ and $M_2=10M_1$, using the form of the rotation matrix given in Eq.~\eqref{eq:rot-mat}, we obtain the following structure for the Yukawa matrix from CI parametrization

\begin{align}
y_N =
\begin{pmatrix}
0.0079+0.0011 i & -0.0109+0.0108i\\
0.0174+0.0017i & 0.0517-0.0020i\\
-0.0009+0.0022i & 0.0707-0.0004i
\end{pmatrix}
\end{align} 

%with $a=0.3,b=0.04$ this gives rise to the desired baryon asymmetry.

%%%%%%%%%%%%%%%%%%%%%%%%%%
\section{Dark matter production via UV freeze-in}
\label{sec:dm-uv-yld}
%%%%%%%%%%%%%%%%%%%%%%%%%%

As the SM bath is getting produced from the decay of the lightest right handed neutrino, the DM can be produced from the thermal bath assuming effective interactions of the form $\overline{\chi}\chi H^\dagger H$ in dim.5. This essentially indicates that the DM is getting produced out-of-equilibrium via {\it freeze-in} from the thermal bath through a dim.5 interaction, while at dim.6 it can also be produced directly from RHNs following the interaction in Eq.~\eqref{eq:dm-rhn-op}. Since in the presence of dim.5 operators, higher dimensional interactions (of dimension $d>5$) are naturally suppressed hence we will show the effect of dim.5 and dim.6 interactions on DM production individually. Now, the key requirement for freeze-in~\cite{Hall:2009bx,Bernal:2017kxu} DM production is to assume that DM was not present in the early universe. The DM is then produced via annihilation and/or decay of the particles in the thermal bath \footnote{As we discuss in details, for dim.6 interactions, DM is produced from RHNs which are not part of the thermal bath but is present in the universe due to inflaton field decaying into them.}. Due to the extremely feeble coupling of the DM with the visible sector particles the DM never really enters into thermal equilibrium. Such feeble interaction is ensured due to higher dimensional operators suppressed by a large cut-off scale, as mentioned earlier. Now, the Boltzmann equation (BEQ) for DM production via annihilation or $2\to 2$ processes of the form $a,b\to1,2$ is then given by~\cite{Edsjo:1997bg}:

\begin{align}
\dot{n_1}+3 \mathcal{\mathcal{H}} n_1 & = \int d\Pi_1 d\Pi_2 d\Pi_a d\Pi_b \left(2\pi\right)^4 \delta^4\left(p_1+p_2-p_a-p_b\right)\left|\mathcal{M}\right|_{a,b\to 1,2}^2 f_a f_b,     
\label{eq:beq-ann}
\end{align}

where $d\Pi_j=\frac{d^3 p_j}{2E_j\left(2\pi\right)^3}$ are Lorentz invariant phase space elements, and $f_i$ is the phase space density of particle $i$ with corresponding number density being

\bea
n_i = \frac{g_i}{\left(2\pi\right)^3}\int d^3p f_i, 
\label{eq:num-density}
\eea

with $g_i$ is the internal degrees of freedom (DOF). Here we have made two crucial assumptions: the initial abundance for particle $1$ is negligible such that $f_1\approx 0$ initially and we also neglect Pauli-blocking/stimulated emission effects, i.e. approximating $1\pm f_i\approx 1$. The BEQ in Eq.~\eqref{eq:beq-ann} can be rewritten as an integral with respect to the centre of mass energy as~\cite{Hall:2009bx,Elahi:2014fsa}

\begin{align}
\dot{n_1}+3 \mathcal{\mathcal{H}} n_1 &\approx \frac{T}{512\pi^6}\int_{s=4m_\chi^2}^\infty ds~ d\Omega P_{ab} P_{12}\left|\mathcal{M}\right|_{ab\to 12}^2\frac{1}{\sqrt{s}} K_1\left(\frac{\sqrt{s}}{T}\right),     
\label{eq:beq-ann-1}
\end{align}

where $P_{ij}=\frac{1}{2\sqrt{s}}\sqrt{s-(m_i+m_j)^2}\sqrt{s-(m_i-m_j)^2}\to \frac{\sqrt{s}}{2}$ in the limit $m_{i,j}\to 0$. The BEQ in terms of the yield or comoving number density $Y_\chi=n_1/s$ ($s$ being the entropy density) can be written in the differential form as

\bea
\begin{aligned}
& -s(T).\mathcal{\mathcal{H}}(T).T.\frac{dY_\chi^{\text{ann}}}{dT}= \frac{T}{512\pi^6}\int_{s=4m_\chi^2}^\infty ds d\Omega P_{ab} P_{12}\left|\mathcal{M}\right|_{ab\to 12}^2\frac{1}{\sqrt{s}} K_1\left(\frac{\sqrt{s}}{T}\right). 
\end{aligned}
\eea
\label{eq:beq-ann-yld}

The total yield then turns out to be

\begin{align}
Y_\chi\left(T\right) &= \frac{1}{512\pi^6}\int_{T}^{T_\text{max}}\frac{dT^{'}}{s(T^{'}).\mathcal{\mathcal{H}}(T^{'})}\int_{s=4m_\chi^2}^\infty ds d\Omega \left(\frac{\sqrt{s}}{2}\right)^2\left|\mathcal{M}\right|_{ab\to 12}^2 \frac{1}{\sqrt{s}} K_1\left(\frac{\sqrt{s}}{T^{'}}\right).
\label{eq:totyld2}
\end{align}

Since we are dealing with pure UV freeze-in case where the DM production is sensitive to high temperature, hence we will ignore masses of all the SM states (as well as the DM) associated with the processes contributing to DM abundance. This is legitimate as long as the reheat temperature and the cut-off scale is large compared to the masses involved. In Eq.~\eqref{eq:totyld2}, we consider the maximum temperature of the universe $T_\text{max}$ to be the reheat temperature $T_\text{RH}$. Now, the reheat temperature is very loosely bounded, as mentioned before. Typically, the lower bound on $T_\text{RH}$ comes from the measurement of light element abundance during BBN, which requires $T_\text{RH}\gtrsim 1.8~\rm MeV$~\cite{Hasegawa:2019jsa}. The upper bound, on the other hand, may emerge from (i) cosmological gravitino problem~\cite{Moroi:1993mb,Kawasaki:1994af} in the context of supersymmetric framework that demands $T_\text{RH}\lesssim 10^{10}~\rm GeV$ to prohibit thermal gravitino overproduction and (ii) simple inflationary scenarios that require at most $T_\text{RH}\sim 10^{16}~\rm GeV$~\cite{Kofman:1997yn,Linde:2005ht} for a successful inflation. In view of this, the reheat temperature of the universe can thus be regarded as a free parameter. 

% Before moving on we would like to mention that even in the absence of direct inflaton-DM coupling, DM can be produced from the radiative inflaton decay accounting for the total observed DM density~\cite{Kaneta:2019zgw}.

%%%%%%%%%%%%%%%%%%%%%%%%%%%%
\section{The Coupled Boltzmann Equations}\label{sec:numeric}
%%%%%%%%%%%%%%%%%%%%%%%%%%%%

% %%%%%%%%%%%%%%%%%%%
% \subsection{The Coupled Equations}
% %%%%%%%%%%%%%%%%%%%

\begin{figure}[htb!]
$$
\includegraphics[scale=0.32]{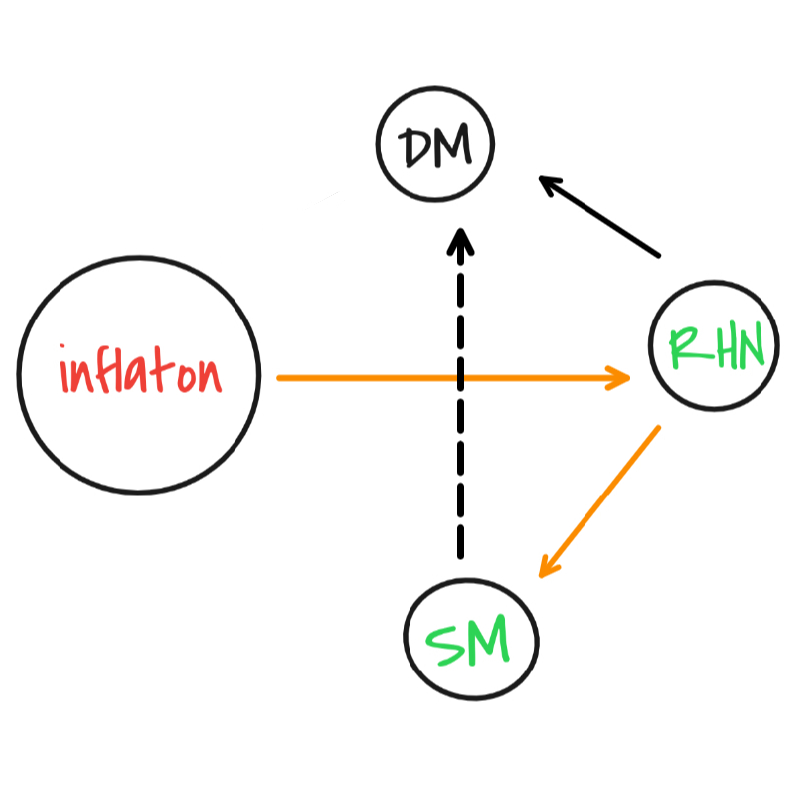}
$$
\caption{A schematic view of the framework for production of different components. In dim.5 there is no interaction between the heavy neutrinos and the DM, while DM-SM interaction is always present for dimensions $d>4$.}\label{fig:scheme}
\end{figure}

Let us recapitulate our scenario so far. For successful leptogenesis, we rely upon the out-of-equilibrium decay of the heavy neutrino $N_1$ that is being produced from the decay of inflaton. The out-of-equilibrium decay eventually produces the observed baryon asymmetry via non-thermal leptogenesis. Since inflaton does not decay into radiation directly, right handed neutrino decay into the SM particles is also responsible for reheating the universe. The DM, on the other hand, is produced from the SM bath via dim.5 DM-SM interaction through the standard UV freeze-in mechanism. In dim.6 the DM can be produced both from the SM bath and from the right handed neutrinos via 4-fermion interaction. Similar to the previous work \cite{Barman:2020plp}, we consider only scalar bilinears for DM and hence DM-SM operators do not appear at dim.6 level. Fig.~\ref{fig:scheme} shows a cartoon depicting the sources of different components of the universe in our scenario based on which the coupled BEQ needs to be formulated. Note that, since there is no direct inflaton-DM coupling, hence the DM can not be produced directly from the inflaton decay. A loop-induced radiative decay of inflaton to a pair of DM is possible in dim.6 due to the presence of $NN\chi\chi$ operator. Such a process will be suppressed by a factor of $\sim y_\varphi^2/16\pi^2\Lambda^4$ and can be neglected as $\Lambda\sim\mathcal{O}\left(10^{14}\right)~\rm GeV$ and $y_\varphi\sim\mathcal{O}\left(10^{-7}\right)$, required to satisfy relic criteria of DM in TeV mass range or below. At dim.5 level no such radiative inflaton decay can arise since we are considering the Higgs-inflaton coupling to be absent. \footnote{DM production via $2\to2$ annihilation of the inflaton mediated via gravitons are studied in~\cite{Ema:2016hlw,Ema:2018ucl,Mambrini:2021zpp,Barman:2021ugy}.} With this, the set of coupled BEQ reads

\bea\begin{aligned}
& \dot{\rho_\varphi} = -3\mathcal{\mathcal{H}}\rho_\varphi-\Gamma_\varphi\rho_\varphi\\&
\dot{\rho_N} = -3\mathcal{\mathcal{H}}\rho_N+\Gamma_\varphi\rho_\varphi-\Gamma_N\rho_N\\&
\dot{n}_{B-L} = -3\mathcal{\mathcal{H}}n_{B-L}+\epsilon\Gamma_N n_N-\Gamma_\text{ID}n_{B-L}\\&
\dot{\rho}_R = -4H\rho_R+\Gamma_N\rho_N\\&
\dot{n}_\chi = -3\mathcal{\mathcal{H}}n_\chi+\langle\sigma v\rangle n_\text{eq}^2
    \end{aligned}\label{eq:cpld-beq1}
\eea

where $\epsilon$ is the CP-asymmetry generated by the lightest RHN decay (as in Eq.~\eqref{eq:cpasym}) and $\Gamma_\text{ID}$ is the inverse decay rate and the DM number density evolution follows from Sec.~\ref{sec:dm-uv-yld}. Here onward we are going to ignore the inverse decay rate. Since the RHNs are assumed to be non-relativistic hence $n_N=\rho_N/m_N$ where in our case $m_N\equiv M_1$. Note that in the BEQ governing the DM abundance we have the equilibrium number density $n_\text{eq}^2$ which is applicable for dim.5 as the DM is produced only from the thermal bath. In case of dim.6 (where we ignore the dim.5 operators), as we will see later, one has to replace the equilibrium number density with the heavy neutrino number density as the RHNs become the source of DM production via Eq.~\eqref{eq:dm-rhn-op}. Next we are going to make appropriate variable transformation in order to solve the set of coupled equations.

%%%%%%%%%%%%%%%%%%%%%%%%
\subsection{Variable transformation \& initial conditions}
%%%%%%%%%%%%%%%%%%%%%%%%

It is convenient to make suitable variable transformation while solving Eq.~\eqref{eq:cpld-beq1}. We make the following set of transformations by scaling the energy and number densities with the scale factor $a$~\cite{HahnWoernle:2008pq,Giudice:1999fb}

\bea\begin{aligned}
& E_\varphi = \rho_\varphi a^3,\\&
E_N = \rho_N a^3,\\&
\widetilde{N}_{B-L} = n_{B-L}a^3,\\&
R = \rho_R a^4,\\&
X = n_\chi a^3.
    \end{aligned}\label{eq:var}
\eea

We also define $$y=\frac{a}{a_I}$$ as the ratio of scale factors and assume $a_I=1$. The factor $a_I$ is chosen as the initial value of the scale factor while $y$ is equivalent to the time variable. With this the Hubble parameter reads

\bea
\mathcal{\mathcal{H}} = \sqrt{\frac{8\pi\left(E_\varphi a_I y+E_N a_I y+R\right)}{3M_\text{pl}^2 a_I^4 y^4}}.
\eea

Instead of the temperature $T$ we use a dimensionless quantity $z=M_1/T$ as a variable and define it as a function of $R$ mentioned in Eq.~\eqref{CI}

\bea
z = \frac{M_1}{T} = \frac{M_1}{\left(30\rho_R/\pi^2 g_\star\right)^\frac{1}{4}} = M_1\Biggl(\frac{\pi^2 g_\star}{30R}\Biggr)^\frac{1}{4}ya_I.
\eea

In terms of theses rescaled variables Eq.~\eqref{eq:cpld-beq1} can be written as

\bea\begin{aligned}
& E_\varphi^{'} = -\frac{\Gamma_\varphi}{\mathcal{H}}\frac{E_\varphi}{y},\\&
E_N^{'} = \frac{\Gamma_\varphi}{\mathcal{H}}\frac{E_\varphi}{y}-\frac{\Gamma_N}{\mathcal{H}y}E_N,\\&
\widetilde{N}_{B-L}^{'} = \frac{\Gamma_N}{\mathcal{H}y}\epsilon\frac{E_N}{M_1},\\&
R^{'} = \frac{\Gamma_N a_I}{\mathcal{H}}E_N,\\&
X^{'} = \frac{1}{16\pi^3\Lambda^2}\frac{y^2}{\mathcal{H}}\Biggl(\frac{M_1}{z}\Biggr)^6.
    \end{aligned}\label{eq:cpld-beq2}
\eea

% +\frac{\pi^2 n_N^\text{eq}}{2\zeta\left(3\right)\left(M_1/z\right)^3}\widetilde{N}_{B-L}\Biggr

% There are several comments in order on the set of equations given in Eq.~\eqref{eq:cpld-beq2}. The denominator in the second term in $\widetilde{N}_{B-L}$ equation appears to quantify the radiation number density that goes as $\sim T^3$ with some prefactors as stated. With the negative sign in front on the right hand side of the $\widetilde{N}_{B-L}$ equation we get a negative asymmetry. \textcolor{red}{These comments don't seem to be in sync with equation (33)}

In the above set of equations all the primes correspond to derivative with respect to $y=a/a_I$.  In obtaining the initial energy density for the inflaton we use the condition $$\Gamma_\varphi=\mathcal{H}\left(a_I\right)$$ that leads to

\bea
E_{\varphi_I} = \frac{3}{8\pi}M_\varphi^2 M_\text{pl}^2. 
\label{eq:inf-init}
\eea

% \noindent where we have made use of the fact that $\Gamma_{\varphi_I}\approx M_\varphi$ at $a=a_I$. 
Excepting for the inflaton, which has an initial energy density, all other quantities are being produced from different sources. Hence we choose the initial values of the variables appearing in Eq.~\eqref{eq:cpld-beq2} as:

\bea\begin{aligned}
& \widetilde{N}_{(B-L)_I}=0;~E_\varphi = E_{\varphi_I};\\& R_I=0;~E_{N_I}=0;~X_I=0.     
    \end{aligned}
\eea

The observed baryon asymmetry depends on the final $B-L$ abundance which can be derived from the redefined $\widetilde{N}_{B-L}$ via

\bea\begin{aligned}
& N_{B-L} =\frac{n_{B-L}}{n_\gamma}=\Biggl[\frac{\pi^4}{60\zeta(3)}\Biggr] \Biggl(\frac{30}{\pi^2 g_{\star\rho}}\Biggr)^{1/4}g_{\star\rho}\tilde{N}_{B-L} R^{-3/4}
    \end{aligned}
\eea

where we have recasted  $n_\gamma=g_\star\zeta(3)T^3/\pi^2$ in terms of $\rho_\gamma=\left(\pi^2/30 \right)g_\star T^4$ and also assumed

\bea
T = \Biggl(\frac{30\rho_R}{\pi^2 g_\star}\Biggr)^{1/4}
\eea

which is allowed since the heavy neutrinos are non-relativistic\footnote{If the produced heavy neutrinos were relativistic then $T = \Bigl(\frac{30\left(\rho_R+\rho_N\right)}{\pi^2 g_\star}\Bigr)^{1/4}$.}. The predicted baryon to photon number ratio has to be compared with the value $\eta_B$ measured at the recombination epoch. It is related to $N_{B-L}$ by~\cite{Buchmuller:2004nz,HahnWoernle:2008pq}

\bea
\eta_B = \Bigl(\frac{a_\text{sph}}{f}\Bigr)N_{B-L}\label{eq:bar-asym}
\eea

where $a_\text{sph}=28/79$ is the fraction of $B-L$ asymmetry converted into a baryon asymmetry by sphaleron processes and $f=N_\gamma^\text{rec}/N_\gamma^\star=2387/86$  is the dilution factor calculated assuming standard photon production from the onset of leptogenesis till recombination~\cite{Buchmuller:2004nz}. The DM abundance can be calculated via

% = 2.75\times 10^8 \frac{m_\chi}{\text{GeV}}\left(n_\chi/s\right)

\bea\begin{aligned}
& \Omega_\chi h^2 =\Biggl(\frac{1.53\times 10^8}{g_{\star s}}\Biggr)\Biggl(\frac{30}{\pi^2g_{\star\rho}}\Biggr)^{1/4}\Biggl(\frac{g_{\star\rho}\pi^4}{60 \zeta\left(3\right)}\Biggr)g_{\star s}m_\chi X\left(y_\infty\right)R\left(y_\infty\right)^{-3/4}
% 2.75\times 10^8m_\chi \Biggl(\frac{30}{\pi^2g_{\star\rho}}\Biggr)^{1/4}\Biggl(\frac{\pi^4}{g_\star\zeta\left(3\right)}\Biggr)XR^{-3/4}.     
    \end{aligned}\label{eq:dm-relic}
\eea

where we have used $s\approx 1.8 g_{\star s}n_\gamma
$~\cite{Kolb:1990vq} to calculate the co-moving entropy density and $y_\infty$ corresponds to the asymptotic value of $y=a/a_I$.

%%%%%%%%%%%%%%%%%%
\section{Results and Discussions}\label{sec:results}
%%%%%%%%%%%%%%%%%%
% Since we are interested in non-thermal leptogenesis, hence we would like to have $T_\text{RH}<M_1$. As a consequence of this, $\Gamma_\varphi\ll\Gamma_N$. Therefore, $N_1$'s always decay instantaneously right after they are produced. Also, in this case $T_\text{RH}$ becomes the real physical reheating temperature and not $T_\text{RH}^N$. This also means that for all practical purposes we can assume $\rho_R=\rho_{N_1}=\Bigl(\pi^2/30\Bigr)g_\star T^4$ as energy density of the heavy neutrino instantaneously gets converted into relativistic degrees of freedom (DOF). 

In this section we would like to discuss and summarize our findings. We mainly focus on dim.5 DM-SM interaction where the DM is produced via UV freeze-in from interaction of the form in Eq.~\eqref{eq:op-dim5}. Later we shall briefly discuss the effect of dim.6 $N_R$-DM operators on DM yield. As clarified in the beginning, we investigate the instantaneous and non-instantaneous inflaton decay cases separately and point out the differences thereof. We begin with the instantaneous inflaton decay scenario and illustrate how the resulting parameter space can give rise to the observed DM relic abundance, together with the correct baryon asymmetry. Then we will move on to the more general non-instantaneous decay case and make a comparative analysis between the two.  

%%%%%%%%%%%%%%
\subsection{Instantaneous inflaton decay}\label{sec:inst-inf-decay}
%%%%%%%%%%%%%%

In the instantaneous inflaton decay approximation Eq.~\eqref{eq:inst-inf-decy} holds and hence the inflaton-heavy neutrino coupling is automatically determined once $T_\text{RH}$ is chosen. Keeping this in mind, let us first look into the dependence of the yield of $\{E_N,E_\varphi,R\}$ on the heavy neutrino mass $M_1$ and reheat temperature $T_\text{RH}$ as these two parameters mainly control the above three quantities while the effective cut-off scale becomes important for the DM relic abundance. As mentioned before, we are interested in pure UV freeze-in production of the DM. This entails to only four-point contact interaction for the DM production via $HH\to\chi\chi$ process where $H$ is the SM-like Higgs doublet. However, one has to take into account of the fact that before electroweak symmetry breaking we have

\bea
H = \begin{pmatrix}
\phi^+ \\ \phi^0      
    \end{pmatrix}
\eea

where the massless Goldstone bosons (GB) $\phi^{\pm,0}$ are the propagating DOFs. One can also note that in dim.5 although we have $NN\to HH$ process via the operator in Eq.~\eqref{eq:op-dim5} which also contributes to the thermal bath. But the out-of-equilibrium $N\to\ell H$ decay will always dominate over pair annihilation of $N$ with the latter being $\Lambda$ suppressed. Hence, heavy neutrino decay is considered to be the dominant source of entropy production. 

% We will now move on to elucidating the effects of heavy neutrino mass and reheat temperature on the inflaton, heavy neutrino, radiation and DM yield.

\begin{figure}[htb!]
  $$
  \includegraphics[scale=0.35]{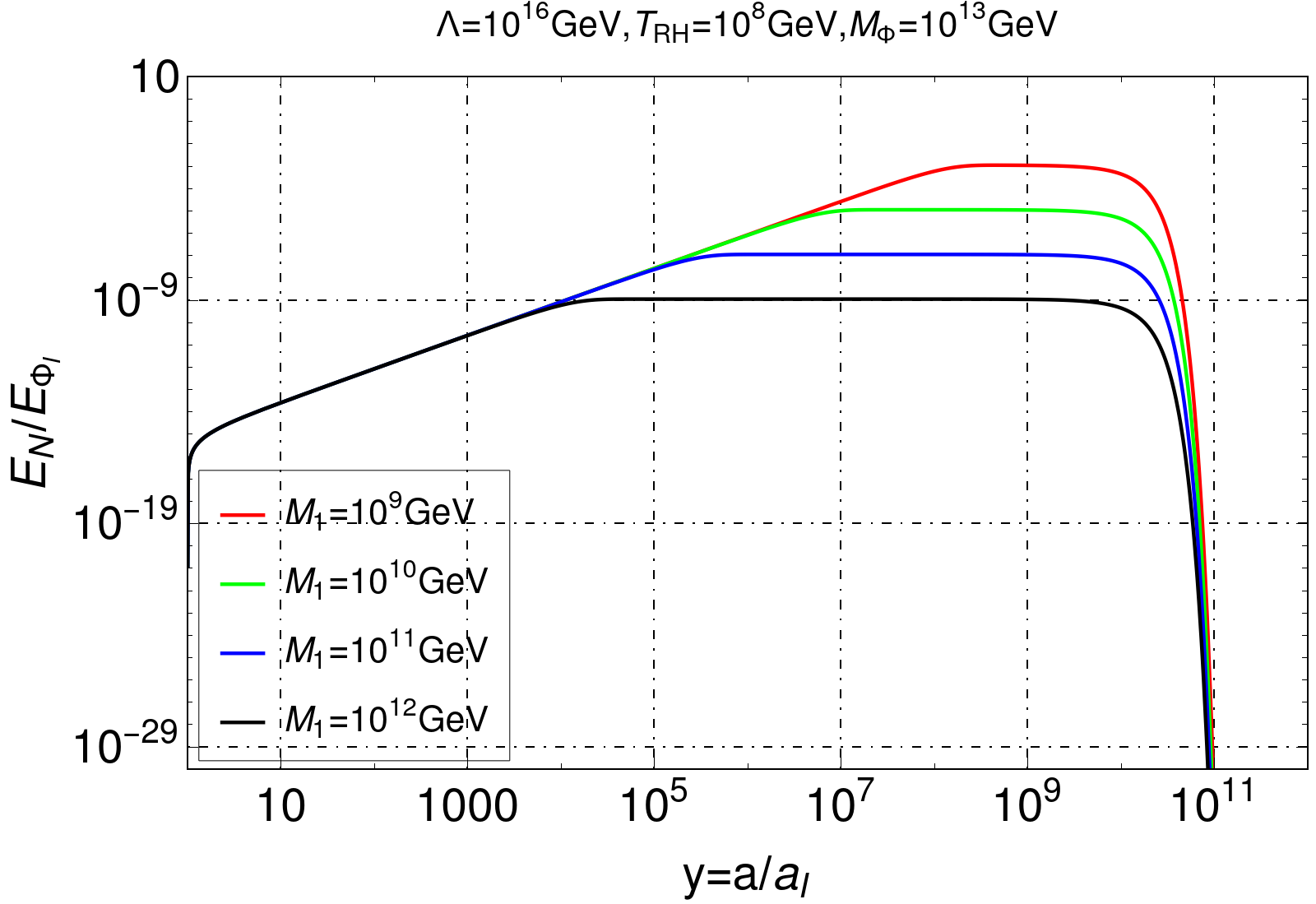}~~
  \includegraphics[scale=0.35]{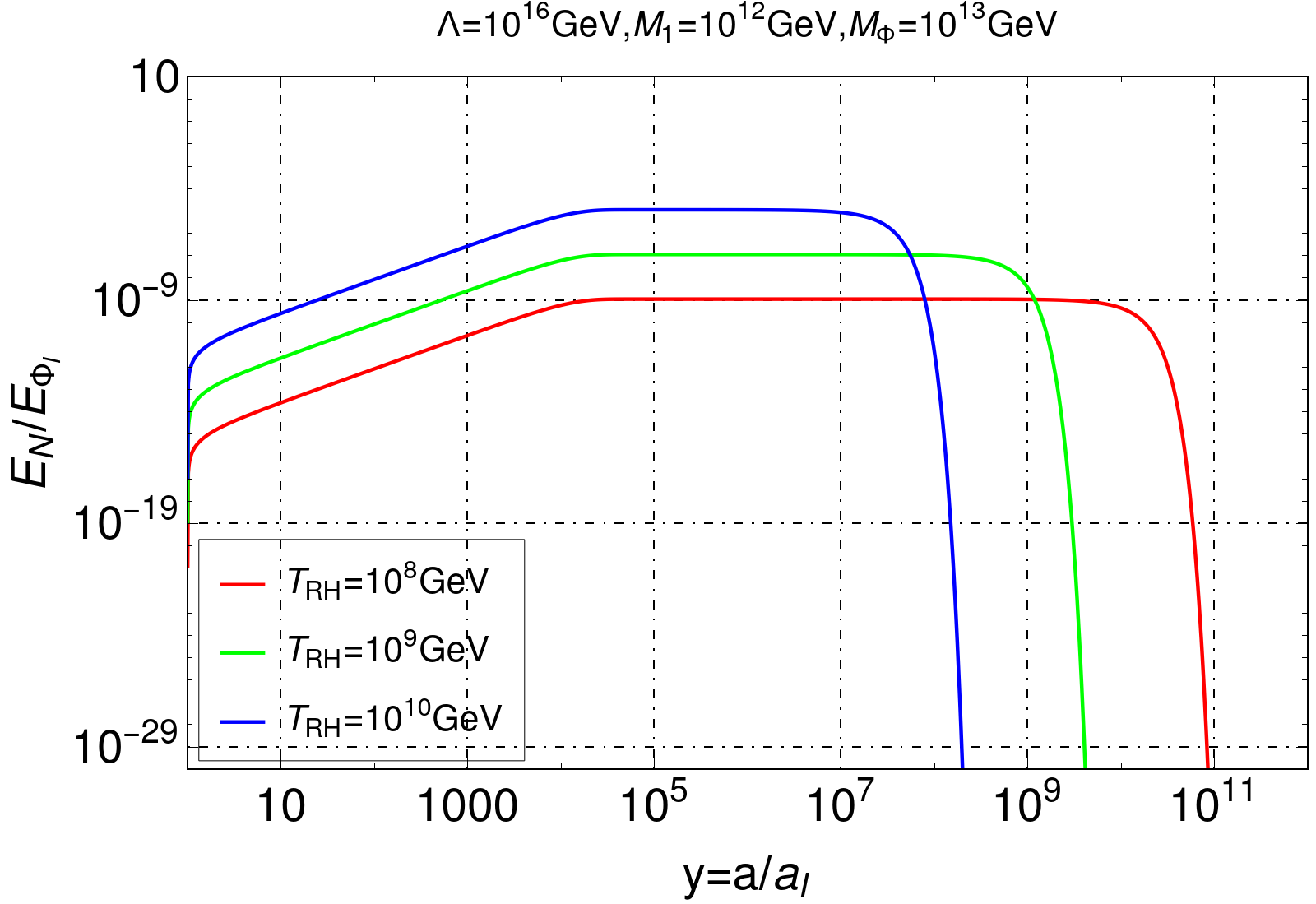}
  $$
  $$
  \includegraphics[scale=0.35]{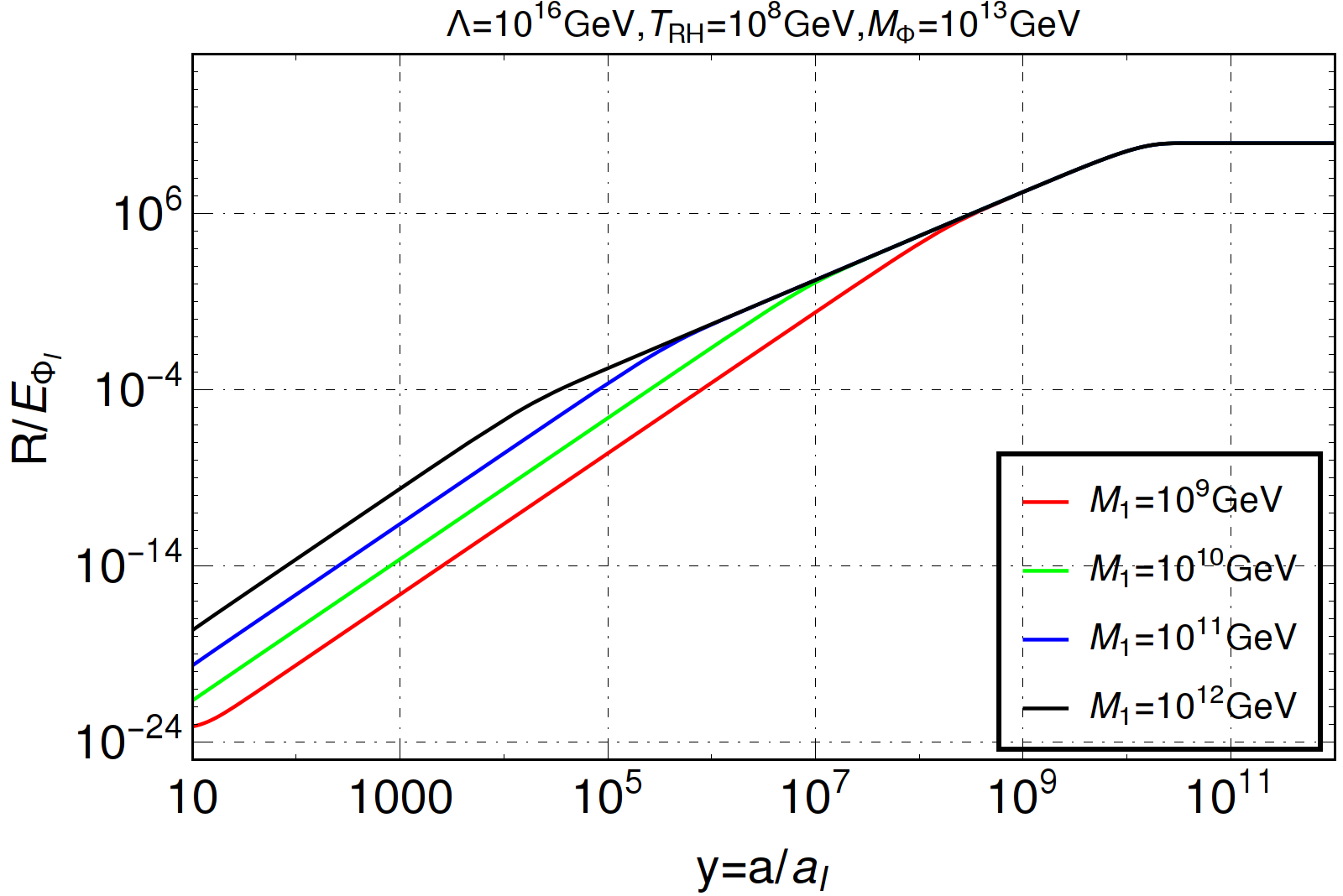}~~
  \includegraphics[scale=0.35]{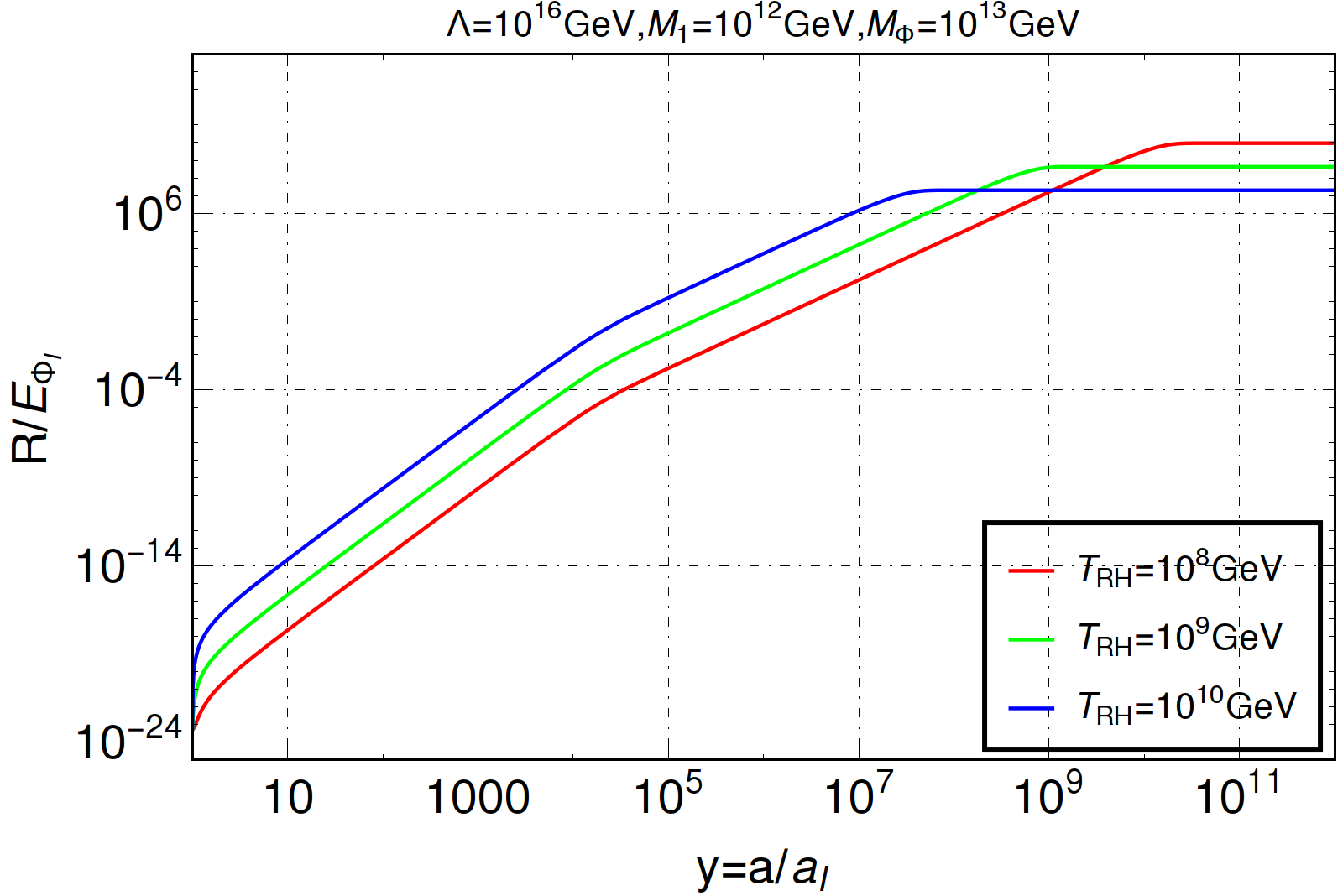}
  $$
  $$
  \includegraphics[scale=0.35]{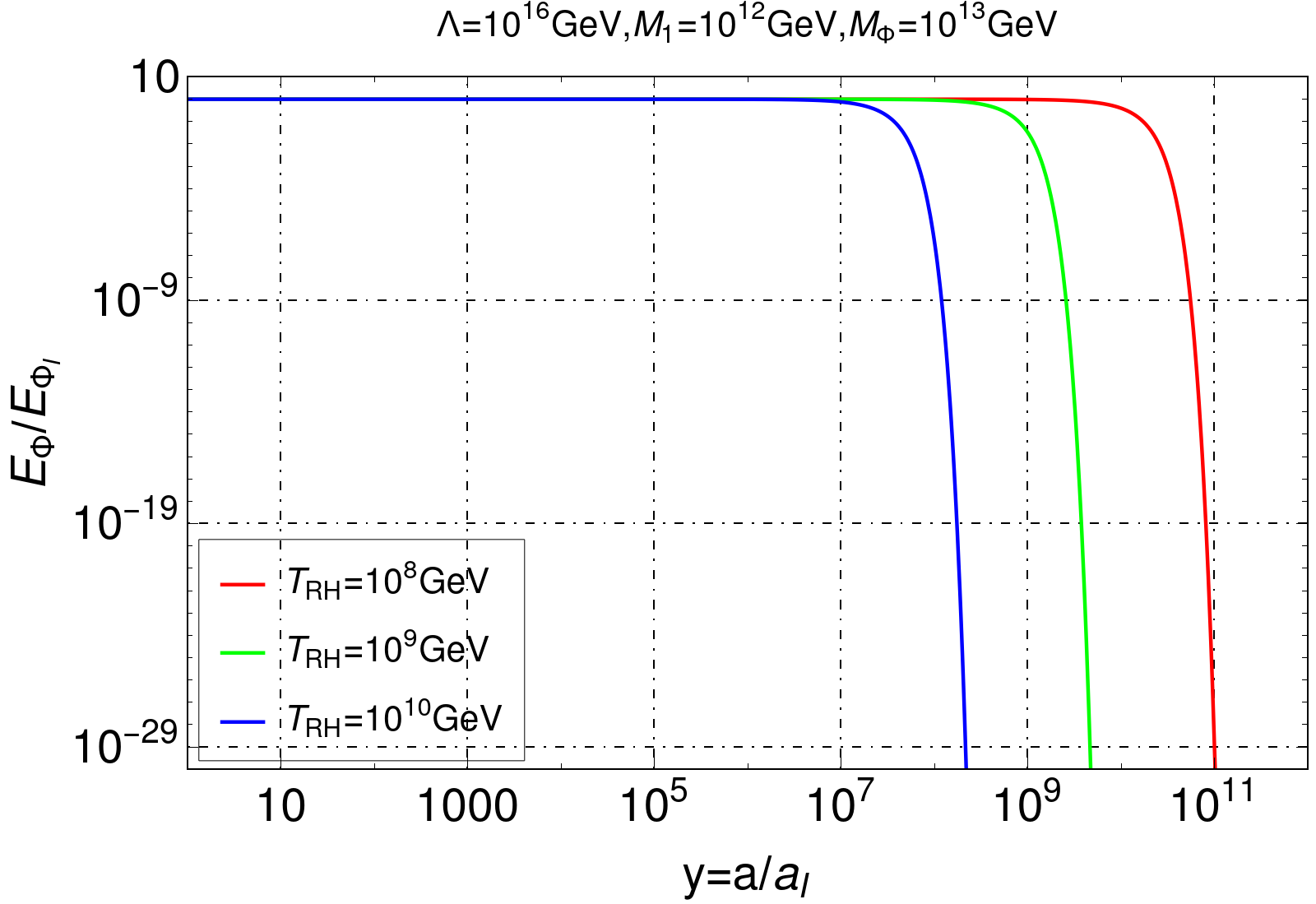}
  $$
  \caption{Top left: Variation of $E_N$ with the ratio of scale factors $y=a/a_I$ for different choices of the heavy neutrino mass $M_1$ for a fixed reheat temperature $T_\text{RH}=10^8~\rm GeV$. Top right: Same as left but for a fixed $M_1=10^{12}~\rm GeV$ with different reheat temperature. Middle left: Radiation $R$ as a function of $y=a/a_I$ for different choices of $M_1$ at a fixed reheat temperature $T_\text{RH}=10^8~\rm GeV$. Middle right: Same as middle left but for a fixed $M_1=10^{12}~\rm GeV$ with different choices of $T_\text{RH}$. Bottom: Variation of $E_\varphi$ with $y=a/a_I$ for a fixed $M_1$ with different choices of the reheat temperature. In all these plots the effective scale is considered to be $\Lambda=10^{16}~\rm GeV$ and an inflaton mass of $M_\Phi=10^{13}$ GeV for illustration.}\label{fig:yldplt}
\end{figure}

\begin{figure}[htb!]
$$
\includegraphics[scale=0.5]{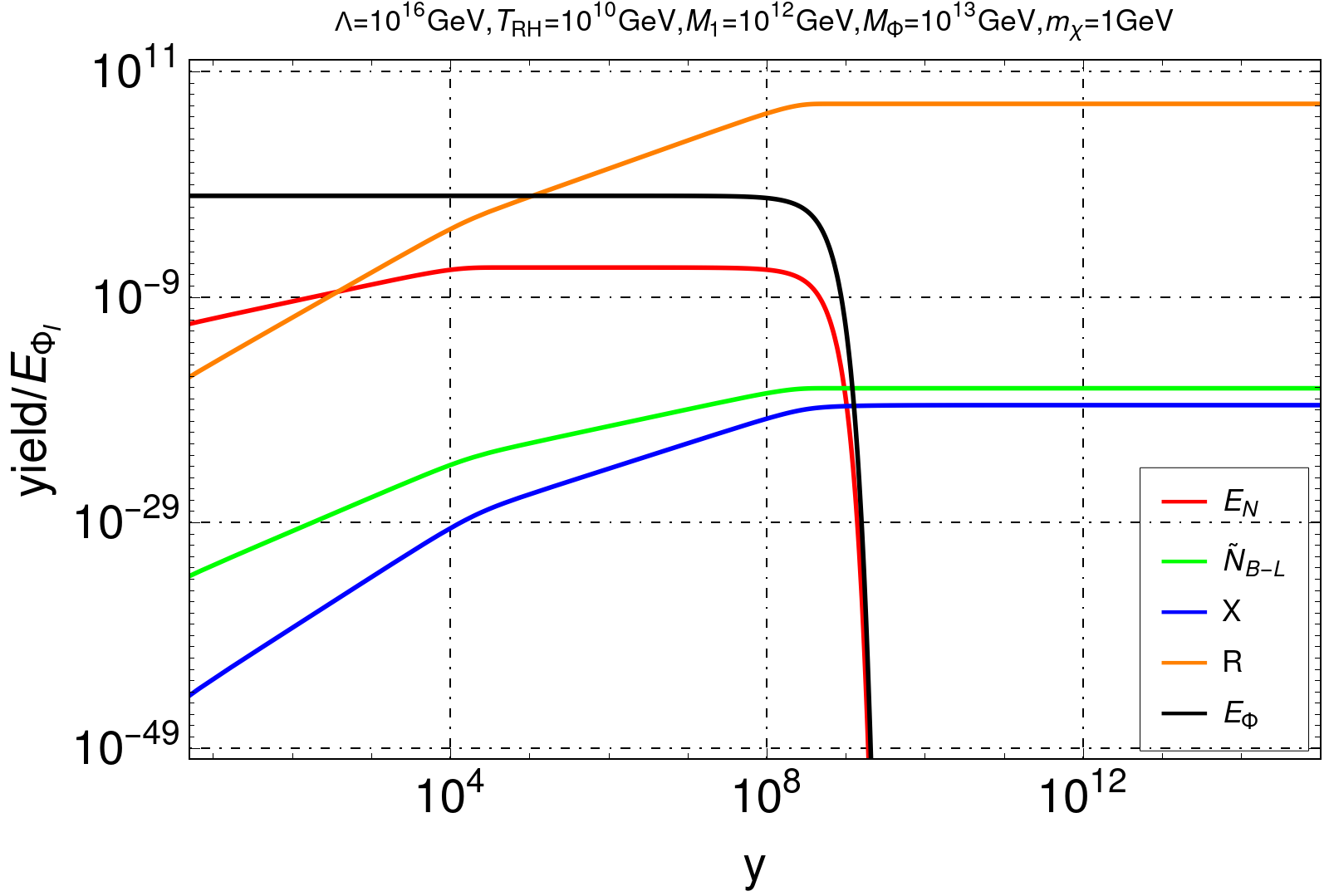}~~
$$
\caption{The evolution of $E_N$, $E_\varphi$, $R$, $\tilde{N}_{B-L}$ and $X$ are shown for $\{\Lambda,M_1,T_\text{RH}\}=\{10^{16},10^{10},10^9\}~\rm GeV$ for a fixed inflaton mass $M_\varphi=10^{13}~\rm GeV$ and DM mass of $m_\chi=1~\rm GeV$. The DM abundance and baryon asymmetry are satisfied at the asymptotic value of the corresponding yield for $X$ and $N_{B-L}$ respectively.}\label{fig:yldplt1}
\end{figure}

% Also, we have chosen $M_2=5M_1$ as the mass of $N_2$ to obey the mass hierarchy amongst the heavy neutrinos.

The variation of $\{E_N,E_\varphi,R\}$ as function of $y=a/a_I$ are demonstrated in Fig.~\ref{fig:yldplt}. We normalize all the yields with the initial inflaton energy such that it turns out to be a dimensionless quantity. In all these plots we have kept the cut-off scale $\Lambda$ and inflaton mass $M_\varphi$ fixed at values $10^{16}~\rm GeV$ and $10^{13}~\rm GeV$ respectively. These choices are purely ad-hoc but ensuring the fact that $\varphi\to N_1 N_1$ decay channel is always open. Later we shall discuss the possible choices of the parameters which are consistent with the observed baryon asymmetry and DM abundance. In all these plots we ensure that the measured light neutrino mass is obtained (by employing Casas-Ibarra parametrization). At dim.5 the thermally averaged $2\to2$ cross section for DM production is given by

\bea
\langle\sigma v\rangle\simeq\frac{1}{\Lambda^2}\label{eq:2to2dim5}
\eea

considering a pure UV freeze-in scenario where we ignore masses of the DM and the SM fields with respect to the cut-off scale and the reheat temperature\footnote{In a more general way one can define~\cite{Garcia:2017tuj,Bernal:2020qyu} the thermally averaged cross-sections as: $\langle\sigma v\rangle=\frac{T^n}{\Lambda^{n+2}}$ with the operator mass dimension $d=5+n/2$ for $n\geq0$ and even.}. With this assumptions incorporated here we see the following features:  

\begin{itemize}

\item In the top left panel of Fig.~\ref{fig:yldplt} we show the variation of $E_N$ with $y=a/a_I$ for four different choices of the heavy neutrino mass $M_1=\{10^9,10^{10},10^{11},10^{12}\}$ GeV in red, green, blue and black respectively for a fixed reheat temperature of $10^8~\rm GeV$. Here we see for a larger $M_1$ the plateau region is wider. This is because for heavier RHN, corresponding Dirac Yukawa coupling is comparatively large, hence its decay starts competing with the production from a much earlier epoch leading to a wider plateau region. Note that all these curves start falling sharply around the same $y=a/a_I$ because of fixed reheat temperature that keeps the inflaton-heavy neutrino decay unchanged as $\Gamma_\varphi\propto T_\text{RH}^2$. As a consequence, when the inflaton decay is completed, the heavy neutrino production stops and $E_N$ falls. For a low mass $N_1$ the yield is larger simply because a smaller $M_1$ gives rise to a comparatively smaller decay width for the RHN. As a result, the decay rate starts competing with the production at a much later epoch leading to a larger $N_1$ yield (and a smaller plateau).   

% Finally, all the curves merge at $y\sim 10^{11}$ because of the fixed Yukawa coupling that gives rise to observed active neutrino mass. For a low mass $N_1$ the yield is larger simply because $\Gamma_\varphi$ is larger and hence more heavy neutrinos can be produced from the inflaton decay. \textcolor{red}{Several contradicting statements: Yukawa coupling is not fixed, but changing with change in $M_1$ due to CI parametrization. $\Gamma_\varphi$ is stated to be constant once due to fixed reheat temperature and then again said to be varying.}

\item The effect of reheat temperature on the heavy neutrino yield is shown in the top right panel of Fig.~\ref{fig:yldplt}. For a fixed $M_1$  the heavy neutrino decay width $\Gamma_N$ is fixed. Now, a larger $T_\text{RH}$ results in larger $\Gamma_\varphi$. This implies that the inflaton decay $\varphi\to NN$ completes earlier for a large $T_\text{RH}$. Thus, for large reheat temperature we see the curves start falling at a lower $y$ showing a smaller plateau since the production of RHN becomes comparable with its decay over a shorter period of time. For large reheat temperature, inflaton decay width is larger and hence its decay into RHNs gets completed at an earlier epoch. For smaller $T_\text{RH}$, on the other hand, the plateau is also wider as the inflaton decay takes place over a longer epoch. Therefore, in this case, the RHN decay can remain comparable to its production over a longer epoch. For large $T_\text{RH}$ the heavy neutrino yield is also larger as $E_N\propto\Gamma_\varphi$.

% {\color{red} This is again because the inflaton decay completes earlier and hence the heavy neutrino production finishes early compared to small $T_\text{RH}$}.

% Since radiation is produced from the decay of the heavy right handed neutrino hence the behaviour of radiation with $y$ for a fixed reheat temperature, as shown i

\item In the middle left panel of Fig.~\ref{fig:yldplt} we show the behaviour of radiation with $y$ for a fixed reheat temperature. Here we see for larger $M_1$ the radiation yield is large as $R\propto\Gamma_N\propto M_1$. However, as the reheat temperature is fixed and so is the inflaton decay width, hence the heavy neutrino production stops as soon as the inflaton decay is completed. This results in saturation of the radiation yield at $y\gtrsim 10^9$. Also, as the entire energy contained in the inflaton field (whose mass is fixed) gets transferred into radiation eventually, all the curves corresponding to different values of $M_1$ merge for higher values of $y$. For a particular  $M_1=10^{12}~\rm GeV$ we see $R$ increases initially upto $y\sim 10^4$ during which the rate of production of the RHN dominates over its decay. Then the radiation yield rises linearly though at a rate slower than initial, as RHN starts to decay till $y\sim 10^{10}$ where the decay rate of the RHN starts overcoming its rate of production. Finally, at $y\gtrsim 10^{10}$ the yield saturates when practically the RHN production has stopped as the inflaton decay is over and the RHN decay into the radiation is also completed. It is possible to make an one-to-one correspondence with the top left panel for the same choice of $M_1$. In the middle right panel we show the effect of varying reheat temperature on the radiation yield. As for large $T_\text{RH}$ the inflaton decays faster to the RHN. This results in an earlier drop in the heavy neutrino yield. As a result, for a fixed $M_1$, the radiation yield from the RHN decay is also smaller. The fact that the heavy neutrino yield drops earlier for larger $T_\text{RH}$, is exactly reflected in the blue curve where we see the radiation starts saturating earlier than the other two. Hence the asymptotic yield of radiation for small $T_\text{RH}$ (and hence smaller $\Gamma_\varphi$) wins over those due to comparatively larger $T_\text{RH}$ (and hence larger $\Gamma_\varphi$).

\item Finally in the bottom panel of Fig.~\ref{fig:yldplt} we show the evolution of the inflaton yield for different choices of the reheat temperature at a fixed inflaton mass. Since $\Gamma_\varphi\propto T_\text{RH}^2$ hence a larger choice of the reheat temperature causes a quick decay of the inflaton to a pair of $N_1$ as shown by the blue curve. The consequence of this on the heavy neutrino yield (and hence on the radiation yield) are already shown in the top and middle panels. As expected, changing $M_1$, keeping $M_1<m_\varphi/2$, without changing $T_\text{RH}$ does not affect the inflaton yield. 

% For the rest of our analysis we are not going to change the inflaton mass but shall always ensure the on-shell decay of the inflaton to heavy neutrinos. 
  
\end{itemize}

\begin{figure}[htb!]
  $$
    \includegraphics[scale=0.4]{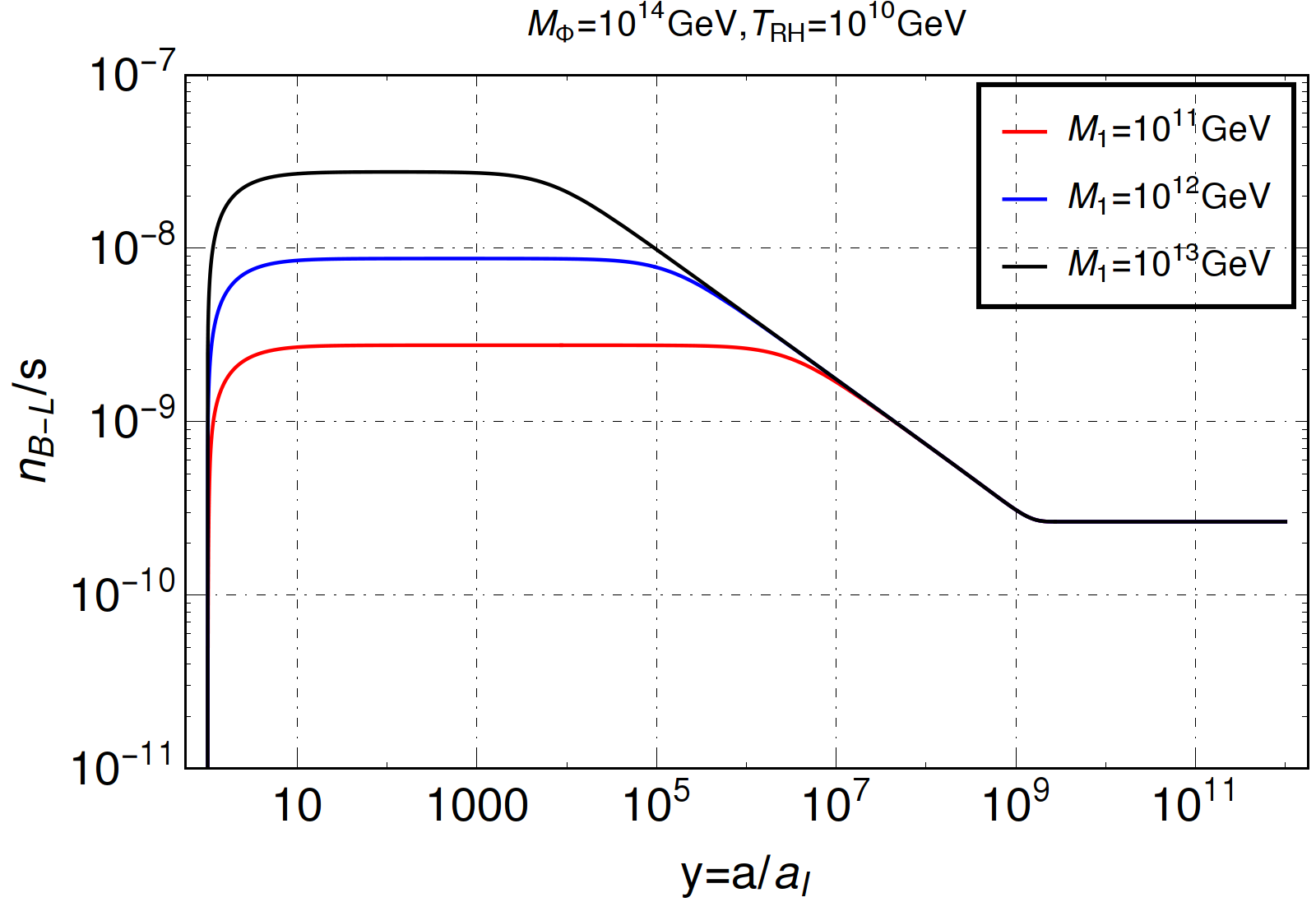}~~\includegraphics[scale=0.4]{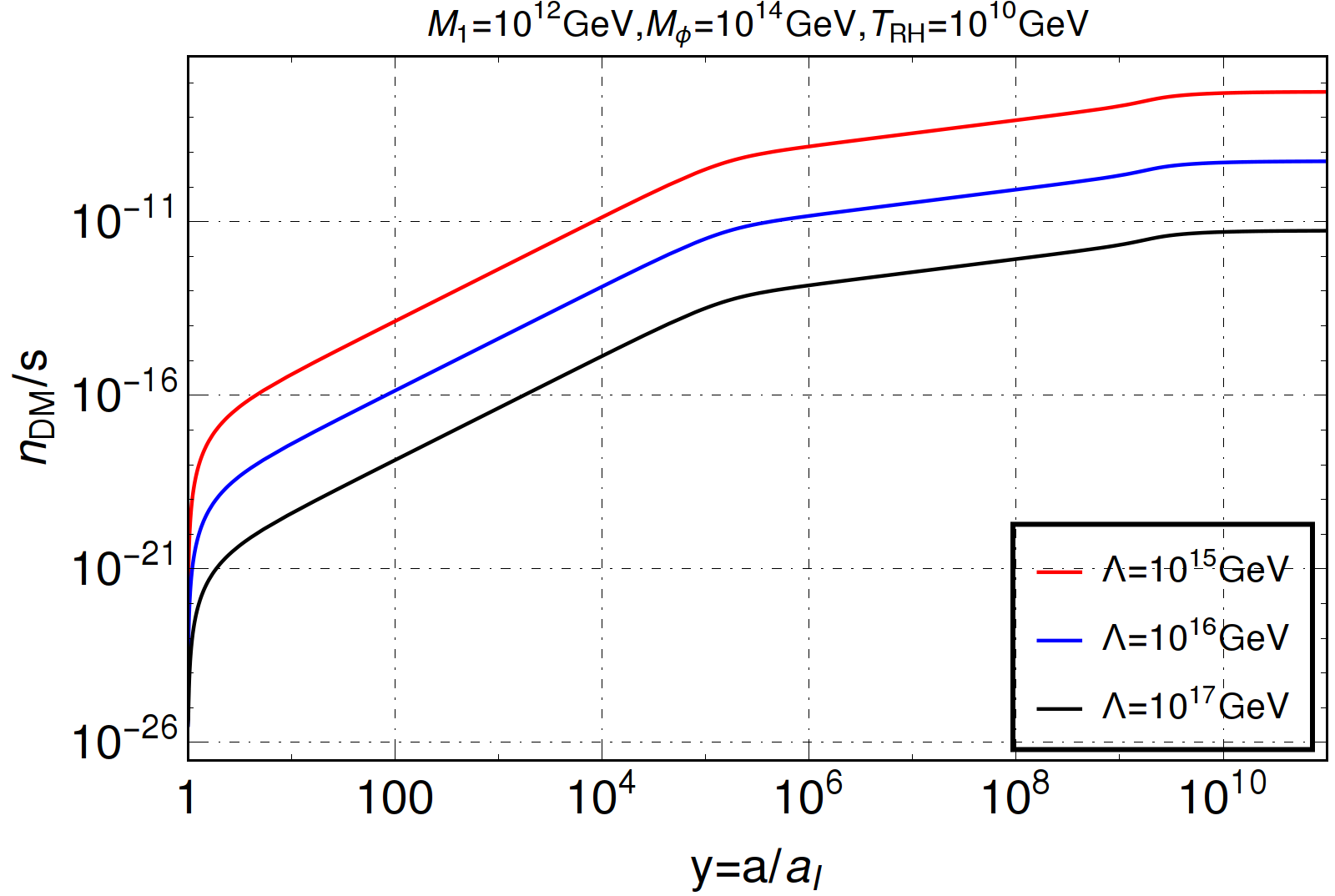}
  $$
  $$
    \includegraphics[scale=0.47]{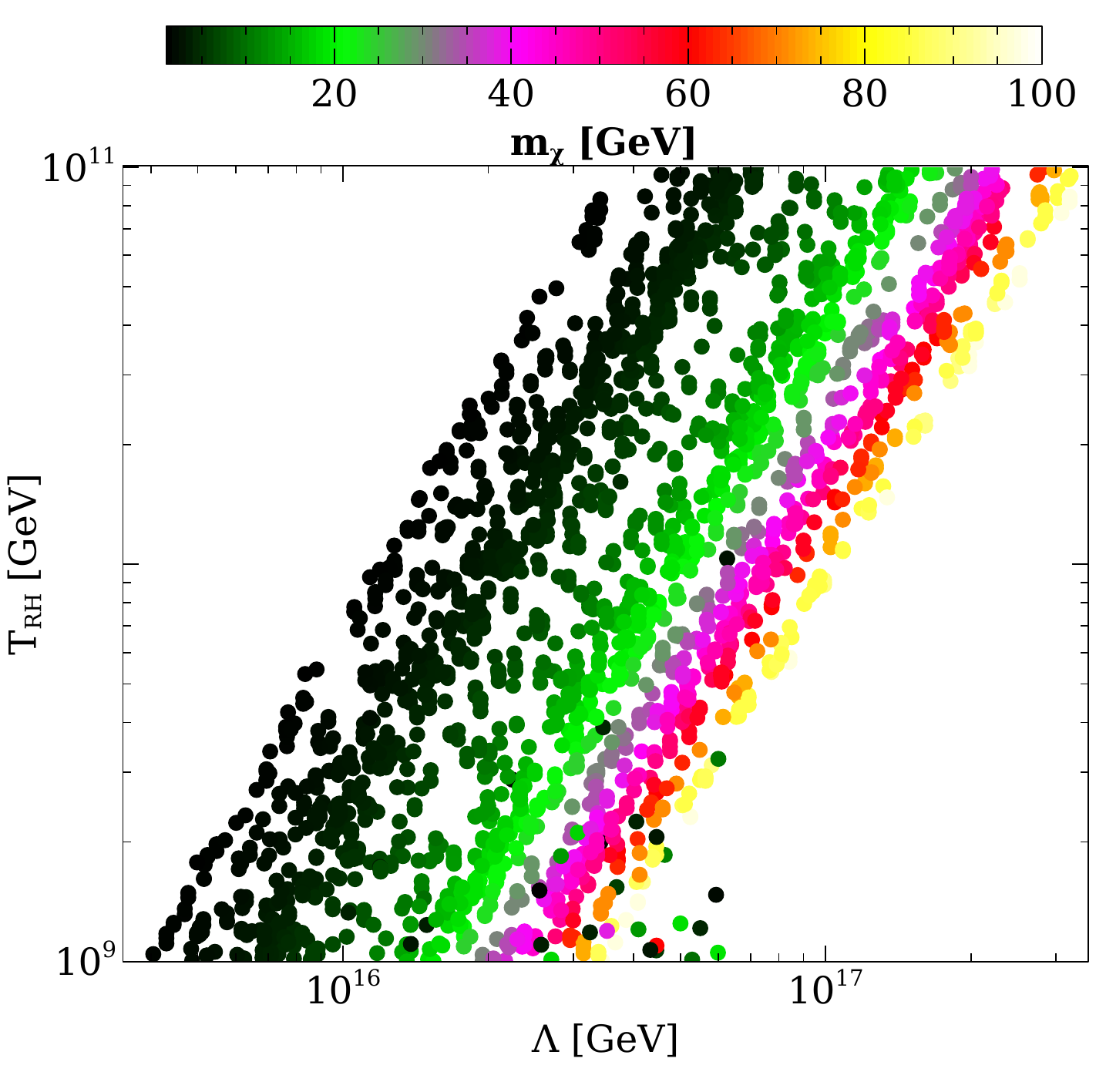}~~\includegraphics[scale=0.45]{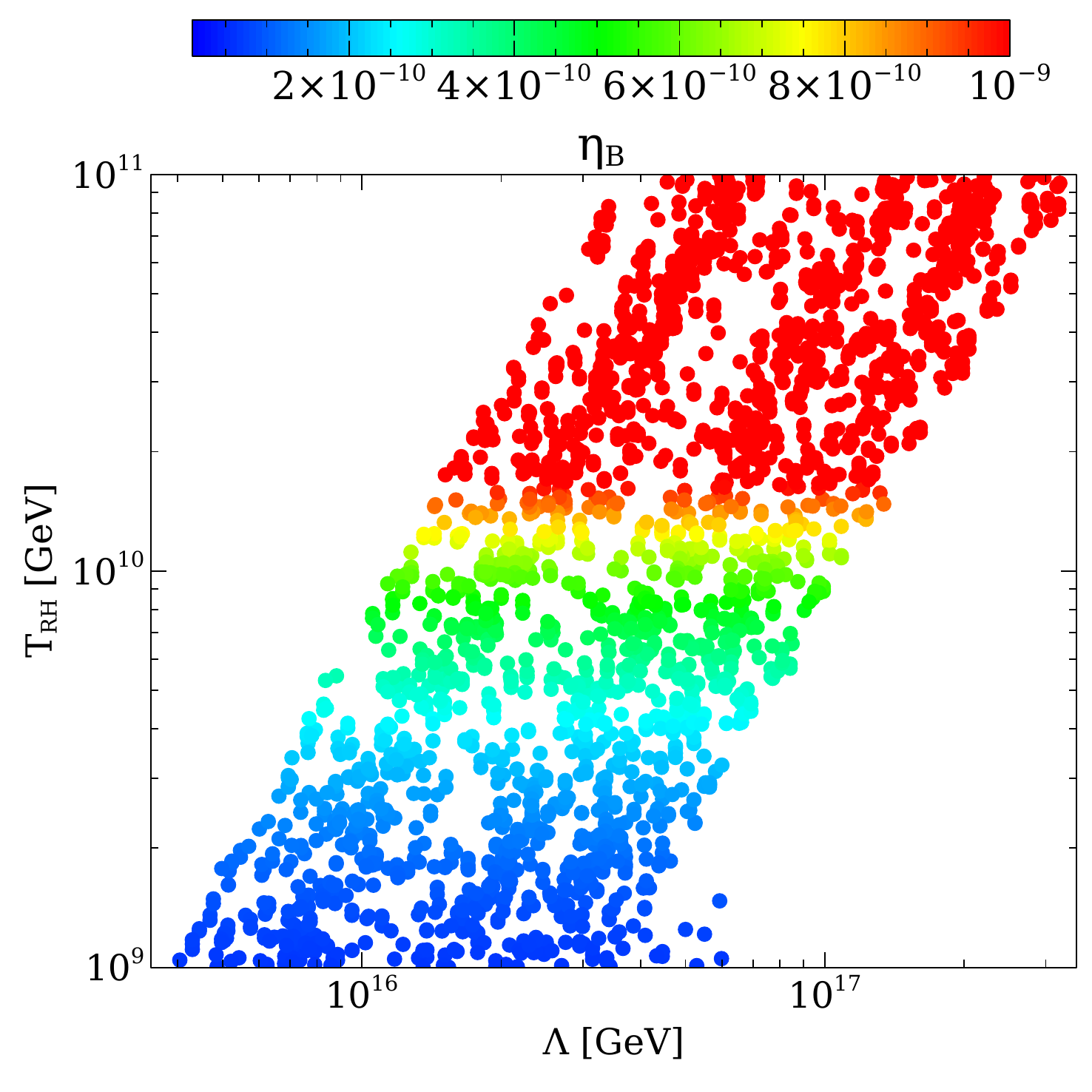}
  $$
\caption{Top Left: Evolution of $n_{B-L}/s$ with $y=a/a_I$ is shown for different choices of the heavy nautrino mass $M_1:\{10^{11},10^{12},10^{13}\}$ GeV in black, red and blue respectively for a fixed reheat temperature $T_\text{RH}=10^{10}$ GeV and inflaton mass $M_\varphi=10^{14}~\rm GeV$. The thick black dashed line shows the $N_{B-L}$ required to produce observed baryon asymmetry via Eq.~\eqref{eq:bar-asym}. Top Right: For the same set of parameters as in the left panel the evolution of DM number density is shown for $\Lambda:\{10^{15},10^{16},10^{17}\}$ GeV in red, blue and black respectively. Bottom Left: Relic density allowed parameter space in $T_\text{RH}-\Lambda$ bi-dimensional plane for $M_1=10^{12}~\rm GeV$ and $M_\varphi=10^{13}~\rm GeV$  where different DM masses are shown in different colours. Bottom Right: Relic density allowed parameter space in $T_\text{RH}-\Lambda$ bi-dimensional plane for $M_1=10^{12}~\rm GeV$ and $M_\varphi=10^{13}~\rm GeV$  where different colours correspond to the baryon asymmetry $\eta_B$.}\label{fig:asym-rel}
\end{figure}

The evolution of the different components with $y=a/a_I$ are shown together in Fig.~\ref{fig:yldplt1}. Here we see trends similar to what we already have observed in Fig.~\ref{fig:yldplt}. The asymptotic value of the yields corresponding to $\tilde{N}_{B-L}$ and $X$ give rise to correct baryon asymmetry and DM abundance (for a DM mass of 1 GeV) at present epoch. For a large reheat temperature one has to choose a larger $\Lambda$ in order to satisfy the DM relic abundance for a fixed DM mass as for pure UV freeze-in the DM yield goes as $\sim M_\text{pl} T_\text{RH}/\Lambda^2$ for the case of dim.5 DM-SM interactions. For DM production via dim.5 DM-SM interaction it has already been shown that $\Lambda\simeq 10^{16}~\rm GeV$ can satisfy the right relic density for $T_\text{RH}\gtrsim 10^9~\rm GeV$~\cite{Barman:2020plp}. As the DM yield is built solely from the thermal bath (radiation) at dim.5 level hence the orange and blue curves, corresponding to radiation and DM yield respectively show the same characteristic evolution. However, as radiation starts to get produced earlier from $N_1$ decay, hence the initial yield for radiation is larger compared to that of the DM. As $\widetilde{N}_{B-L}$ is generated from the heavy neutrino decays, hence the final asymmetry becomes saturated exactly at the point when all the $N_1$ decay is completed. This is visible by comparing the black and green curves. Inflaton decay being the only source of heavy neutrino production the red and black curves fall at the same point. Since RHN has sizeable Yukawa coupling for such high scale Type-I seesaw realization, they can decay quickly leading to quick depletion in their number density once their production from inflaton decay stops. One can also notice that all the curves except for the black one (which corresponds to the inflaton) has a `knee' around $y\sim 10^4$. This occurs due to the entropy injection from the decay of the heavy neutrino when the RHN decay starts becoming comparable to its production. In all these plots the The effect of maximum temperature during reheating has been taken into account as the maximum temperature $T_\text{max}$ corresponds to $y_\text{max}\simeq 1.4$ (see Appendix.~\ref{sec:maxtemp}).
Production of the right amount of $N_{B-L}$ will lead to the observed baryon asymmetry $\eta_B$ following Eq.~\eqref{eq:bar-asym}. The evolution of the $B-L$ asymmetry for dim.5 interaction is shown in the top left panel of Fig.~\ref{fig:asym-rel}. Here we have depicted how $n_{B-L}/s$ evolves with the dimensionless quantity $y=a/a_I$ for three different masses of the heavy neutrino $N_1$. We notice a very interesting dynamics of the yield in this case. As $\Gamma_N\gg\Gamma_\varphi$, hence the heavy neutrinos immediately start decaying after they get produced from the inflaton. As a result, $n_{B-L}/s$ builds up with $y$ at the very beginning $y\sim 1$. The $B-L$ yield remains fixed up to $y\sim 10^6$ (for the blue curve) since in this period the radiation yield is small as one can confirm from the middle left panel plot of Fig.~\ref{fig:yldplt}. Beyond $y\sim 10^6$, as the RHN decay becomes comparable to its production, the radiation builds up faster due to entropy injection from the RHN decay. This results in rapid fall of $n_{B-L}/s$. At $y\sim 10^9$ the RHN decay is completed resulting in a constant $B-L$ whence no more entropy injection is possible. 
% during which the production of $N_1$ from inflaton balances its decay to the SM giving rise to the plateau. This can also be realized from the top left panel plot of Fig.~\ref{fig:yldplt} where we see similar feature. At $y\lesssim 10^7$ the asymmetry falls as the rate of production of $N_1$ from inflaton decay can not compete with the rate of $N_1$ decay into radiation. This again corresponds to the `knee' region in the top left panel of Fig.~\ref{fig:yldplt}. Finally, as all of the $N_1$ decays completely into radiation, $n_{B-L}$ saturates to a constant value which gives rise to the observed baryon asymmetry $\eta_B$. Since $N_{B-L}\propto \Gamma_N$, hence for a larger $M_1$ the $n_{B-L}$ (and hence $\eta_B$) yield is enhanced but the plateau is rather small showing that $N_1$ lifetime is even shorter. \textcolor{blue}{As different values of $M_1$ lead to same asymptotic value of radiation yield with fixed inflaton mass and reheat temperature, here also we notice that the asymptotic value of $B-L$ asymmetry is same.} (\textcolor{red}{Will very different or small choice of $M_1$ lead to a different lepton asymmetry? CP asymmetry parameter will be small in such a case even though reheat temperature and inflaton mass, which govern non-thermal leptogenesis usually, is kept fixed. Also the reason behind falling B-L asymmetry after y=10000 or so is due to entropy injection?}) 
In the top right panel of Fig.~\ref{fig:asym-rel} we show how the DM yield $n_\text{DM}/s$ evolves with $y=a/a_I$ for the same set of parameters as in the top left panel. As the DM number density is a function of $\propto 1/\Lambda^2$, hence a larger $\Lambda$ results in a suppressed DM yield. For a fixed DM mass, this implies, a larger $\Lambda$ calls for a smaller $T_\text{RH}$ in order to satisfy the observed relic abundance since in dim.5 the DM abundance for pure UV freeze-in can approximately be given by: $\Omega_\chi h^2\sim m_\chi T_\text{RH}/\Lambda^2$. One can also notice the bulge at $y\sim 10^5$ which we also observed in the DM yield $X$ in Fig.~\ref{fig:yldplt1} due to entropy injection in the thermal bath from heavy neutrino decay. The DM and the $n_{B-L}$ yield shows an exact complementary behaviour as the DM is always being produced from the thermal bath and finally becomes constant while $n_{B-L}$ is produced from the decay of the RHN which then saturates at $y\to\infty$ once the decay is completed. To produce the right relic abundance for the DM via UV freeze-in at dim.5 level, together with a successful non-thermal leptogenesis, one has to choose the free parameters accordingly. In the bottom panel of Fig.~\ref{fig:asym-rel} we show the parameter space that satisfies correct relic abundance for different DM masses. Here we fix the inflaton mass $M_\varphi=10^{13}~\rm GeV$ and heavy RHN mass $M_1=10^{11}~\rm GeV$, and vary rest of the parameters in the following range

$$
m_\chi: \{1-100\}~~\text{GeV};~~\Lambda:\{10^{15}-10^{17}\}~\text{GeV};~~ T_\text{RH}:\{10^9-10^{11}\}~\text{GeV}.
$$

In the bottom left panel we show the DM relic density allowed region in $T_\text{RH}-\Lambda$ plane where the colouring is done over the DM mass. Here we see, for a fixed cut-off scale $\Lambda$, a smaller DM mass requires larger $T_\text{RH}$ in order to satisfy the observed abundance. This is expected as the relic abundance in dim.5, as we mentioned earlier, varies as $\Omega_\chi h^2\sim m_\chi T_\text{RH}/\Lambda^2$. Hence a large DM mass needs a smaller reheat temperature to attain the right relic density. This also implies, for DM mass $\sim\mathcal{O}\left(\text{TeV}\right)$, $T_\text{RH}<10^{10}~\rm GeV$ to satisfy the relic abundance in order to keep $\Lambda<M_\text{pl}$. It is possible to achieve the observed baryon asymmetry $\eta_B$ within this range of $T_\text{RH}$ as evident from the bottom right panel of Fig.~\ref{fig:asym-rel}. Here we show the relic density allowed points in $T_\text{RH}-\Lambda$ plane where the colouring is with $\eta_B$. We see correct baryon asymmetry is obtained for a very narrow range of $T_\text{RH}\sim 10^{10}~\rm GeV$. As $T_\text{RH}$ determines the inflaton decay width, hence a large $T_\text{RH}$ causes a larger inflaton decay width resulting in larger $N_1$ abundance (as one can also notice in the top left panel of Fig.~\ref{fig:yldplt}). This results in a larger $\eta_B$ in turn. Varying $\Lambda$ for a fixed cut-off scale does not affect the baryon asymmetry as one should expect. Note that we have chosen DM mass in the GeV-TeV window, both in benchmark cases as well as in scans. Since DM relic is generated via UV freeze-in, it is possible to produce supermassive non-thermal DM within the present framework by carefully tuning $T_\text{RH}$ and $\Lambda$ such that with the same choice of parameters one can also satisfy the observed baryon asymmetry. For example, in case of dim.5, it is possible to produce 1 PeV DM for $\Lambda\sim 10^{17}~\rm GeV$ and $T_\text{RH}\sim 10^7~\rm GeV$ with $\eta_B$ of the correct order\footnote{Production of supermassive FIMPs together with light neutrino mass generation and leptogenesis have been explored in the context of UV-complete models~\cite{Chianese:2018dsz,Chianese:2019epo}}. 

% This is because $\eta_B\propto\Gamma_N$ which depends on $M_1$, and for a fixed $M_1$ changing $T_\text{RH}$ does not affect $\eta_B$. Thus, we see, for the choice of the cut-off scale, DM mass and reheat temperature, it is possible to get right relic abundance together with the observed baryon asymmetry over a large parameter space for the dim.5 DM-SM interaction.     

% We take two representative values of the DM mass $m_\chi=1$ MeV and 1 GeV to illustrtate. As for a larger DM mass the abundance is larger as expected because $\Omega_\chi h^2\propto m_\chi$ but that can be tamed down to the PLANCK observed value by suitably choosing the effective scale. Hence, one has to choose a larger $\Lambda$ to obtain the correct relic for a larger DM mass. One can also notice the bulge at $y\sim 10^5$ which we also observed in the DM yield $X$ in Fig.~\ref{fig:yldplt1} due to entropy injection in the thermal bath from heavy neutrino decay which is also visible in Fig.~\ref{fig:yldplt1}. 

\begin{figure}
    \centering
    \includegraphics[scale=0.5]{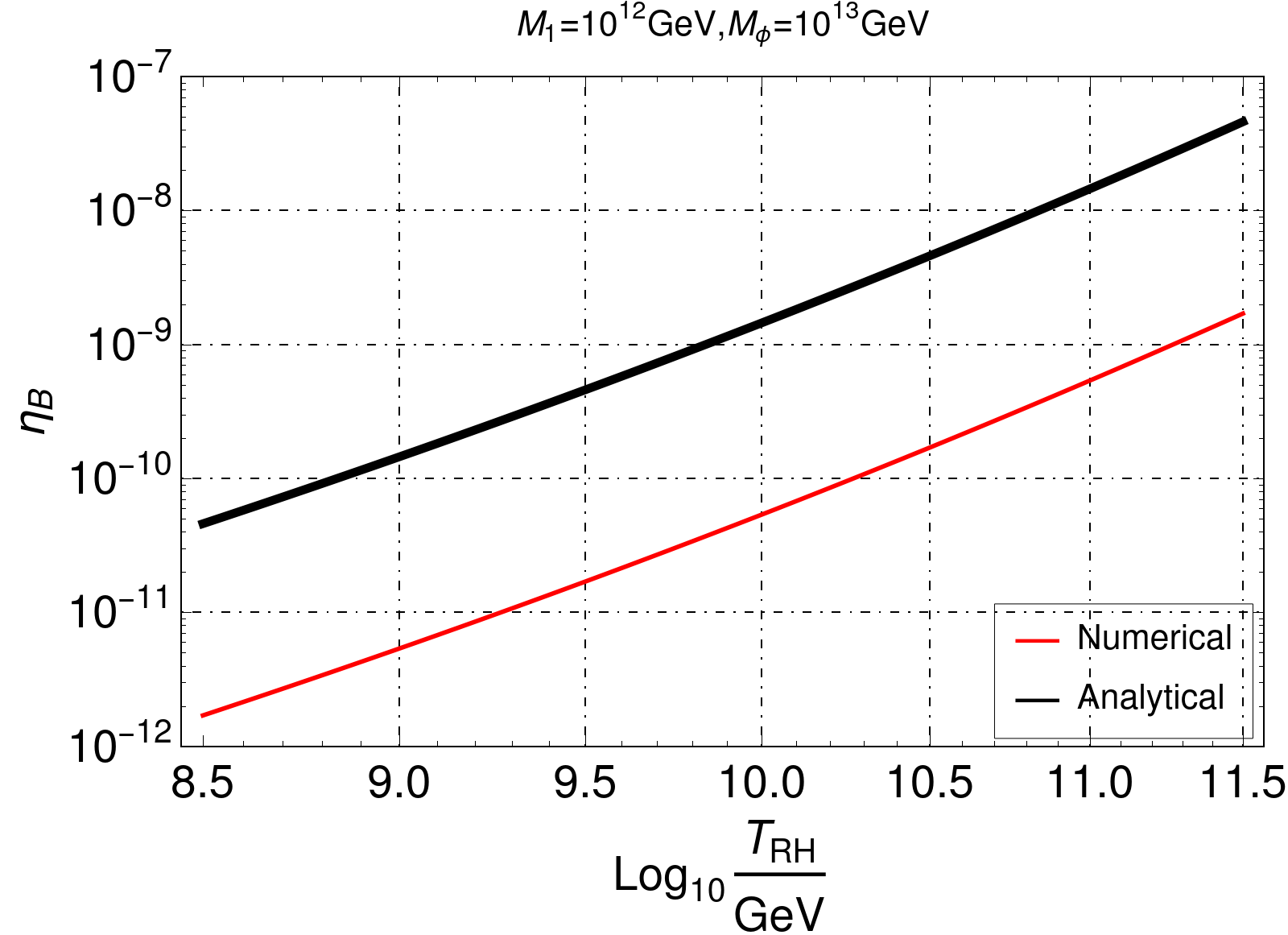}
    \caption{A comparison of analytically obtained baryon asymmetry (black) with that obtained numerically (red) by solving the coupled BEQ (Eq.~\eqref{eq:cpld-beq2}) for a fixed heavy neutrino and inflaton mass.}
    \label{fig:numvsanalyt}
\end{figure}

Before moving on to the next section we would like to comment on the consistency of our rigorous numerical analysis in determining the final (observed) baryon asymmetry $\eta_B$ following Eq.~\eqref{eq:bar-asym}, with simple analytical estimates. Under the assumption that right handed neutrinos decay almost instantaneously after getting produced from inflaton decay, the baryon asymmetry has the well-known analytical expression~\cite{SENOGUZ20046,Giudice:1999fb,Asaka:1999jb,Hamaguchi:2001gw}

\begin{equation}
\frac{n_B}{s}\simeq -\frac{28}{79}\frac{3}{2}\frac{T_\text{RH}}{m_\varphi}\cdot\mathcal{B}\cdot\epsilon_1
\label{eq:analyt}
\end{equation}

\noindent which was derived in the context of inflaton field and leptogenesis without the presence of DM in the same framework. The branching ratio of the inflaton decay into RHNs in Eq.~\eqref{eq:analyt} namely, $\mathcal{B}=1$ since the inflaton decays only into the RHNs. In Fig.~\ref{fig:numvsanalyt} we have plotted the baryon asymmetry $\eta_B$ as a function of the reheating temperature for a fixed RHN and inflaton mass. We find, the asymmetry obtained from the exact numerical solution, (shown by the red curve) is slightly less than the one obtained from approximate analytical solution (shown by the black curve). To quantify, for $T_\text{RH}=10^9~\rm GeV$ the analytical estimation gives $\eta_{B}^\text{analyt}\approx 1.46 \times 10^{-10}$ while doing the explicit numerical calculation we find $\eta_{B}^\text{numr} \approx 5\times 10^{-12}$. The smallness in baryon asymmetry obtained in exact numerical treatment is expected as entropy release is properly incorporated, which is absent in instantaneous right handed neutrino decay approximation. While this shows the overall agreement in the behavior of baryon asymmetry obtained from analytical and numerical calculations, it also points out the necessity of performing a full numerical analysis to get a better estimate of baryon asymmetry. While the rise in baryon asymmetry with the reheating temperature for analytical solution is expected from Eq.~\eqref{eq:analyt}, similar rise in the numerical solution can be explained from the definition of reheat temperature in instantaneous inflaton decay approximation in Eq.~\eqref{eq:inst-inf-decy} and the coupled BEQ in~\eqref{eq:cpld-beq2}. A larger $T_{\text{RH}}$, will result in a larger decay width of the inflaton (as per Eq.~\eqref{eq:inst-inf-decy}), leading to larger abundance of $N_1$, as we can see from the top right panel of Fig.~\ref{fig:yldplt}. With larger $N_1$ abundance we will have a larger lepton asymmetry, which, in turn, shall give rise to an increase in the baryon asymmetry.

%%%%%%%%%%%%%%%%%%
\subsubsection{Effect of dimension 6 operator on DM yield and leptogenesis}
%%%%%%%%%%%%%%%%%%

In the presence of dim.5 interactions, the operators which arise at the level of dim.6 become sub-dominant. Therefore, in order to single out the effect of dim.6 interactions here we consider dim.5 DM-SM operators to be absent. Since we are considering only scalar DM bilinears, therefore gauge invariant dim.6 DM-SM operators do not arise. This leaves pair annihilation of the heavy neutrinos via $NN\chi\chi$ interaction as the only source for DM. On incorporating this in the yield equation, one ends up with the following set of BEQs

\bea\begin{aligned}
& E_\varphi^{'} = -\frac{\Gamma_\varphi}{\mathcal{H}}\frac{E_\varphi}{y}\\&
E_N^{'} = \frac{\Gamma_\varphi}{\mathcal{H}}\frac{E_\varphi}{y}-\frac{\Gamma_N}{\mathcal{H}y}E_N\\&
\widetilde{N}_{B-L}^{'} = \frac{\Gamma_N}{\mathcal{H}y}\epsilon\frac{E_N}{M_1}\\&
R^{'} = \frac{\Gamma_N a_I}{\mathcal{H}}E_N\\&
X^{'} = \frac{\langle\sigma v\rangle_{NN\to\chi\chi}}{\mathcal{H}}\frac{E_N^2}{M_1^2}y^{-4}.
    \end{aligned}\label{eq:cpld-beq3}
\eea

where $\langle\sigma v\rangle_{NN\to\chi\chi}$ is the pair annihilation cross-section of the heavy neutrinos to DM. One should note here the presence of $E_N^2/M_1^2$ term which accounts for the DM production from the heavy neutrino. As the heavy neutrinos are always out-of-equilibrium, hence we can write

\bea
\left(\sigma v\right)_{NN\to\chi\chi}\approx\frac{M_1^2}{2\pi\Lambda^4}.
\eea

\begin{figure}[htb!]
  $$
    \includegraphics[scale=0.5]{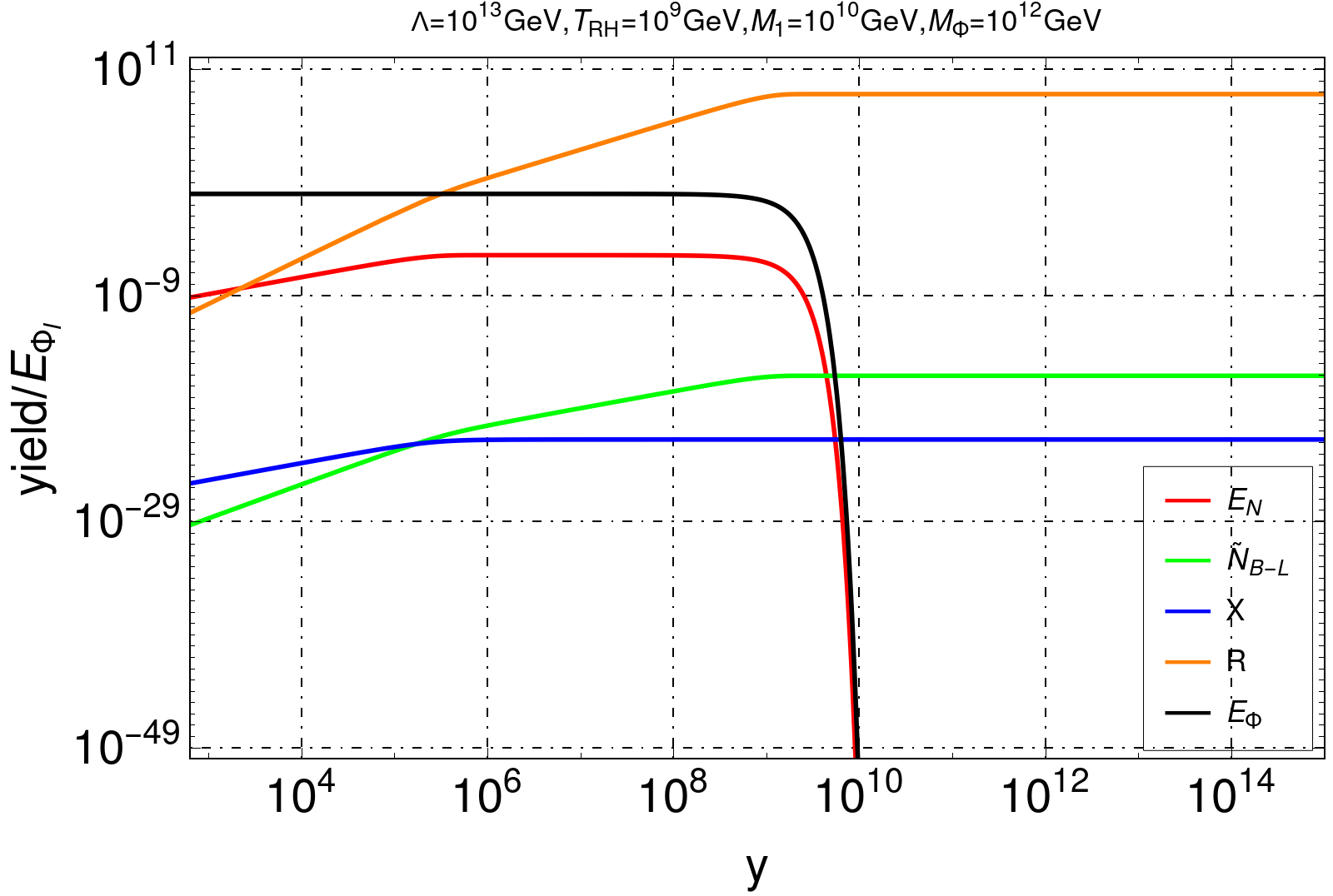}
  $$
\caption{Yield of different components as a function of $y=a/a_I$ is shown for a fixed $\Lambda=10^{13}~\rm GeV$, $T_\text{RH}=10^9~\rm GeV$, $M_\varphi=10^{12}~\rm GeV$ and $M_1=10^{10}~\rm GeV$. The yields corresponding to $\tilde{N}_{B-L}$ and $X$ produce observed baryon asymmetry and DM abundance at $y\to\infty$ for a DM mass of 10 GeV.}\label{fig:yld-dim6}
\end{figure}

\begin{figure}[htb!]
  $$
    \includegraphics[scale=0.4]{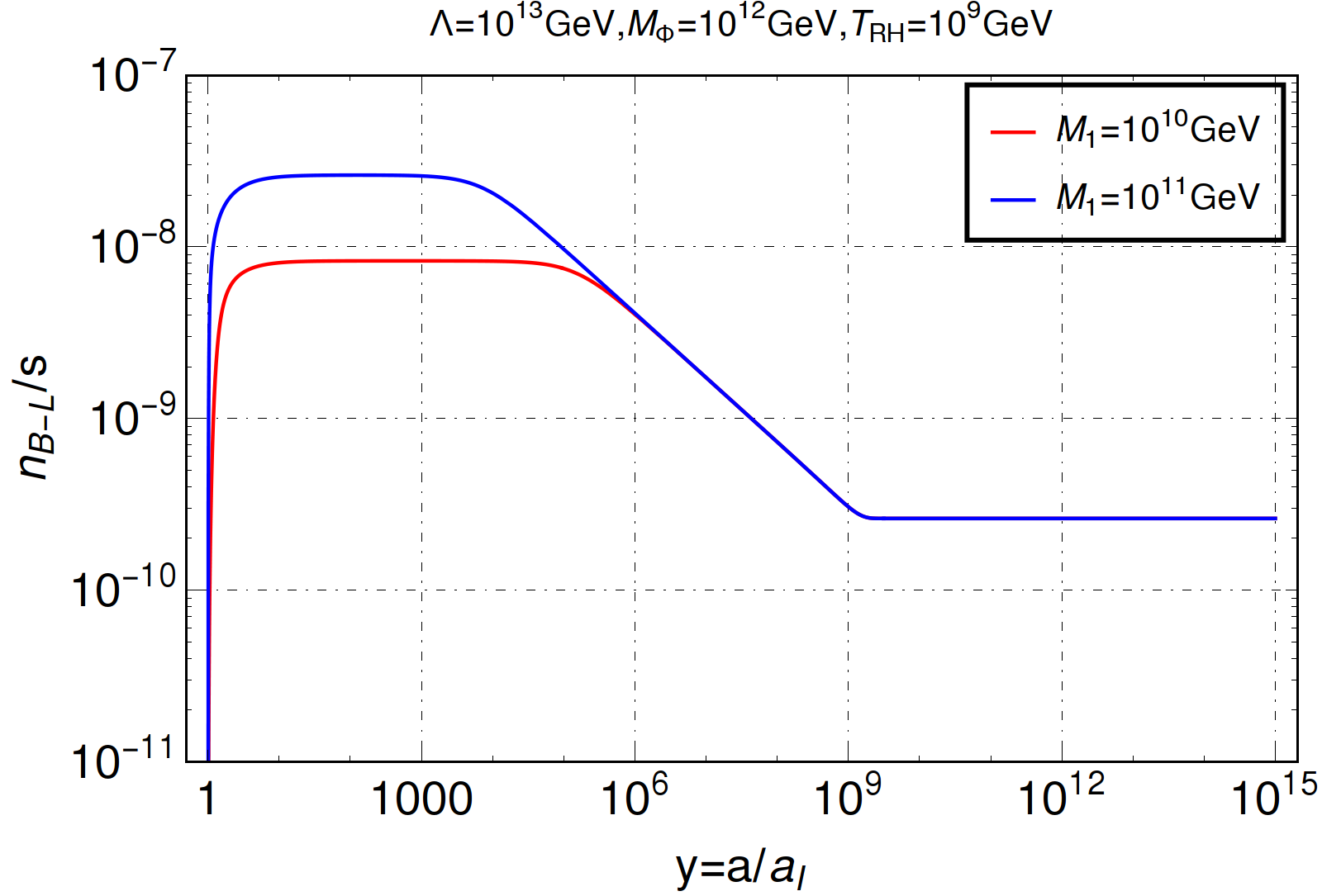}~~\includegraphics[scale=0.4]{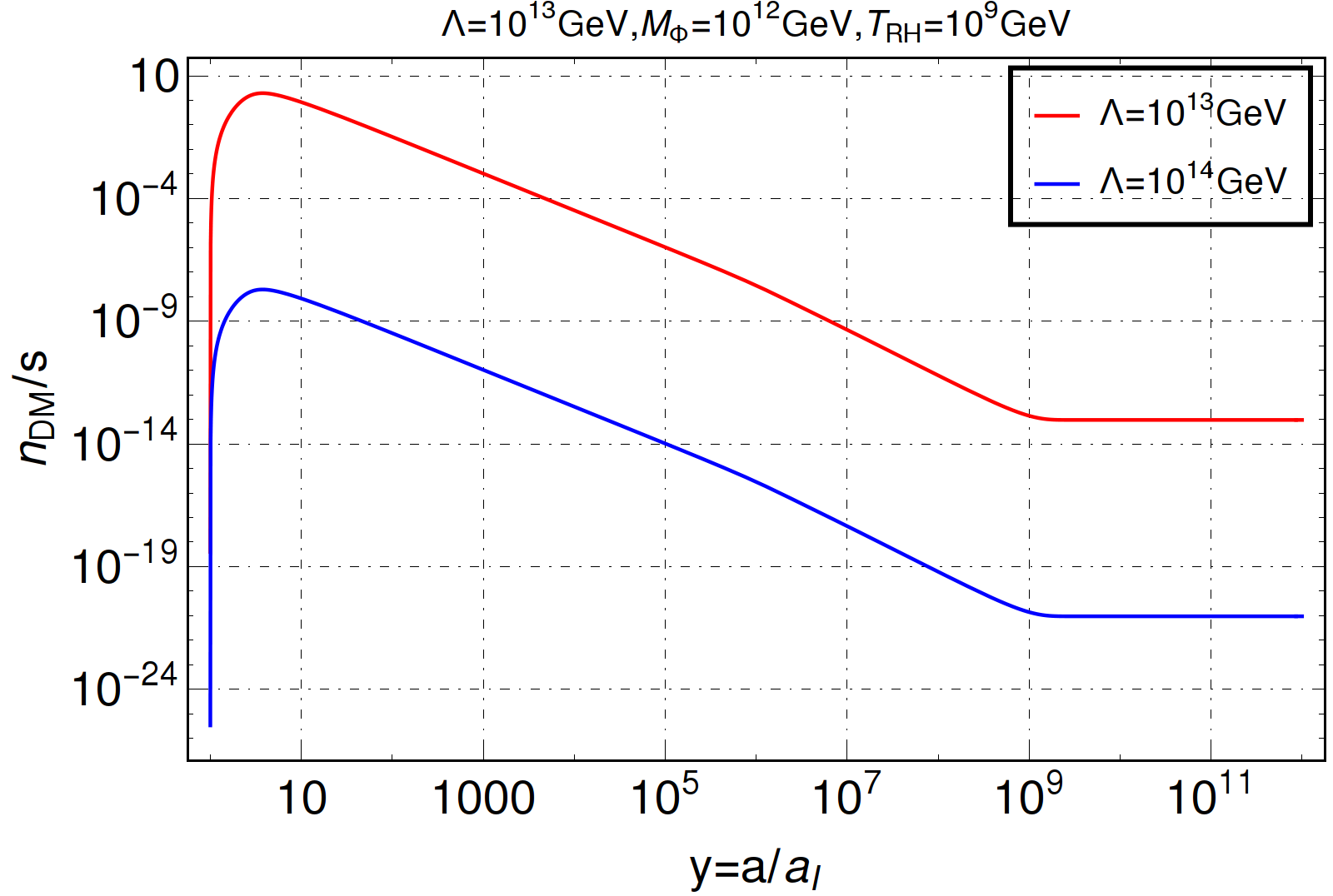}
  $$
\caption{Left: Evolution of $B-L$ asymmetry with $y=a/a_I$ for a fixed inflaton mass $M_\varphi=10^{12}~\rm GeV$ and reheat temperature $T_\text{RH}=10^9~\rm GeV$ for two choices of $M_1:\{10^{10},10^{11}\}~\rm GeV$ shown in red and blue respectively. Right: DM yield $n_\text{DM}/s$ as a function of $y=a/a_I$ for different choices of $\Lambda=\{10^{13},10^{14}\}~\rm GeV$ and $T_\text{RH}=10^9~\rm GeV$.}\label{fig:asym-rel-d6}
\end{figure}

On the other hand, several operators connecting the heavy neutrinos with the SM fields appear at dim.6 as mentioned in Sec.~\ref{sec:model}. However, the dominant source of leptogenesis can still be considered to be due to $N_1\to\ell H$ decay as $2\to2$ scattering processes involving $N_1$ in the initial or the final states are suppressed for $M_1>T_\text{RH}$. With this, we can now solve the set of BEQs in Eq.~\eqref{eq:cpld-beq3}. The evolution of yield for dim.6  interactions with $y=a/a_I$ is shown in Fig.~\ref{fig:yld-dim6} for a fixed cut-off scale, reheat temperature, inflaton mass and lightest right handed neutrino mass. For a DM mass of 10 GeV, the DM yield (shown by the blue curve) produces the right relic abundance. Also, an observed baryon asymmetry is obtained corresponding to $\tilde{N}_{B-L}\left(y\to\infty\right)$. We see, the yield of the different components, excepting for the DM, follow the pattern of dim.5 interaction as we have seen in Fig.~\ref{fig:yldplt1}. The DM yield flattens and becomes constant at a very early time $y\sim 10^5$. This happens at the same time where the decay of the RHN becomes comparable to its production as we can see comparing the red curve with the blue one. Since the DM yield depends on the pair annihilation of the heavy neutrinos, hence the moment the heavy neutrino decay starts becoming comparable to its production, the DM abundance freezes in and becomes constant. In Fig.~\ref{fig:asym-rel-d6} we show the evolution of $n_{B-L}/s$ (left) and DM number density (right) with $y=a/a_I$. In the left panel we show the variation of $n_{B-L}/s$ for some benchmark choices of $M_1$. Here we see $n_{B-L}/s$ evolves in the same way as we have already seen for the dim.5 case in the top left panel of Fig.~\ref{fig:asym-rel}. A more interesting dynamics however appears for the case of DM number density evolution as shown in the right panel. Here we see that the DM number density sharply increases at $y\sim 10$ which falls afterwards and becomes saturated at $y\sim 10^9$. The sharp increase of DM number density at very high temperature is a consequence of heavy neutrino annihilation into DM. As the heavy neutrino starts decaying, the DM number density rapidly falls due to entropy injection in the thermal bath, and finally reaching constant at $y\sim 10^9$. The $B-L$ asymmetry also saturates at around the same epoch. This happens because at dim.6 the heavy neutrino becomes the common origin of $B-L$ asymmetry as well as DM genesis and hence the $B-L$ and DM yield build up in a similar fashion. Since at dim.6 the DM abundance via UV freeze-in approximately becomes $\Omega_\chi h^2\propto m_\chi T_\text{RH}^3/\Lambda^4$ hence compared to the dim.5 case, for a much lower $\Lambda$ observed DM abundance can be achieved for a given DM mass and reheat temperature. We find that for a DM mass of 10 GeV, in order to satisfy the observed relic abundance, one can have $\Lambda=10^{13}~\rm GeV$ for $T_\text{RH}=10^9~\rm GeV$ which is at least four orders of magnitude smaller than the dim.5 case for a DM mass of the same order. This also establishes our earlier claim that dim.6 interactions become ineffective in the presence of dim.5 operators.

\subsection{Non-instantaneous inflaton decay}\label{sec:non-inst-decay}
%%%%%%%%%%%%%%%%%%

\begin{figure}[htb!]
$$
\includegraphics[scale=0.42]{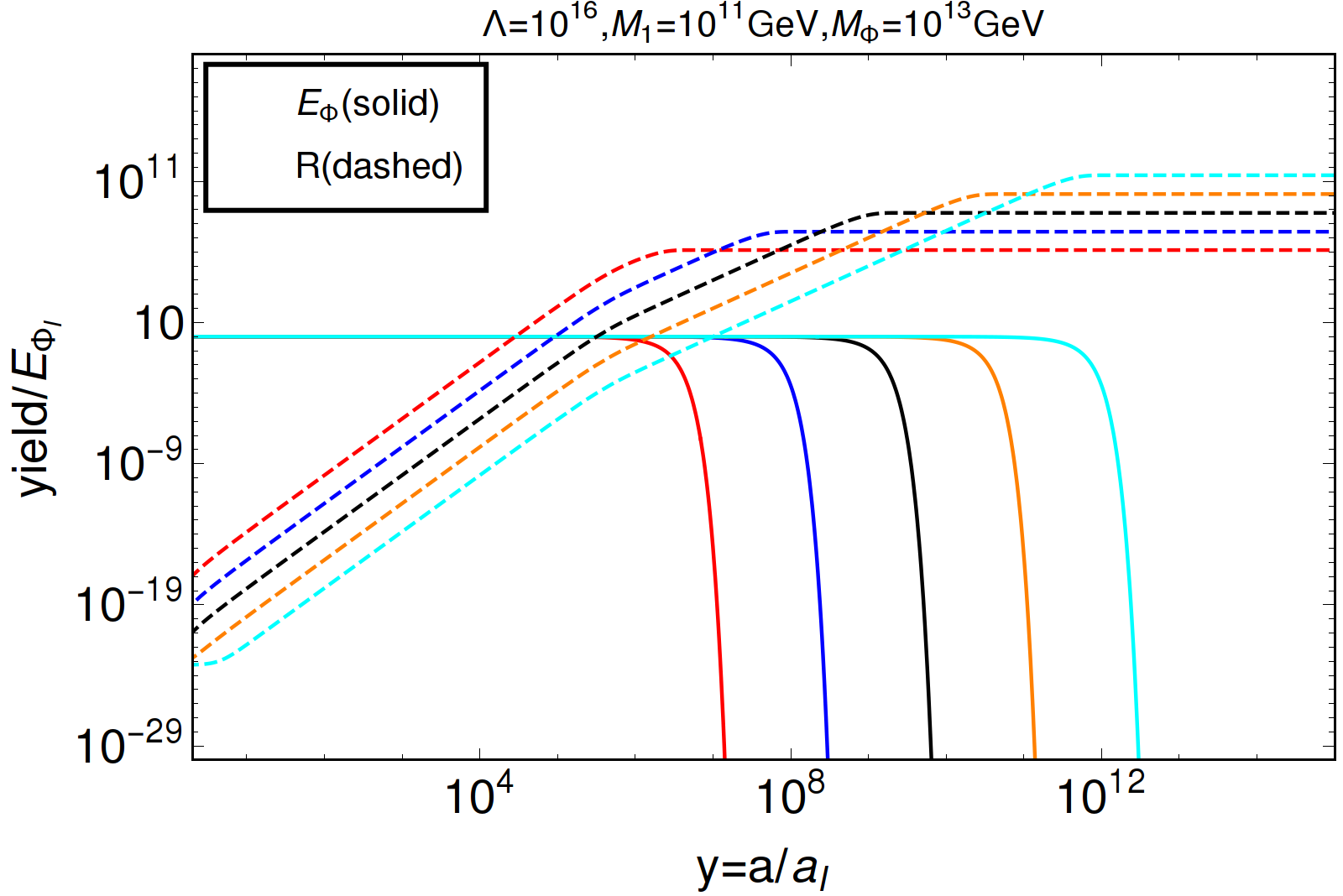}~~
\includegraphics[scale=0.4]{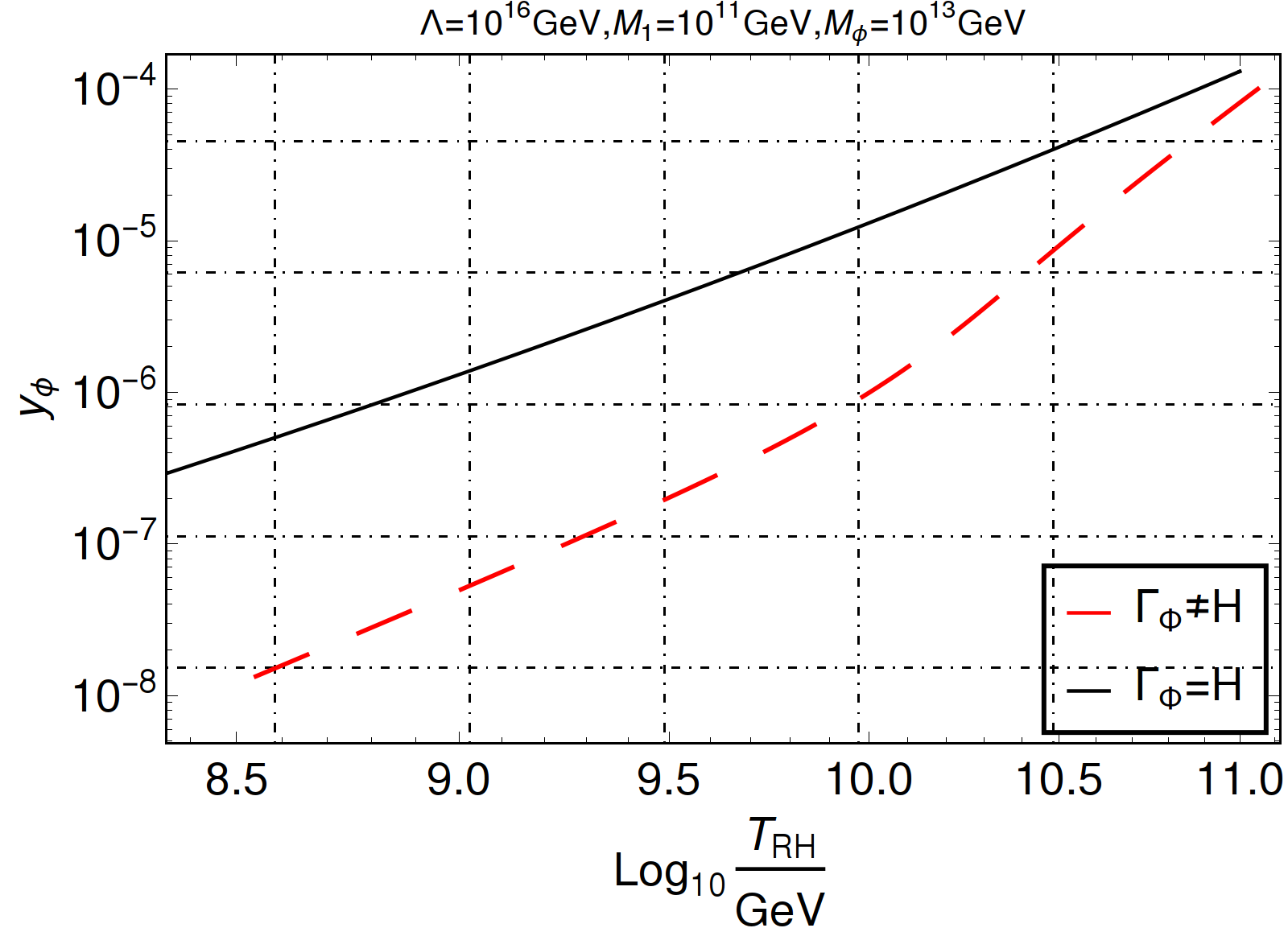}
$$
$$
\includegraphics[scale=0.42]{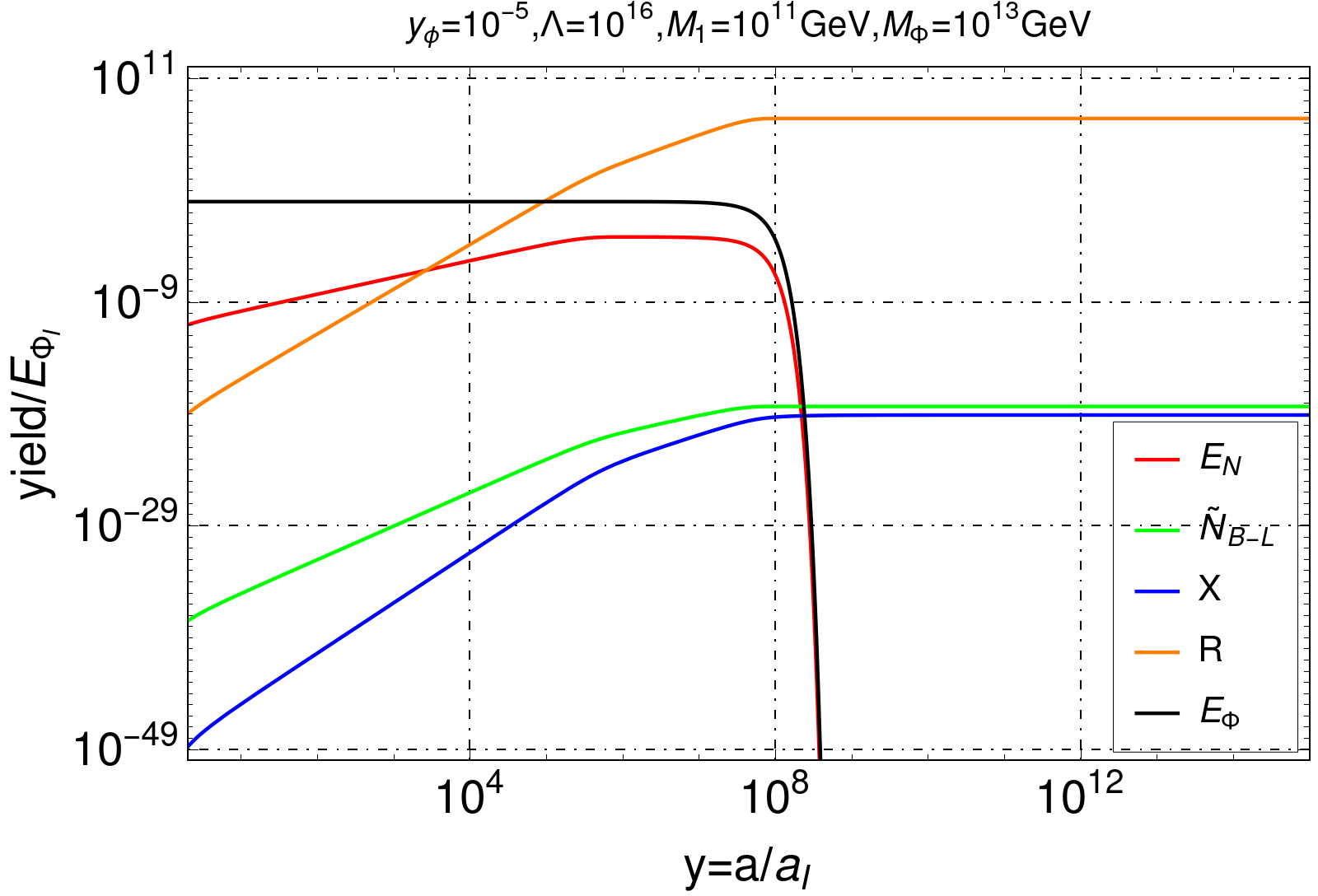}
$$
\caption{Top Left: Evolution of $R$ and $E_\varphi$ with $y=a/a_I$. Here all the solid lines correspond to the variation of $E_\varphi$ while all the dashed lines correspond to $R$. Different coloured lines indicate different choices of $y_\varphi:\{10^{-4},10^{-5},10^{-6},10^{-7},10^{-8}\}$ shown in red, blue, black, orange and cyan respectively. The cut-off scale, inflaton mass and heavy neutrino mass are fixed at the values mentioned in the plot label. Top Right: A comparison of instantaneous vs non-instantaneous decay approximation in $y_\varphi-T_\text{RH}$ plane for a fixed choice of other parameters as in the left plot. The reheat temperature for the non-instantaneous case has been obtained from the LHS plot. Bottom: Variation of yield of different components with $y=a/a_I$ for fixed values of $y_\varphi$, $\Lambda$, $M_\varphi$ and $M_1$ that gives rise to the observed baryon asymmetry and correct relic density for a DM of mass 1 GeV.}\label{fig:yld-non-inst}
\end{figure}

In case where the instantaneous inflaton decay approximation breaks down, one can no more make the assumption $\Gamma_\varphi=\mathcal{H}$ to parametrize the inflaton-heavy neutrino Yukawa coupling in terms of the reheat temperature $T_\text{RH}$. In this case the inflaton-heavy neutrino coupling becomes a free parameter. The reheat temperature, on the other hand, can be obtained by defining the temperature at which $E_\Phi\left(y_\text{RH}\right)=R\left(y_\text{RH}\right)$  where $y_\text{RH}$ defines the corresponding scale factor ratio\footnote{For the non-instantaneous case $T_\text{RH}$ is  defined to be the temperature of the thermal bath when it starts to dominate over the residual energy of the inflaton. This is the point when the energy density of the inflaton starts transferring into the energy density of radiation~\cite{Davidson:2000er,Garcia:2017tuj,Garcia:2020eof}.}. Similar definition of reheating temperature for such non-instantaneous inflaton decay scenario have also been used in earlier works, for example~\cite{Garcia:2020eof}. In fact, we have checked, by taking larger $R/E_\Phi$ ratio at $y=y_\text{RH}$, there is only a slight change in $T_{\rm RH}$ that can be neglected. Thus, we stick to the conventional definition already used in earlier works. We first make an estimation of the inflaton-heavy neutrino Yukawa coupling $y_\varphi$ and establish a relation between $y_\varphi$ and $T_\text{RH}$ for dim.5 when the instantaneous decay is no longer an approximation. In the top right panel of Fig.~\ref{fig:yld-non-inst} we show the dependence of the Yukawa coupling $y_\varphi$ on the reheat temperature during instantaneous decay (in black) and non-instantaneous decay (red dashed) of the inflaton field.  In the non-instantaneous decay limit the coupling $y_\varphi$ becomes a free parameter, independent of the reheating temperature unlike the instantaneous decay limit where $T_\text{RH}$ determines $y_\varphi$. Thus, under the non-instantaneous decay approximation one can choose a much lower $y_\varphi$ with respect to the instantaneous case (corresponding to a given $T_\text{RH}$) as evident from the top right panel of Fig.~\ref{fig:yld-non-inst}. However, for large $y_\varphi$ the two limiting cases converge as the decay always happens instantaneously. We thus consider $$\{\Lambda,y_\varphi,m_\chi\}$$ to be the relevant free parameters in the context of non-instantaneous decay scenario. In the top left panel of Fig.~\ref{fig:yld-non-inst} we show the evolution of $E_\varphi$ and $R$ with $y=a/a_I$ for different choices of $y_\varphi$ (shown by different colours). For a fixed $y_\varphi$ all the solid curves show the evolution of $E_\varphi$ while the dashed curves show how the radiation $R$ builds up. By applying the condition $E_\varphi=R$ (at each $y=y_\text{RH}$) we determine $T_\text{RH}$ corresponding to each $y_\varphi$ which is shown by the red dashed lines in the top right panel. % {\color{red} The dependence of $y_\varphi$ on $T_\text{RH}$ for the non-instantaneous case is obtained } {\sout{The red points in the plot on the top right panel of Fig.~\ref{fig:yld-non-inst} corresponding to different $y_\varphi$ are obtained}} from the top left panel plot of of Fig.~\ref{fig:yld-non-inst} where we  We then determine $T_\text{RH}$ corresponding to each $y_\varphi$ by applying the condition $E_\varphi=R$ at $y=y_\text{RH}$. 
From the top right panel it is also clear that at very high values of $T_\text{RH}$ the non-instantaneous approximation merges with the instantaneous one as large $T_\text{RH}$ becomes equivalent to large $y_\varphi$. One has to, however, be careful about the fact that for a given $M_\varphi$ and $M_1$ arbitrarily large Yukawa coupling (even if within the perturbative limit) needs to be avoided as that might correspond to a reheat temperature $T_\text{RH}>M_1$ contrary to the requirement for non-thermal leptogenesis. It is also important to note that changing $\Lambda$ keeping everything else fixed does not change $y_\text{RH}$. Therefore, the top right panel plot remains unaltered irrespective of the choices of the cut-off scale.   We now show the yields of the different components as before with $y=a/a_I$ in the bottom panel. The nature of the yields remain the same as one can see by comapring with Fig.~\ref{fig:yldplt1}. For $y_\varphi=10^{-5}$ and $\Lambda=10^{17}~\rm GeV$ we can obtain observed DM abundance for a DM of mass $m_\chi=1~\rm GeV$ and baryon asymmetry of the correct order in the limit $y\to\infty$. 

%One can note from the top right panel of Fig.~\ref{fig:yld-non-inst}, $y_\varphi=10^{-5}$ corresponds to a reheat temperature of $T_\text{RH}=10^9~\text{ GeV}<M_1=10^{12}~\rm GeV$. This is largely in agreement with the instantaneous decay approximation case as one can see from Fig.~\ref{fig:yldplt1}.

\begin{figure}[htb!]
$$
\includegraphics[scale=0.4]{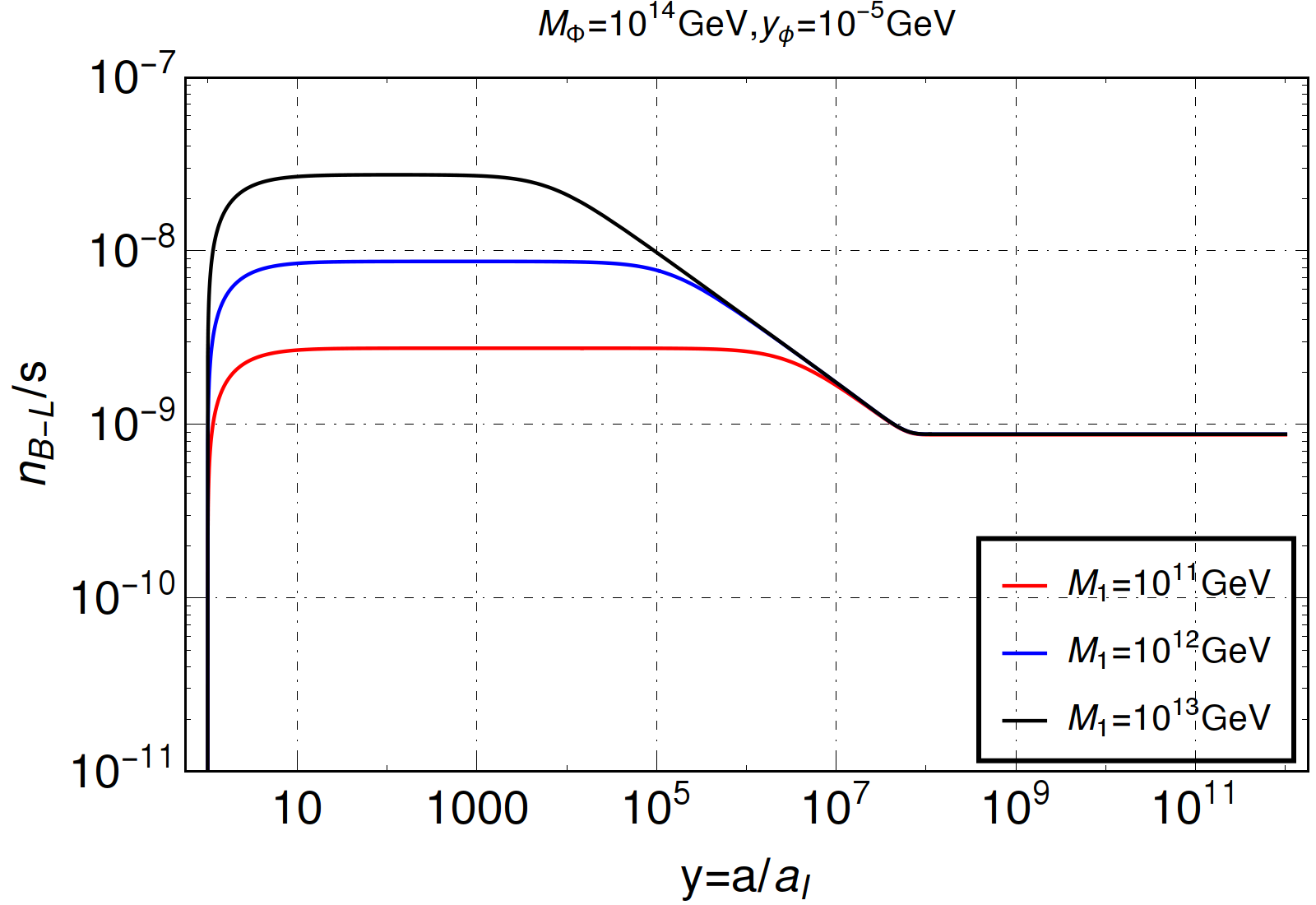}~~
\includegraphics[scale=0.4]{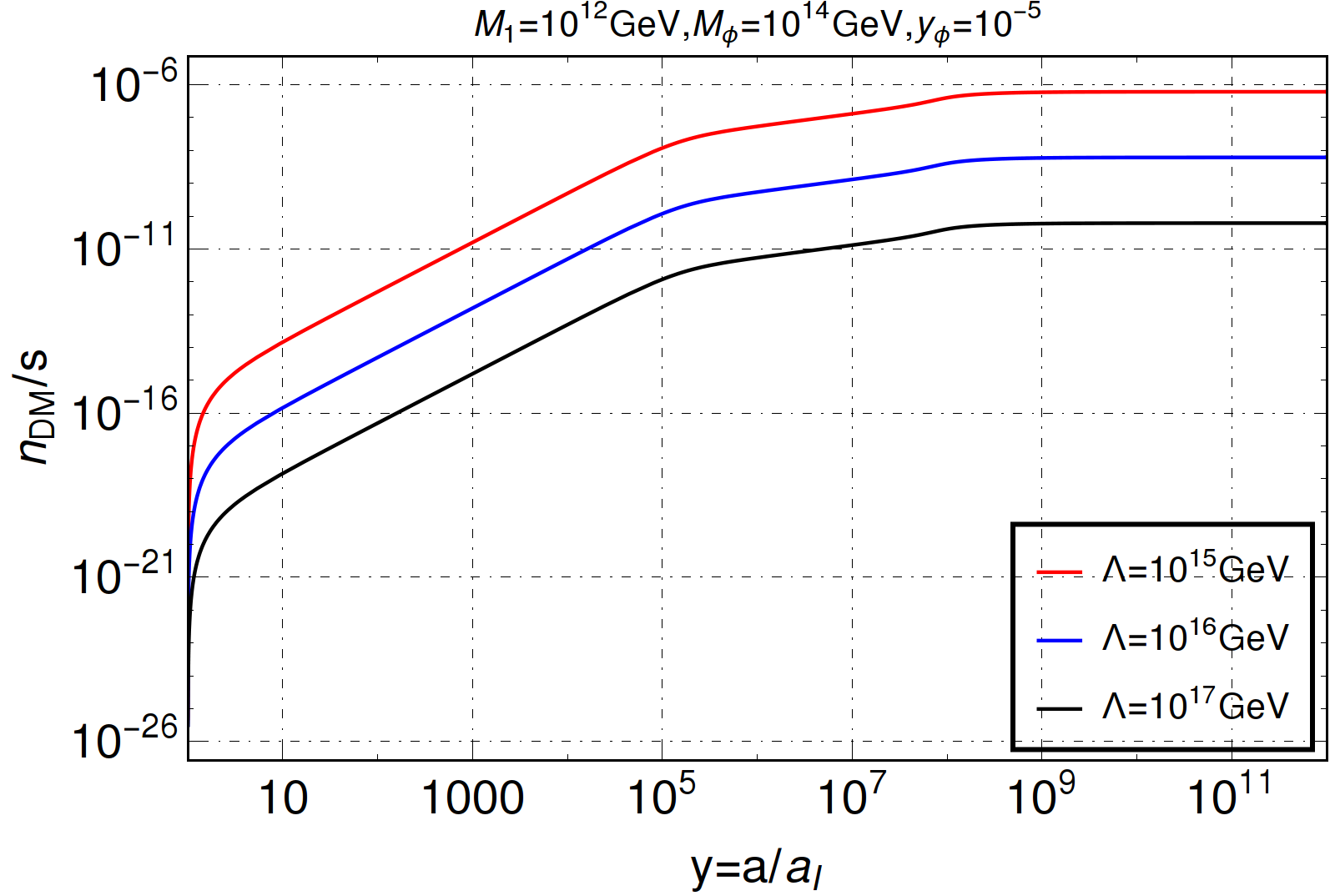}
$$
$$
\includegraphics[scale=0.5]{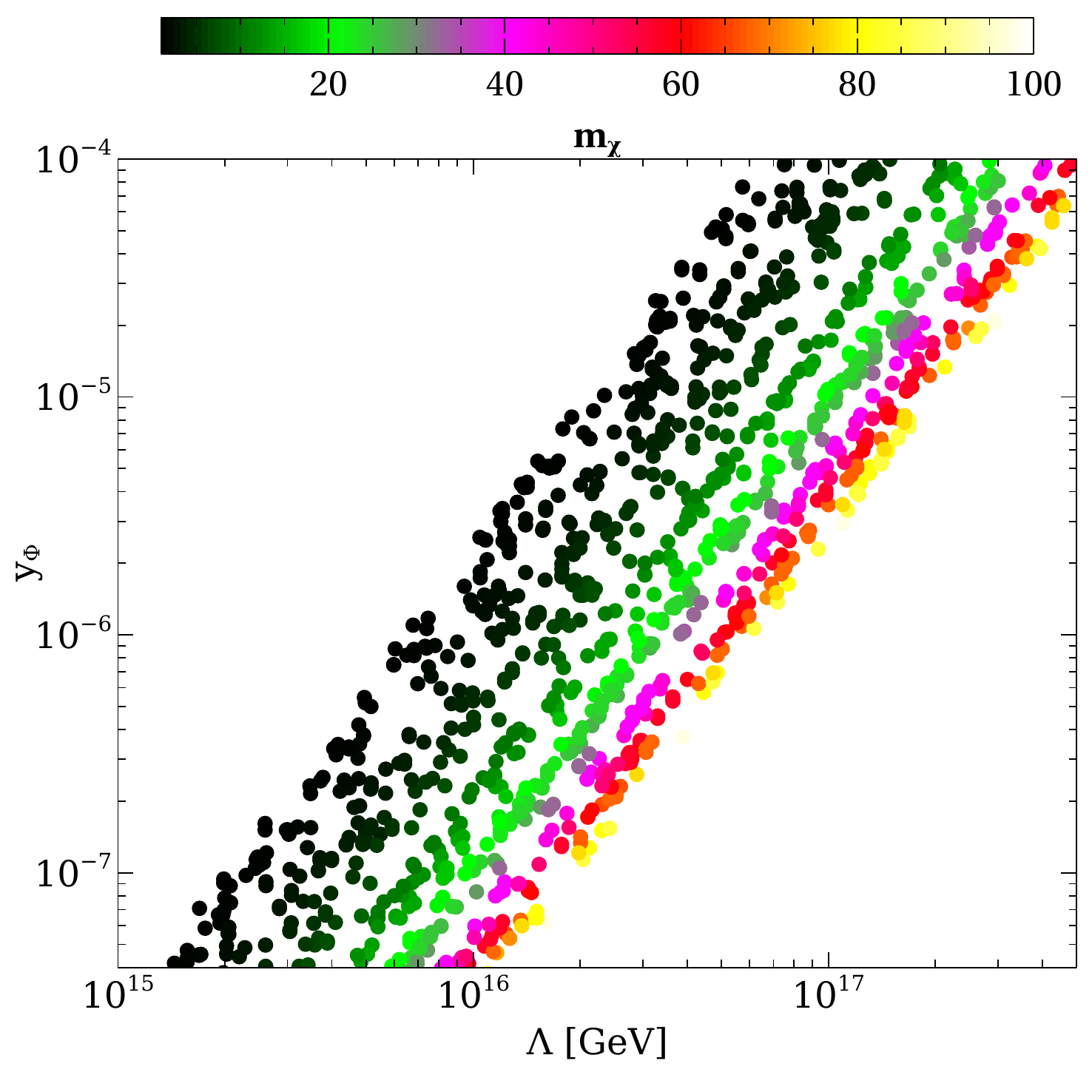}~~
\includegraphics[scale=0.5]{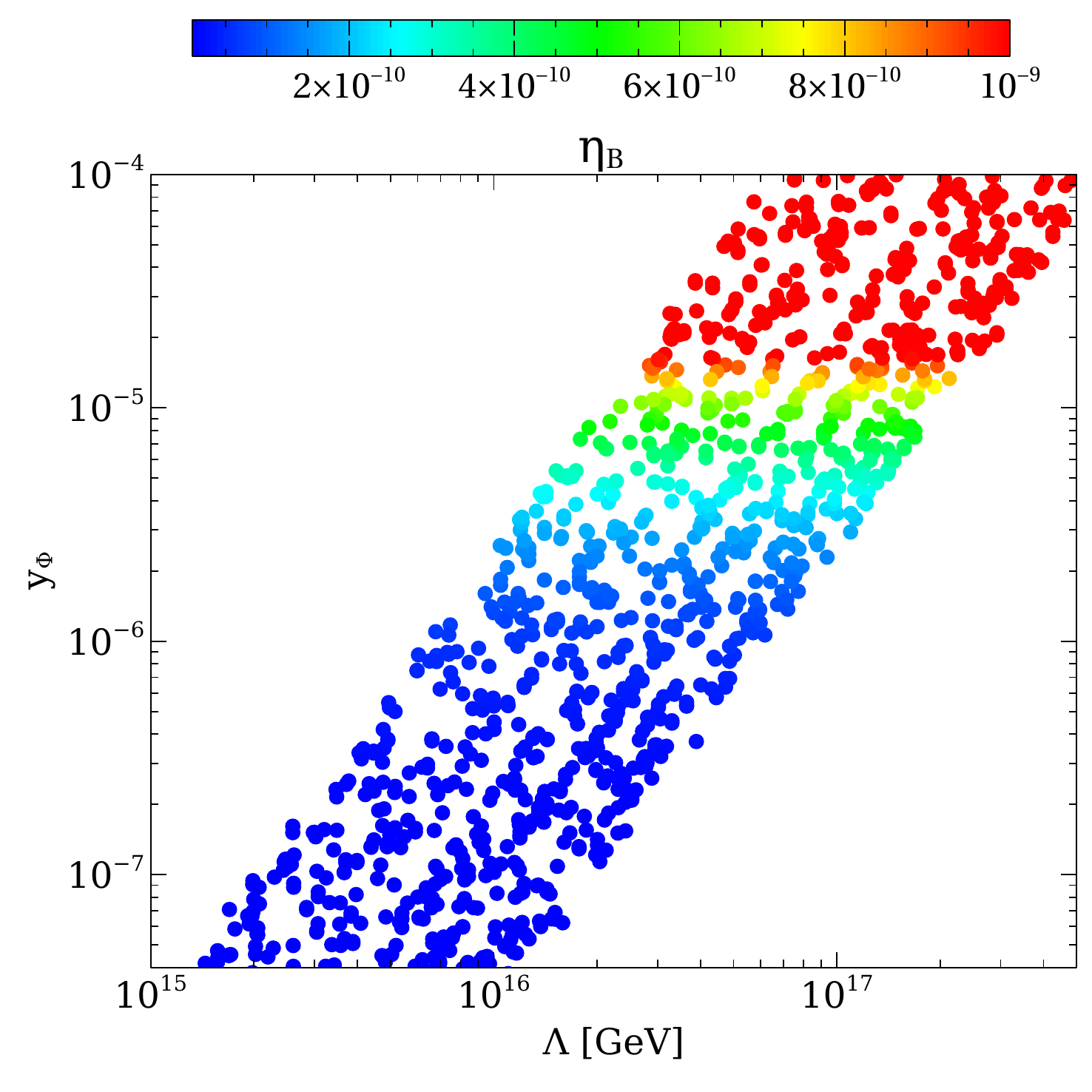}
$$
\caption{Top Left: $n_{B-L}/s$ as a function of $y=a/a_I$ is shown for a fixed $y_\varphi=10^{-5}$, $M_\varphi=10^{13}~\rm GeV$ and for three different choices of $M_1:\{10^{10},10^{11},10^{12}\}~\rm GeV$ in red, blue and black respectively. Top Right: Variation of $n_\text{DM}/s$ with $y=a/a_I$ for a fixed DM mass and different choices of $\Lambda=\{10^{15},10^{16},10^{17}\}~\rm GeV$ shown in red, blue and black respectively. Bottom Left: Relic density allowed parameter space in $y_\varphi-\Lambda$ plane where the colouring is with respect to DM mass. Bottom Right: Variation of baryon asymmetry $\eta_B$ in $y_\varphi-\Lambda$ plane. In both cases the inflaton and heavy neutrino masses are fixed at $M_\varphi=10^{13}~\rm GeV$ and $M_1=10^{11}~\rm GeV$ respectively.}\label{fig:rel-asym-non-inst}
\end{figure}

The variation of $n_{B-L}/s$ with $y=a/a_I$ for the non-instantaneous scenario is shown in the top left panel of Fig.~\ref{fig:rel-asym-non-inst} where we see a larger $M_1$ leads to a larger asymmetry for a given $y_\varphi$ following the earlier trend in Fig.~\ref{fig:asym-rel}. On the top right panel of Fig.~\ref{fig:rel-asym-non-inst} we show how the DM yield $n_\text{DM}/s$ evolves with $y=a/a_I$ for the same choice of $M_\varphi$ and $y_\varphi$ as in the top left panel. Here we see the same trend as in Fig.~\ref{fig:asym-rel} {\it viz.} a larger $\Lambda$ leads to smaller final DM abundance for a fixed $y_\varphi$. This is again same as in the top right panel of Fig.~\ref{fig:asym-rel} where $T_\text{RH}$ plays the same role as $y_\varphi$. Thus, for a given DM mass, a larger $\Lambda$ requires a smaller $y_\varphi$ to produce the observed relic abundance. In the bottom left panel of Fig.~\ref{fig:rel-asym-non-inst} we show the the region of parameter space where Planck observed relic density is satisfied for DM mass ranging $m_\chi:\{1-100\}~\rm GeV$. The inflaton-heavy neutrino Yukawa $y_\varphi$, which is now a free parameter, is varied in the range $y_\varphi:\{10^{-8}-10^{-4}\}$. Here we again see, similar to the bottom left panel of Fig.~\ref{fig:asym-rel}, a larger DM mass needs comparatively larger $\Lambda$ to satisfy the observed DM abundance. Since in the case of non-instantaneous inflaton decay, as discussed earlier, a larger $y_\varphi$ indicates a larger reheat temperature, hence for a fixed $\Lambda$ increasing $y_\varphi$ results in DM over-abundance. Thus, a large $y_\varphi$ at a fixed $\Lambda$ requires a smaller DM mass to produce observed DM abundance. Again, since $\Gamma_\varphi\propto y_\varphi^2$ hence increasing $y_\varphi$ increases $N_1$ abundance (as $E_N^{'}\sim\Gamma_\varphi$) which in turn results in excess in the baryon asymmetry. This can exactly be compared with the effect of increasing $T_\text{RH}$ in case of instantaneous decay scenario. The observed asymmetry can be obtained for $y_\varphi\sim 10^{-5}$ which certainly produces right amount of DM as seen in the bottom left panel. For the case of dim.6, as one can understand, it is possible to obtain the observed DM abundance for a smaller $\Lambda$ as we mentioned earlier in the case of instantaneous decay approximation. Apart from that other conclusions remain same as the instantaneous decay case and hence we do not elaborate them further.  

%%%%%%%%%%%%%%%%%%%%%%%%%%

\section{Possible UV Completion}
\label{sec:uvcomplete}
We have realized the possibility of non-thermal leptogenesis and UV freeze-in for a singlet fermionic DM with the inflaton field playing a non-trivial role. Since we are considering hierarchical right handed neutrinos, the scale of leptogenesis is high $M_1 > 10^9$ GeV. As we have found, to realize the freeze-in scenario, the cut-off scale needs to be $\Lambda\gtrsim 10^{15}$ GeV, considering dimension five operators connecting DM and SM bilinears. Although, for DM-SM operators with dimension $d>5$ one can realize UV freeze-in at a lower cut-off scale, the requirement of keeping it above the scale of leptogenesis forces us to consider $\Lambda>M_1$ such that the effective theory remains valid. While we stick to this minimal setup to justify UV freeze-in, where the DM is required to have non-renormalizable interaction with the SM, such operators can well be realized in a UV complete theory by incorporating additional degrees of freedom (DOFs). If such additional DOFs have mass $M$, we can generate effective DM-SM operators at a scale $\mu\ll M$, by integrating out those heavy fields. For example, the dim.5 operator $ (\overline{\chi}\chi) (H^{\dagger} H)/\Lambda$ can be generated by extending the model with additional vector like lepton doublet $\psi_{L,R}$. We also need $\psi_{L,R}$ to be odd under the discrete $Z_2$ symmetry, such that Yukawa interactions of the type $$-\mathcal{L} \supset m_{\psi} \overline{\psi} \psi + y_1 \overline{\psi_L} \widetilde{H} \chi + y_2 \overline{\psi^c_R} H \chi + h.c.,
$$ can be generated. At a scale $\mu \ll m_{\psi}$, the heavy vector like leptons can be integrated out, resulting in an operator $ y_1 y_2\left(\overline{\chi}\chi\right)\left(H^{\dagger} H\right)/\Lambda$, where $\Lambda$ can be identified with the masses of heavy fermions which are integrated out. Note that, by introducing a heavy singlet scalar $\phi$, {\it even} under $Z_2$, one can have an interaction of the form $$-\mathcal{L}\supset y_N\overline{N}N\phi+y_\chi\overline{\chi}\chi\phi+h.c.,$$ that can lead to dim.6 operators $\overline{N}N\overline{\chi}\chi/\Lambda^2$ on integrating out $\phi$ at a scale $\mu\ll\mu_\phi$ with $\Lambda\equiv\mu_\phi$. In fact, a heavy $Z_2$ even scalar can also provide dim.5 operators of the form $\overline{\chi}\chi H^\dagger H$ at a scale $\mu\ll\mu_\phi$. While we do not pursue the details of such UV completions, one may refer to several earlier works in this direction. For example, the authors of~\cite{Biswas:2018aib} considered loop suppressions as the origin of FIMP interactions, the possibility of DM-SM interactions via superheavy neutral gauge bosons were discussed in~\cite{Mambrini:2013iaa}, while the authors of~\cite{Kim:2017mtc, Kim:2018xsp} considered clockwork origin of FIMP couplings. 

% {\color{black} in all these cases we also have $NN\chi\chi$ int.?}.}

%%%%%%%%%%%%%%%
\section{Conclusion}\label{sec:concl}
%%%%%%%%%%%%%%%

In this work we have addressed two long-standing problems in particle cosmology, namely the matter-antimatter asymmetry and the dark matter relic abundance in a minimal setup while keeping the DM analysis model independent. In order to explain both we have considered the SM effective theory being augmented by two right handed neutrino $N_i$ leading to what is known as $\nu$SMEFT which explains light neutrino masses via Type-I seesaw mechanism. We further consider a SM gauge singlet $Z_2$-odd fermion that is absolutely stable over the age of the universe and hence plays the role of a potential DM candidate. While RHNs, in spite of being gauge singlets, can interact with the SM due to the presence of tree level Yukawa couplings, the $Z_2$-odd fermion singlet DM can communicate with the SM only via non-renormalizable interactions with dimension $d>4$. We consider a high scale Type-I seesaw which favours leptogenesis with hierarchical RHNs. Therefore, we assume new physics at a scale $\Lambda \gg \text{TeV}$, which is also the mass scale of the heavy fields integrated out and write down the operators below $\Lambda$. In the absence of extra fields, the RHN-DM interaction also arises due to 4-fermion operators that appear only in dimension 6 or higher. We take up only scalar DM bilinears for simplicity and thus in dim.6 no SM gauge invariant operator appears connecting the DM with the SM field. At dim.5 level, however, DM-SM interactions can arise involving SM Higgs. Since we focus on non-thermal DM as well as non-thermal leptogenesis, we consider a singlet scalar field mimicking the inflaton field after slow-roll whose decay into RHNs play a crucial role in DM production, leptogenesis as well as overall reheating of the universe.

We consider baryogenesis via non-thermal leptogenesis where the heavy neutrino masses are larger than the reheat temperature of the universe for explaining the observed baryon asymmetry. The inflaton exclusively decays into a pair of lightest of the heavy neutrinos $N_1$ via a renormalizable interaction. The heavy neutrino $N_1$ thus produced further decays into SM bath producing a net $B-L$ asymmetry that is converted into a baryon asymmetry by sphaleron processes. The decay of the inflaton field into $N_1$ pair can be either instantaneous and non-instantaneous where in the former case the inflaton-RHN coupling is parametrized by the reheat temperature while in the latter the coupling can be considered to be a free parameter. We analyse both the cases separately and show it is possible to obtain the observed baryon asymmetry depending on the choice of reheat temperature (for instantaneous decay case) or the inflation-RHN Yukawa coupling (for non-instantaneous decay case) together with the heavy neutrino mass.

The higher dimensional operators mediating DM-SM and DM-RHN interactions motivate us to consider UV freeze-in as the mechanism for DM genesis. At dim.5 level, the DM can get produced from the SM bath itself which is again generated from the out-of-equilibrium decay of $N_1$. At dim.6, on the other hand, the only source of DM is the 4-fermion operator consisting the RHN and the DM fields. Since dim.6 interactions become sub-dominant in presence of dim.5 operators, hence we take up each cases one by one and analyse them separately. Since the inflaton decay triggers the leptogenesis while the DM can be produced both from the SM bath and the heavy neutrinos depending on the dimension of interaction, hence we solve a set of coupled Boltzmann equations to obtain the final yield of different components. Because of the coupled nature of the solutions, a very non-trivial dynamics in the resulting yield shows up. We find that the observed baryon asymmetry together with the right DM abundance can be obtained in dim.5 for $\Lambda\gtrsim 10^{15}~\rm GeV$ for DM mass up to 100 GeV with $T_\text{RH}\gtrsim 10^9~\rm GeV$. This translates into a bound on inflaton-RHN coupling $10^{-7}\lesssim y_\varphi\lesssim 10^{-4}$ for the non-instantaneous reheating scenario. In case of dim.6 interaction (where dim.5 operators are absent) the bound on the cut-off scale is significantly lifted up to $\Lambda\gtrsim 10^{14}~\rm GeV$ for the same set of parameters. In all such cases we make sure that the inflaton mass $M_\varphi>2M_1$ and $M_1<T_\text{RH}$ in compliance with our previous assumptions. As the DM in the case of dim.6 interaction is produced from the heavy neutrino annihilation here we see a very intriguing dynamics in the DM yield. The DM yield shows a peak at early times and then gets diluted with the subsequent entropy injection as the heavy neutrino starts decaying. This behaviour is absent in dim.5 as the DM there gets produced only from the radiation and hence the two, namely the radiation and the DM energy density evolve in the similar way. While we perform a complete numerical analysis by solving the coupled Boltzmann equations, for the sake of completeness, we have also compared our numerical estimation of baryon asymmetry with the one obtained using approximate analytical formula under the assumption of instantaneous RHN decay. Comparing the analytical and numerical results in one of the cases, we found that the overall baryon asymmetry behaves similarly with the reheat temperature under both approaches, although the numerical analysis lead to slightly larger values.

Finally, in presence of heavy neutrinos, inflaton decay alone can provide a common origin of observed baryon asymmetry and DM abundance in the present universe along with successful reheating. While the role of inflaton field in non-thermal leptogenesis and DM production has been studied separately in earlier works, we have presented an analysis combining both and indicating the interesting interplay of inflaton, right handed neutrino and DM leading to the desired phenomenology including light neutrino masses and mixing. This framework though numerically challenging because of the presence of five coupled Boltzmann equations but economical in terms of the free parameters. Since we have separated the inflationary or exponential expansion phase from post-inflationary reheating, there certainly remain prospects of connecting such a set-up to more exotic physics or model-dependent scenarios where specific form of inflationary potentials, both during and after the slow-roll phase can dictate the dynamics of DM and lepton asymmetry yields. We leave such detailed studies for future endeavours.

% The other two neutrinos are considered to be heavy (typically $m_\varphi<2M_{2,3}$) with $N_3$ decoupled at a very early stage.

%%%%%%%%%%%%%%%%%
\section*{Acknowledgement}
%%%%%%%%%%%%%%%%%
DB acknowledges the support from Early Career Research Award from DST-SERB, Government of India (reference number: ECR/2017/001873). RR would like to acknowledge fruitful discussion with Qaisar Shafi and Arunansu Sil. BB would like to acknowledge the uniterrupted and fast internet service from BSNL (Bharat Sanchar Nigam Limited) which helped to work smoothly and efficiently. 

%%%%%%%%%%%%%%%
%%%%%%%%%%%%%%%
\appendix
%%%%%%%%%%%%%%%%%%
%%%%%%%%%%%%%%%%%%
\section{Maximum temperature during reheating}\label{sec:maxtemp}
%%%%%%%%%%%%%%%%%%

% The reheat temperature $(T_\text{RH})$ is usually calculated by assuming an instantaneous conversion of the energy density in the inflaton field into radiation when the decay width of the inflaton is equal to the Hubble expansion rate. In reality, however, reheating is not an instantaneous process~\cite{}. For example, if the inflaton is described by a simple model with quadratic potential, the radiation dominated phase follows a prolonged stage of matter domination during which the energy density of the universe is dominated by the coherent oscillations of the inflaton field. For different choices of inflaton potential, the equation of state during this pre-reheating or preheating phase can be different from the one in matter dominated phase. The temperature of the thermal bath during this phase can reach a value much larger than $T_\text{RH}$ and is typically denoted by: $T_\text{max}\sim\left(H_I M_\text{pl}\right)^{1/4}\sqrt{T_\text{RH}}$, where for a simple model of inflation $H_I$ can be identical with the inflaton mass~\cite{}. The temperature subsequently comes down from $T_\text{max}$ to  $T_\text{RH}$ when the inflaton decay is completed. The DM, thus can be produced at a temperature $T=T_\text{max}$ ({\it i.e.,} prior to the inflation ends), which is higher than $T_\text{RH}$. In our case, as we have see, the inflaton decay is non-instantaneous and therefore it is possible obtain an analytical expression for $T_\text{max}$.
% \textcolor{red}{This appendix is not being referred to from the text if I am not wrong.}

Assuming the inflaton energy density dominates in early times $E_{\varphi_I}\gg H_I$ we can write from the radiation evolution equation 

\bea\begin{aligned}
&\frac{dR}{dy}\Biggr|_\text{early} = \frac{\Gamma_N}{H}E_N\approx \frac{\Gamma_N}{\sqrt{\frac{16\pi E_{\varphi_I}}{3M_\text{pl}^2 y^3}}}E_{\varphi_I}= \Gamma_N E_{\varphi_I}\Biggl(\frac{3M_\text{pl}^2}{16\pi}\Biggr)^{1/2}\frac{y^{3/2}}{\sqrt{E_{\varphi_I}}}
    \end{aligned}
\eea

where $H_I$ is determined via

\bea\begin{aligned}
& E_{\varphi_I}=\frac{3M_\text{pl}^2}{8\pi}H_I^2.     
    \end{aligned}
\eea

as $E_{\varphi_I}=\rho_{\varphi_I} a_I^3$. Now, at $y=1,R=0$ which gives

% R =\frac{2}{5}\Gamma_N \Biggl(\frac{3M_\text{pl}^2}{16\pi}\Biggr)^{1/2}\sqrt{E_{\varphi_I}}\left(y^{5/2}-1\right)=\frac{3}{20\sqrt{2}\pi}\Gamma_N H_I M_\text{pl}^2\\&\implies 

\bea\begin{aligned}
& T=\Biggl(\frac{30}{\pi^2 g_\star\left(T\right)}\Biggr)^{1/4}\Biggl(\frac{3}{20\sqrt{2}\pi}\Biggr)^{1/4}\Gamma_N^{1/4}M_\text{pl}^{1/2}H_I^{1/4}\left(y^{-3/2}-y^{-4}\right)^{1/4}.     
    \end{aligned}
\eea

Thus, we can write

\bea
T = T_\text{max}\xi\left(y\right)
\eea

where $\xi\left(y\right)=\left(y^{-3/2}-y^{-4}\right)^{1/4}$ that starts from zero and grows until $y_\text{max}=\left(8/3\right)^{2/5}$ where it reaches a maximum $\xi_\text{max}=1$ that corresponds to $T=T_\text{max}$ and then decreases as $y^{-3/8}$. Now, as $\Gamma_N\sim M_1/8\pi>T_\text{RH}$, therefore,

\bea\begin{aligned}
& T_\text{max} >\left(8/3\right)^{2/5} \Biggl(\frac{30}{\pi^2 g_\star\left(T\right)}\Biggr)^{1/4}\Biggl(\frac{3}{20\sqrt{2}\pi}\Biggr)^{1/4}M_1^{1/4}M_\text{pl}^{1/2}H_I^{1/4}.     
    \end{aligned}
\eea

For a model of chaotic inflation $H_I\sim M_\varphi$. Then

\bea\begin{aligned}
& T_\text{max} \gtrsim 10^{14}~\text{GeV}\Biggl(\frac{30}{\pi^2 g_\star\left(T_\text{max}\right)}\Biggr)^{1/4}M_\text{pl}^{1/2}\Biggl(\frac{H_I}{10^{14}~\text{GeV}}\Biggr)^{1/4} \Biggl(\frac{M_1}{10^{12}~\text{GeV}}\Biggr)^{1/4}.    
    \end{aligned}
\eea

\bibliographystyle{JHEP}
\bibliography{Bibliography}

\providecommand{\href}[2]{#2}\begingroup\raggedright\begin{thebibliography}{100}

\bibitem{Zyla:2020zbs}
{\scshape Particle Data Group} collaboration, \emph{{Review of Particle
  Physics}}, \href{https://doi.org/10.1093/ptep/ptaa104}{\emph{PTEP} {\bfseries
  2020} (2020) 083C01}.

\bibitem{Aghanim:2018eyx}
{\scshape Planck} collaboration, \emph{{Planck 2018 results. VI. Cosmological
  parameters}},  \href{https://arxiv.org/abs/1807.06209}{{\ttfamily
  1807.06209}}.

\bibitem{Sakharov:1967dj}
A.D.~Sakharov, \emph{{Violation of CP Invariance, C asymmetry, and baryon
  asymmetry of the universe}},
  \href{https://doi.org/10.1070/PU1991v034n05ABEH002497}{\emph{Pisma Zh. Eksp.
  Teor. Fiz.} {\bfseries 5} (1967) 32}.

\bibitem{Weinberg:1979bt}
S.~Weinberg, \emph{{Cosmological Production of Baryons}},
  \href{https://doi.org/10.1103/PhysRevLett.42.850}{\emph{Phys. Rev. Lett.}
  {\bfseries 42} (1979) 850}.

\bibitem{Kolb:1979qa}
E.W.~Kolb and S.~Wolfram, \emph{{Baryon Number Generation in the Early
  Universe}}, \href{https://doi.org/10.1016/0550-3213(80)90167-4,
  10.1016/0550-3213(82)90012-8}{\emph{Nucl. Phys.} {\bfseries B172} (1980)
  224}.

\bibitem{Fukugita:1986hr}
M.~Fukugita and T.~Yanagida, \emph{{Baryogenesis Without Grand Unification}},
  \href{https://doi.org/10.1016/0370-2693(86)91126-3}{\emph{Phys. Lett.}
  {\bfseries B174} (1986) 45}.

\bibitem{Kuzmin:1985mm}
V.A.~Kuzmin, V.A.~Rubakov and M.E.~Shaposhnikov, \emph{{On the Anomalous
  Electroweak Baryon Number Nonconservation in the Early Universe}},
  \href{https://doi.org/10.1016/0370-2693(85)91028-7}{\emph{Phys. Lett.}
  {\bfseries 155B} (1985) 36}.

\bibitem{Fong:2013wr}
C.S.~Fong, E.~Nardi and A.~Riotto, \emph{{Leptogenesis in the Universe}},
  \href{https://doi.org/10.1155/2012/158303}{\emph{Adv. High Energy Phys.}
  {\bfseries 2012} (2012) 158303}
  [\href{https://arxiv.org/abs/1301.3062}{{\ttfamily 1301.3062}}].

\bibitem{Mohapatra:1979ia}
R.N.~Mohapatra and G.~Senjanovic, \emph{{Neutrino Mass and Spontaneous Parity
  Violation}}, \href{https://doi.org/10.1103/PhysRevLett.44.912}{\emph{Phys.
  Rev. Lett.} {\bfseries 44} (1980) 912}.

\bibitem{Yanagida:1979as}
T.~Yanagida, \emph{{HORIZONTAL SYMMETRY AND MASSES OF NEUTRINOS}}, {\emph{Conf.
  Proc.} {\bfseries C7902131} (1979) 95}.

\bibitem{GellMann:1980vs}
M.~Gell-Mann, P.~Ramond and R.~Slansky, \emph{{Complex Spinors and Unified
  Theories}}, {\emph{Conf. Proc.} {\bfseries C790927} (1979) 315}
  [\href{https://arxiv.org/abs/1306.4669}{{\ttfamily 1306.4669}}].

\bibitem{Glashow:1979nm}
S.L.~Glashow, \emph{{The Future of Elementary Particle Physics}},
  \href{https://doi.org/10.1007/978-1-4684-7197-7_15}{\emph{NATO Sci. Ser. B}
  {\bfseries 61} (1980) 687}.

\bibitem{Schechter:1980gr}
J.~Schechter and J.W.F.~Valle, \emph{{Neutrino Masses in SU(2) x U(1)
  Theories}}, \href{https://doi.org/10.1103/PhysRevD.22.2227}{\emph{Phys. Rev.}
  {\bfseries D22} (1980) 2227}.

\bibitem{Buchmuller:2004nz}
W.~Buchmuller, P.~Di~Bari and M.~Plumacher, \emph{{Leptogenesis for
  pedestrians}}, \href{https://doi.org/10.1016/j.aop.2004.02.003}{\emph{Annals
  Phys.} {\bfseries 315} (2005) 305}
  [\href{https://arxiv.org/abs/hep-ph/0401240}{{\ttfamily hep-ph/0401240}}].

\bibitem{DuttaBanik:2020vfr}
A.~Dutta~Banik, R.~Roshan and A.~Sil, \emph{{Neutrino mass and asymmetric dark
  matter: study with inert Higgs doublet and high scale validity}},
  \href{https://arxiv.org/abs/2011.04371}{{\ttfamily 2011.04371}}.

\bibitem{Davidson:2002qv}
S.~Davidson and A.~Ibarra, \emph{{A Lower bound on the right-handed neutrino
  mass from leptogenesis}},
  \href{https://doi.org/10.1016/S0370-2693(02)01735-5}{\emph{Phys. Lett.}
  {\bfseries B535} (2002) 25}
  [\href{https://arxiv.org/abs/hep-ph/0202239}{{\ttfamily hep-ph/0202239}}].

\bibitem{Pilaftsis:2003gt}
A.~Pilaftsis and T.E.J.~Underwood, \emph{{Resonant leptogenesis}},
  \href{https://doi.org/10.1016/j.nuclphysb.2004.05.029}{\emph{Nucl. Phys.}
  {\bfseries B692} (2004) 303}
  [\href{https://arxiv.org/abs/hep-ph/0309342}{{\ttfamily hep-ph/0309342}}].

\bibitem{Dev:2017wwc}
P.S.B.~Dev, M.~Garny, J.~Klaric, P.~Millington and D.~Teresi, \emph{{Resonant
  enhancement in leptogenesis}},
  \href{https://doi.org/10.1142/S0217751X18420034}{\emph{Int. J. Mod. Phys.}
  {\bfseries A33} (2018) 1842003}
  [\href{https://arxiv.org/abs/1711.02863}{{\ttfamily 1711.02863}}].

\bibitem{Lazarides:1991wu}
G.~Lazarides and Q.~Shafi, \emph{{Origin of matter in the inflationary
  cosmology}}, \href{https://doi.org/10.1016/0370-2693(91)91090-I}{\emph{Phys.
  Lett. B} {\bfseries 258} (1991) 305}.

\bibitem{Murayama:1992ua}
H.~Murayama, H.~Suzuki, T.~Yanagida and J.~Yokoyama, \emph{{Chaotic inflation
  and baryogenesis by right-handed sneutrinos}},
  \href{https://doi.org/10.1103/PhysRevLett.70.1912}{\emph{Phys. Rev. Lett.}
  {\bfseries 70} (1993) 1912}.

\bibitem{Kolb:1996jt}
E.W.~Kolb, A.D.~Linde and A.~Riotto, \emph{{GUT baryogenesis after
  preheating}}, \href{https://doi.org/10.1103/PhysRevLett.77.4290}{\emph{Phys.
  Rev. Lett.} {\bfseries 77} (1996) 4290}
  [\href{https://arxiv.org/abs/hep-ph/9606260}{{\ttfamily hep-ph/9606260}}].

\bibitem{Giudice:1999fb}
G.~Giudice, M.~Peloso, A.~Riotto and I.~Tkachev, \emph{{Production of massive
  fermions at preheating and leptogenesis}},
  \href{https://doi.org/10.1088/1126-6708/1999/08/014}{\emph{JHEP} {\bfseries
  08} (1999) 014} [\href{https://arxiv.org/abs/hep-ph/9905242}{{\ttfamily
  hep-ph/9905242}}].

\bibitem{Asaka:1999yd}
T.~Asaka, K.~Hamaguchi, M.~Kawasaki and T.~Yanagida, \emph{{Leptogenesis in
  inflaton decay}},
  \href{https://doi.org/10.1016/S0370-2693(99)01020-5}{\emph{Phys. Lett. B}
  {\bfseries 464} (1999) 12}
  [\href{https://arxiv.org/abs/hep-ph/9906366}{{\ttfamily hep-ph/9906366}}].

\bibitem{Asaka:1999jb}
T.~Asaka, K.~Hamaguchi, M.~Kawasaki and T.~Yanagida, \emph{{Leptogenesis in
  inflationary universe}},
  \href{https://doi.org/10.1103/PhysRevD.61.083512}{\emph{Phys. Rev. D}
  {\bfseries 61} (2000) 083512}
  [\href{https://arxiv.org/abs/hep-ph/9907559}{{\ttfamily hep-ph/9907559}}].

\bibitem{Hamaguchi:2001gw}
K.~Hamaguchi, H.~Murayama and T.~Yanagida, \emph{{Leptogenesis from N dominated
  early universe}},
  \href{https://doi.org/10.1103/PhysRevD.65.043512}{\emph{Phys. Rev. D}
  {\bfseries 65} (2002) 043512}
  [\href{https://arxiv.org/abs/hep-ph/0109030}{{\ttfamily hep-ph/0109030}}].

\bibitem{Jeannerot:2001qu}
R.~Jeannerot, S.~Khalil and G.~Lazarides, \emph{{Leptogenesis in smooth hybrid
  inflation}}, \href{https://doi.org/10.1016/S0370-2693(01)00429-4}{\emph{Phys.
  Lett. B} {\bfseries 506} (2001) 344}
  [\href{https://arxiv.org/abs/hep-ph/0103229}{{\ttfamily hep-ph/0103229}}].

\bibitem{Fujii:2002jw}
M.~Fujii, K.~Hamaguchi and T.~Yanagida, \emph{{Leptogenesis with almost
  degenerate majorana neutrinos}},
  \href{https://doi.org/10.1103/PhysRevD.65.115012}{\emph{Phys. Rev. D}
  {\bfseries 65} (2002) 115012}
  [\href{https://arxiv.org/abs/hep-ph/0202210}{{\ttfamily hep-ph/0202210}}].

\bibitem{Giudice:2003jh}
G.~Giudice, A.~Notari, M.~Raidal, A.~Riotto and A.~Strumia, \emph{{Towards a
  complete theory of thermal leptogenesis in the SM and MSSM}},
  \href{https://doi.org/10.1016/j.nuclphysb.2004.02.019}{\emph{Nucl. Phys. B}
  {\bfseries 685} (2004) 89}
  [\href{https://arxiv.org/abs/hep-ph/0310123}{{\ttfamily hep-ph/0310123}}].

\bibitem{Pascoli:2003rq}
S.~Pascoli, S.~Petcov and C.~Yaguna, \emph{{Quasidegenerate neutrino mass
  spectrum, mu ---> e + gamma decay and leptogenesis}},
  \href{https://doi.org/10.1016/S0370-2693(03)00698-1}{\emph{Phys. Lett. B}
  {\bfseries 564} (2003) 241}
  [\href{https://arxiv.org/abs/hep-ph/0301095}{{\ttfamily hep-ph/0301095}}].

\bibitem{Asaka:2002zu}
T.~Asaka, H.~Nielsen and Y.~Takanishi, \emph{{Nonthermal leptogenesis from the
  heavier Majorana neutrinos}},
  \href{https://doi.org/10.1016/S0550-3213(02)00934-3}{\emph{Nucl. Phys. B}
  {\bfseries 647} (2002) 252}
  [\href{https://arxiv.org/abs/hep-ph/0207023}{{\ttfamily hep-ph/0207023}}].

\bibitem{Panotopoulos:2006wj}
G.~Panotopoulos, \emph{{Non-thermal leptogenesis and baryon asymmetry in
  different neutrino mass models}},
  \href{https://doi.org/10.1016/j.physletb.2006.10.052}{\emph{Phys. Lett. B}
  {\bfseries 643} (2006) 279}
  [\href{https://arxiv.org/abs/hep-ph/0606127}{{\ttfamily hep-ph/0606127}}].

\bibitem{HahnWoernle:2008pq}
F.~Hahn-Woernle and M.~Plumacher, \emph{{Effects of reheating on
  leptogenesis}},
  \href{https://doi.org/10.1016/j.nuclphysb.2008.07.032}{\emph{Nucl. Phys. B}
  {\bfseries 806} (2009) 68} [\href{https://arxiv.org/abs/0801.3972}{{\ttfamily
  0801.3972}}].

\bibitem{Buchmuller:2013dja}
W.~Buchmuller, V.~Domcke, K.~Kamada and K.~Schmitz, \emph{{A Minimal
  Supersymmetric Model of Particle Physics and the Early Universe}},
  \href{https://arxiv.org/abs/1309.7788}{{\ttfamily 1309.7788}}.

\bibitem{Croon:2019dfw}
D.~Croon, N.~Fernandez, D.~McKeen and G.~White, \emph{{Stability, reheating and
  leptogenesis}}, \href{https://doi.org/10.1007/JHEP06(2019)098}{\emph{JHEP}
  {\bfseries 06} (2019) 098}
  [\href{https://arxiv.org/abs/1903.08658}{{\ttfamily 1903.08658}}].

\bibitem{Borah:2020wyc}
D.~Borah, S.~Jyoti~Das and A.K.~Saha, \emph{{Cosmic Inflation in Minimal
  $U(1)_{B-L}$ Model: Implications for (Non) Thermal Dark Matter and
  Leptogenesis}},  \href{https://arxiv.org/abs/2005.11328}{{\ttfamily
  2005.11328}}.

\bibitem{Samanta:2020gdw}
R.~Samanta, A.~Biswas and S.~Bhattacharya, \emph{{Non-thermal production of
  lepton asymmetry and dark matter in minimal seesaw with right handed neutrino
  induced Higgs potential}},
  \href{https://arxiv.org/abs/2006.02960}{{\ttfamily 2006.02960}}.

\bibitem{Zwicky:1933gu}
F.~Zwicky, \emph{{Die Rotverschiebung von extragalaktischen Nebeln}},
  \href{https://doi.org/10.1007/s10714-008-0707-4}{\emph{Helv. Phys. Acta}
  {\bfseries 6} (1933) 110}.

\bibitem{Rubin:1970zza}
V.C.~Rubin and W.K.~Ford, Jr., \emph{{Rotation of the Andromeda Nebula from a
  Spectroscopic Survey of Emission Regions}},
  \href{https://doi.org/10.1086/150317}{\emph{Astrophys. J.} {\bfseries 159}
  (1970) 379}.

\bibitem{Clowe:2006eq}
D.~Clowe, M.~Bradac, A.H.~Gonzalez, M.~Markevitch, S.W.~Randall, C.~Jones
  et~al., \emph{{A direct empirical proof of the existence of dark matter}},
  \href{https://doi.org/10.1086/508162}{\emph{Astrophys. J. Lett.} {\bfseries
  648} (2006) L109} [\href{https://arxiv.org/abs/astro-ph/0608407}{{\ttfamily
  astro-ph/0608407}}].

\bibitem{Srednicki:1988ce}
M.~Srednicki, R.~Watkins and K.A.~Olive, \emph{{Calculations of Relic Densities
  in the Early Universe}},
  \href{https://doi.org/10.1016/0550-3213(88)90099-5}{\emph{Nucl. Phys.}
  {\bfseries B310} (1988) 693}.

\bibitem{Gondolo:1990dk}
P.~Gondolo and G.~Gelmini, \emph{{Cosmic abundances of stable particles:
  Improved analysis}},
  \href{https://doi.org/10.1016/0550-3213(91)90438-4}{\emph{Nucl. Phys.}
  {\bfseries B360} (1991) 145}.

\bibitem{Kolb:1990vq}
E.W.~Kolb and M.S.~Turner, \emph{{The Early Universe}}, {\emph{Front. Phys.}
  {\bfseries 69} (1990) 1}.

\bibitem{Aprile:2018dbl}
E.~Aprile et~al., \emph{{Dark Matter Search Results from a One
  Tonne$\times$Year Exposure of XENON1T}},
  \href{https://arxiv.org/abs/1805.12562}{{\ttfamily 1805.12562}}.

\bibitem{Hall:2009bx}
L.J.~Hall, K.~Jedamzik, J.~March-Russell and S.M.~West, \emph{{Freeze-In
  Production of FIMP Dark Matter}},
  \href{https://doi.org/10.1007/JHEP03(2010)080}{\emph{JHEP} {\bfseries 03}
  (2010) 080} [\href{https://arxiv.org/abs/0911.1120}{{\ttfamily 0911.1120}}].

\bibitem{Bernal:2017kxu}
N.~Bernal, M.~Heikinheimo, T.~Tenkanen, K.~Tuominen and V.~Vaskonen, \emph{{The
  Dawn of FIMP Dark Matter: A Review of Models and Constraints}},
  \href{https://doi.org/10.1142/S0217751X1730023X}{\emph{Int. J. Mod. Phys.}
  {\bfseries A32} (2017) 1730023}
  [\href{https://arxiv.org/abs/1706.07442}{{\ttfamily 1706.07442}}].

\bibitem{Arcadi:2013aba}
G.~Arcadi and L.~Covi, \emph{{Minimal Decaying Dark Matter and the LHC}},
  \href{https://doi.org/10.1088/1475-7516/2013/08/005}{\emph{JCAP} {\bfseries
  1308} (2013) 005} [\href{https://arxiv.org/abs/1305.6587}{{\ttfamily
  1305.6587}}].

\bibitem{Yaguna:2011qn}
C.E.~Yaguna, \emph{{The Singlet Scalar as FIMP Dark Matter}},
  \href{https://doi.org/10.1007/JHEP08(2011)060}{\emph{JHEP} {\bfseries 08}
  (2011) 060} [\href{https://arxiv.org/abs/1105.1654}{{\ttfamily 1105.1654}}].

\bibitem{Chu:2011be}
X.~Chu, T.~Hambye and M.H.G.~Tytgat, \emph{{The Four Basic Ways of Creating
  Dark Matter Through a Portal}},
  \href{https://doi.org/10.1088/1475-7516/2012/05/034}{\emph{JCAP} {\bfseries
  1205} (2012) 034} [\href{https://arxiv.org/abs/1112.0493}{{\ttfamily
  1112.0493}}].

\bibitem{Blennow:2013jba}
M.~Blennow, E.~Fernandez-Martinez and B.~Zaldivar, \emph{{Freeze-in through
  portals}}, \href{https://doi.org/10.1088/1475-7516/2014/01/003}{\emph{JCAP}
  {\bfseries 1401} (2014) 003}
  [\href{https://arxiv.org/abs/1309.7348}{{\ttfamily 1309.7348}}].

\bibitem{Merle:2015oja}
A.~Merle and M.~Totzauer, \emph{{keV Sterile Neutrino Dark Matter from Singlet
  Scalar Decays: Basic Concepts and Subtle Features}},
  \href{https://doi.org/10.1088/1475-7516/2015/06/011}{\emph{JCAP} {\bfseries
  1506} (2015) 011} [\href{https://arxiv.org/abs/1502.01011}{{\ttfamily
  1502.01011}}].

\bibitem{Shakya:2015xnx}
B.~Shakya, \emph{{Sterile Neutrino Dark Matter from Freeze-In}},
  \href{https://doi.org/10.1142/S0217732316300056}{\emph{Mod. Phys. Lett.}
  {\bfseries A31} (2016) 1630005}
  [\href{https://arxiv.org/abs/1512.02751}{{\ttfamily 1512.02751}}].

\bibitem{Hessler:2016kwm}
A.G.~Hessler, A.~Ibarra, E.~Molinaro and S.~Vogl, \emph{{Probing the scotogenic
  FIMP at the LHC}}, \href{https://doi.org/10.1007/JHEP01(2017)100}{\emph{JHEP}
  {\bfseries 01} (2017) 100}
  [\href{https://arxiv.org/abs/1611.09540}{{\ttfamily 1611.09540}}].

\bibitem{Biswas:2016bfo}
A.~Biswas and A.~Gupta, \emph{{Freeze-in Production of Sterile Neutrino Dark
  Matter in U(1)$_{\rm B-L}$ Model}},
  \href{https://doi.org/10.1088/1475-7516/2017/05/A01,
  10.1088/1475-7516/2016/09/044}{\emph{JCAP} {\bfseries 1609} (2016) 044}
  [\href{https://arxiv.org/abs/1607.01469}{{\ttfamily 1607.01469}}].

\bibitem{Konig:2016dzg}
J.~König, A.~Merle and M.~Totzauer, \emph{{keV Sterile Neutrino Dark Matter
  from Singlet Scalar Decays: The Most General Case}},
  \href{https://doi.org/10.1088/1475-7516/2016/11/038}{\emph{JCAP} {\bfseries
  1611} (2016) 038} [\href{https://arxiv.org/abs/1609.01289}{{\ttfamily
  1609.01289}}].

\bibitem{Biswas:2016iyh}
A.~Biswas and A.~Gupta, \emph{{Calculation of Momentum Distribution Function of
  a Non-thermal Fermionic Dark Matter}},
  \href{https://doi.org/10.1088/1475-7516/2017/03/033,
  10.1088/1475-7516/2017/05/A02}{\emph{JCAP} {\bfseries 1703} (2017) 033}
  [\href{https://arxiv.org/abs/1612.02793}{{\ttfamily 1612.02793}}].

\bibitem{Biswas:2016yjr}
A.~Biswas, S.~Choubey and S.~Khan, \emph{{FIMP and Muon ($g-2$) in a
  U$(1)_{L_{\mu}-L_{\tau}}$ Model}},
  \href{https://doi.org/10.1007/JHEP02(2017)123}{\emph{JHEP} {\bfseries 02}
  (2017) 123} [\href{https://arxiv.org/abs/1612.03067}{{\ttfamily
  1612.03067}}].

\bibitem{Duch:2017khv}
M.~Duch, B.~Grzadkowski and D.~Huang, \emph{{Strongly self-interacting vector
  dark matter via freeze-in}},
  \href{https://doi.org/10.1007/JHEP01(2018)020}{\emph{JHEP} {\bfseries 01}
  (2018) 020} [\href{https://arxiv.org/abs/1710.00320}{{\ttfamily
  1710.00320}}].

\bibitem{Biswas:2018aib}
A.~Biswas, D.~Borah and A.~Dasgupta, \emph{{UV complete framework of freeze-in
  massive particle dark matter}},
  \href{https://doi.org/10.1103/PhysRevD.99.015033}{\emph{Phys. Rev.}
  {\bfseries D99} (2019) 015033}
  [\href{https://arxiv.org/abs/1805.06903}{{\ttfamily 1805.06903}}].

\bibitem{Heeba:2018wtf}
S.~Heeba, F.~Kahlhoefer and P.~Stöcker, \emph{{Freeze-in production of
  decaying dark matter in five steps}},
  \href{https://doi.org/10.1088/1475-7516/2018/11/048}{\emph{JCAP} {\bfseries
  1811} (2018) 048} [\href{https://arxiv.org/abs/1809.04849}{{\ttfamily
  1809.04849}}].

\bibitem{Zakeri:2018hhe}
S.~Peyman~Zakeri, S.~Mohammad Moosavi~Nejad, M.~Zakeri and S.~Yaser~Ayazi,
  \emph{{A Minimal Model For Two-Component FIMP Dark Matter: A Basic Search}},
  \href{https://doi.org/10.1088/1674-1137/42/7/073101}{\emph{Chin. Phys.}
  {\bfseries C42} (2018) 073101}
  [\href{https://arxiv.org/abs/1801.09115}{{\ttfamily 1801.09115}}].

\bibitem{Becker:2018rve}
M.~Becker, \emph{{Dark Matter from Freeze-In via the Neutrino Portal}},
  \href{https://doi.org/10.1140/epjc/s10052-019-7095-7}{\emph{Eur. Phys. J.}
  {\bfseries C79} (2019) 611}
  [\href{https://arxiv.org/abs/1806.08579}{{\ttfamily 1806.08579}}].

\bibitem{Heeba:2019jho}
S.~Heeba and F.~Kahlhoefer, \emph{{Probing the freeze-in mechanism in dark
  matter models with U(1) gauge extensions}},
  \href{https://doi.org/10.1103/PhysRevD.101.035043}{\emph{Phys. Rev.}
  {\bfseries D101} (2020) 035043}
  [\href{https://arxiv.org/abs/1908.09834}{{\ttfamily 1908.09834}}].

\bibitem{Lebedev:2019ton}
O.~Lebedev and T.~Toma, \emph{{Relativistic Freeze-in}},
  \href{https://doi.org/10.1016/j.physletb.2019.134961}{\emph{Phys. Lett.}
  {\bfseries B798} (2019) 134961}
  [\href{https://arxiv.org/abs/1908.05491}{{\ttfamily 1908.05491}}].

\bibitem{Barman:2019lvm}
B.~Barman, S.~Bhattacharya and M.~Zakeri, \emph{{Non-Abelian Vector Boson as
  FIMP Dark Matter}},
  \href{https://doi.org/10.1088/1475-7516/2020/02/029}{\emph{JCAP} {\bfseries
  2002} (2020) 029} [\href{https://arxiv.org/abs/1905.07236}{{\ttfamily
  1905.07236}}].

\bibitem{Bhattacharya:2019tqq}
S.~Bhattacharya, N.~Chakrabarty, R.~Roshan and A.~Sil, \emph{{Multicomponent
  dark matter in extended $U(1)_{B-L}$: neutrino mass and high scale
  validity}},  \href{https://arxiv.org/abs/1910.00612}{{\ttfamily 1910.00612}}.

\bibitem{Datta:2021elq}
A.~Datta, R.~Roshan and A.~Sil, \emph{{Imprint of the seesaw mechanism on
  feebly interacting dark matter and the baryon asymmetry}},
  \href{https://arxiv.org/abs/2104.02030}{{\ttfamily 2104.02030}}.

\bibitem{Bhattacharya:2021jli}
S.~Bhattacharya, R.~Roshan, A.~Sil and D.~Vatsyayan, \emph{{Symmetry origin of
  Baryon Asymmetry, Dark Matter and Neutrino Mass}},
  \href{https://arxiv.org/abs/2105.06189}{{\ttfamily 2105.06189}}.

\bibitem{Elahi:2014fsa}
F.~Elahi, C.~Kolda and J.~Unwin, \emph{{UltraViolet Freeze-in}},
  \href{https://doi.org/10.1007/JHEP03(2015)048}{\emph{JHEP} {\bfseries 03}
  (2015) 048} [\href{https://arxiv.org/abs/1410.6157}{{\ttfamily 1410.6157}}].

\bibitem{McDonald:2015ljz}
J.~McDonald, \emph{{Warm Dark Matter via Ultra-Violet Freeze-In: Reheating
  Temperature and Non-Thermal Distribution for Fermionic Higgs Portal Dark
  Matter}}, \href{https://doi.org/10.1088/1475-7516/2016/08/035}{\emph{JCAP}
  {\bfseries 1608} (2016) 035}
  [\href{https://arxiv.org/abs/1512.06422}{{\ttfamily 1512.06422}}].

\bibitem{Chen:2017kvz}
S.-L.~Chen and Z.~Kang, \emph{{On UltraViolet Freeze-in Dark Matter during
  Reheating}}, \href{https://doi.org/10.1088/1475-7516/2018/05/036}{\emph{JCAP}
  {\bfseries 05} (2018) 036}
  [\href{https://arxiv.org/abs/1711.02556}{{\ttfamily 1711.02556}}].

\bibitem{Biswas:2019iqm}
A.~Biswas, S.~Ganguly and S.~Roy, \emph{{Fermionic dark matter via UV and IR
  freeze-in and its possible X-ray signature}},
  \href{https://doi.org/10.1088/1475-7516/2020/03/043}{\emph{JCAP} {\bfseries
  03} (2020) 043} [\href{https://arxiv.org/abs/1907.07973}{{\ttfamily
  1907.07973}}].

\bibitem{Bernal:2019mhf}
N.~Bernal, F.~Elahi, C.~Maldonado and J.~Unwin, \emph{{Ultraviolet Freeze-in
  and Non-Standard Cosmologies}},
  \href{https://doi.org/10.1088/1475-7516/2019/11/026}{\emph{JCAP} {\bfseries
  1911} (2019) 026} [\href{https://arxiv.org/abs/1909.07992}{{\ttfamily
  1909.07992}}].

\bibitem{Bernal:2020bfj}
N.~Bernal, J.~Rubio and H.~Veermäe, \emph{{Boosting Ultraviolet Freeze-in in
  NO Models}}, \href{https://doi.org/10.1088/1475-7516/2020/06/047}{\emph{JCAP}
  {\bfseries 06} (2020) 047}
  [\href{https://arxiv.org/abs/2004.13706}{{\ttfamily 2004.13706}}].

\bibitem{Bernal:2020qyu}
N.~Bernal, J.~Rubio and H.~Veerm\"ae, \emph{{UV Freeze-in in Starobinsky
  Inflation}}, \href{https://doi.org/10.1088/1475-7516/2020/10/021}{\emph{JCAP}
  {\bfseries 10} (2020) 021}
  [\href{https://arxiv.org/abs/2006.02442}{{\ttfamily 2006.02442}}].

\bibitem{Barman:2020plp}
B.~Barman, D.~Borah and R.~Roshan, \emph{{Effective Theory of Freeze-in Dark
  Matter}}, \href{https://doi.org/10.1088/1475-7516/2020/11/021}{\emph{JCAP}
  {\bfseries 11} (2020) 021}
  [\href{https://arxiv.org/abs/2007.08768}{{\ttfamily 2007.08768}}].

\bibitem{Barman:2020ifq}
B.~Barman, S.~Bhattacharya and B.~Grzadkowski, \emph{{Feebly coupled vector
  boson dark matter in effective theory}},
  \href{https://arxiv.org/abs/2009.07438}{{\ttfamily 2009.07438}}.

\bibitem{Barman:2020jrf}
B.~Barman, A.~Dutta~Banik and A.~Paul, \emph{{Implications of NANOGrav results
  and UV freeze-in in a fast-expanding Universe}},
  \href{https://arxiv.org/abs/2012.11969}{{\ttfamily 2012.11969}}.

\bibitem{Garcia:2020eof}
M.A.G.~Garcia, K.~Kaneta, Y.~Mambrini and K.A.~Olive, \emph{{Reheating and
  Post-inflationary Production of Dark Matter}},
  \href{https://doi.org/10.1103/PhysRevD.101.123507}{\emph{Phys. Rev. D}
  {\bfseries 101} (2020) 123507}
  [\href{https://arxiv.org/abs/2004.08404}{{\ttfamily 2004.08404}}].

\bibitem{delAguila:2008ir}
F.~del Aguila, S.~Bar-Shalom, A.~Soni and J.~Wudka, \emph{{Heavy Majorana
  Neutrinos in the Effective Lagrangian Description: Application to Hadron
  Colliders}},
  \href{https://doi.org/10.1016/j.physletb.2008.11.031}{\emph{Phys. Lett. B}
  {\bfseries 670} (2009) 399}
  [\href{https://arxiv.org/abs/0806.0876}{{\ttfamily 0806.0876}}].

\bibitem{Aparici:2009fh}
A.~Aparici, K.~Kim, A.~Santamaria and J.~Wudka, \emph{{Right-handed neutrino
  magnetic moments}},
  \href{https://doi.org/10.1103/PhysRevD.80.013010}{\emph{Phys. Rev. D}
  {\bfseries 80} (2009) 013010}
  [\href{https://arxiv.org/abs/0904.3244}{{\ttfamily 0904.3244}}].

\bibitem{Bhattacharya:2015vja}
S.~Bhattacharya and J.~Wudka, \emph{{Dimension-seven operators in the standard
  model with right handed neutrinos}},
  \href{https://doi.org/10.1103/PhysRevD.94.055022}{\emph{Phys. Rev. D}
  {\bfseries 94} (2016) 055022}
  [\href{https://arxiv.org/abs/1505.05264}{{\ttfamily 1505.05264}}].

\bibitem{Liao:2016qyd}
Y.~Liao and X.-D.~Ma, \emph{{Operators up to Dimension Seven in Standard Model
  Effective Field Theory Extended with Sterile Neutrinos}},
  \href{https://doi.org/10.1103/PhysRevD.96.015012}{\emph{Phys. Rev. D}
  {\bfseries 96} (2017) 015012}
  [\href{https://arxiv.org/abs/1612.04527}{{\ttfamily 1612.04527}}].

\bibitem{Chala:2020vqp}
M.~Chala and A.~Titov, \emph{{One-loop matching in the SMEFT extended with a
  sterile neutrino}},
  \href{https://doi.org/10.1007/JHEP05(2020)139}{\emph{JHEP} {\bfseries 05}
  (2020) 139} [\href{https://arxiv.org/abs/2001.07732}{{\ttfamily
  2001.07732}}].

\bibitem{Bischer:2020sop}
I.~Bischer, T.~Plehn and W.~Rodejohann, \emph{{Dark Matter EFT, the Third --
  Neutrino WIMPs}},  \href{https://arxiv.org/abs/2008.04718}{{\ttfamily
  2008.04718}}.

\bibitem{Buchmuller:2005eh}
W.~Buchmuller, R.D.~Peccei and T.~Yanagida, \emph{{Leptogenesis as the origin
  of matter}},
  \href{https://doi.org/10.1146/annurev.nucl.55.090704.151558}{\emph{Ann. Rev.
  Nucl. Part. Sci.} {\bfseries 55} (2005) 311}
  [\href{https://arxiv.org/abs/hep-ph/0502169}{{\ttfamily hep-ph/0502169}}].

\bibitem{Kawasaki:2004qu}
M.~Kawasaki, K.~Kohri and T.~Moroi, \emph{{Big-Bang nucleosynthesis and
  hadronic decay of long-lived massive particles}},
  \href{https://doi.org/10.1103/PhysRevD.71.083502}{\emph{Phys. Rev. D}
  {\bfseries 71} (2005) 083502}
  [\href{https://arxiv.org/abs/astro-ph/0408426}{{\ttfamily
  astro-ph/0408426}}].

\bibitem{SENOGUZ20046}
V.~Senoguz and Q.~Shafi, \emph{Gut scale inflation, non-thermal leptogenesis,
  and atmospheric neutrino oscillations},
  \href{https://doi.org/https://doi.org/10.1016/j.physletb.2003.12.020}{\emph{Physics
  Letters B} {\bfseries 582} (2004) 6}.

\bibitem{Kaneta:2019zgw}
K.~Kaneta, Y.~Mambrini and K.A.~Olive, \emph{{Radiative production of
  nonthermal dark matter}},
  \href{https://doi.org/10.1103/PhysRevD.99.063508}{\emph{Phys. Rev. D}
  {\bfseries 99} (2019) 063508}
  [\href{https://arxiv.org/abs/1901.04449}{{\ttfamily 1901.04449}}].

\bibitem{Casas:2001sr}
J.A.~Casas and A.~Ibarra, \emph{{Oscillating neutrinos and $\mu \to e,
  \gamma$}}, \href{https://doi.org/10.1016/S0550-3213(01)00475-8}{\emph{Nucl.
  Phys. B} {\bfseries 618} (2001) 171}
  [\href{https://arxiv.org/abs/hep-ph/0103065}{{\ttfamily hep-ph/0103065}}].

\bibitem{Davidson:2008bu}
S.~Davidson, E.~Nardi and Y.~Nir, \emph{{Leptogenesis}},
  \href{https://doi.org/10.1016/j.physrep.2008.06.002}{\emph{Phys. Rept.}
  {\bfseries 466} (2008) 105}
  [\href{https://arxiv.org/abs/0802.2962}{{\ttfamily 0802.2962}}].

\bibitem{Edsjo:1997bg}
J.~Edsjo and P.~Gondolo, \emph{{Neutralino relic density including
  coannihilations}},
  \href{https://doi.org/10.1103/PhysRevD.56.1879}{\emph{Phys. Rev. D}
  {\bfseries 56} (1997) 1879}
  [\href{https://arxiv.org/abs/hep-ph/9704361}{{\ttfamily hep-ph/9704361}}].

\bibitem{Hasegawa:2019jsa}
T.~Hasegawa, N.~Hiroshima, K.~Kohri, R.S.L.~Hansen, T.~Tram and S.~Hannestad,
  \emph{{MeV-scale reheating temperature and thermalization of oscillating
  neutrinos by radiative and hadronic decays of massive particles}},
  \href{https://doi.org/10.1088/1475-7516/2019/12/012}{\emph{JCAP} {\bfseries
  12} (2019) 012} [\href{https://arxiv.org/abs/1908.10189}{{\ttfamily
  1908.10189}}].

\bibitem{Moroi:1993mb}
T.~Moroi, H.~Murayama and M.~Yamaguchi, \emph{{Cosmological constraints on the
  light stable gravitino}},
  \href{https://doi.org/10.1016/0370-2693(93)91434-O}{\emph{Phys. Lett. B}
  {\bfseries 303} (1993) 289}.

\bibitem{Kawasaki:1994af}
M.~Kawasaki and T.~Moroi, \emph{{Gravitino production in the inflationary
  universe and the effects on big bang nucleosynthesis}},
  \href{https://doi.org/10.1143/PTP.93.879}{\emph{Prog. Theor. Phys.}
  {\bfseries 93} (1995) 879}
  [\href{https://arxiv.org/abs/hep-ph/9403364}{{\ttfamily hep-ph/9403364}}].

\bibitem{Kofman:1997yn}
L.~Kofman, A.D.~Linde and A.A.~Starobinsky, \emph{{Towards the theory of
  reheating after inflation}},
  \href{https://doi.org/10.1103/PhysRevD.56.3258}{\emph{Phys. Rev. D}
  {\bfseries 56} (1997) 3258}
  [\href{https://arxiv.org/abs/hep-ph/9704452}{{\ttfamily hep-ph/9704452}}].

\bibitem{Linde:2005ht}
A.D.~Linde, \emph{{Particle physics and inflationary cosmology}}, vol.~5
  (1990), [\href{https://arxiv.org/abs/hep-th/0503203}{{\ttfamily
  hep-th/0503203}}].

\bibitem{Ema:2016hlw}
Y.~Ema, R.~Jinno, K.~Mukaida and K.~Nakayama, \emph{{Gravitational particle
  production in oscillating backgrounds and its cosmological implications}},
  \href{https://doi.org/10.1103/PhysRevD.94.063517}{\emph{Phys. Rev. D}
  {\bfseries 94} (2016) 063517}
  [\href{https://arxiv.org/abs/1604.08898}{{\ttfamily 1604.08898}}].

\bibitem{Ema:2018ucl}
Y.~Ema, K.~Nakayama and Y.~Tang, \emph{{Production of Purely Gravitational Dark
  Matter}}, \href{https://doi.org/10.1007/JHEP09(2018)135}{\emph{JHEP}
  {\bfseries 09} (2018) 135}
  [\href{https://arxiv.org/abs/1804.07471}{{\ttfamily 1804.07471}}].

\bibitem{Mambrini:2021zpp}
Y.~Mambrini and K.A.~Olive, \emph{{Gravitational Production of Dark Matter
  during Reheating}},  \href{https://arxiv.org/abs/2102.06214}{{\ttfamily
  2102.06214}}.

\bibitem{Barman:2021ugy}
B.~Barman and N.~Bernal, \emph{{Gravitational SIMPs}},
  \href{https://arxiv.org/abs/2104.10699}{{\ttfamily 2104.10699}}.

\bibitem{Garcia:2017tuj}
M.A.G.~Garcia, Y.~Mambrini, K.A.~Olive and M.~Peloso, \emph{{Enhancement of the
  Dark Matter Abundance Before Reheating: Applications to Gravitino Dark
  Matter}}, \href{https://doi.org/10.1103/PhysRevD.96.103510}{\emph{Phys. Rev.
  D} {\bfseries 96} (2017) 103510}
  [\href{https://arxiv.org/abs/1709.01549}{{\ttfamily 1709.01549}}].

\bibitem{Chianese:2018dsz}
M.~Chianese and S.F.~King, \emph{{The Dark Side of the Littlest Seesaw:
  freeze-in, the two right-handed neutrino portal and leptogenesis-friendly
  fimpzillas}},
  \href{https://doi.org/10.1088/1475-7516/2018/09/027}{\emph{JCAP} {\bfseries
  09} (2018) 027} [\href{https://arxiv.org/abs/1806.10606}{{\ttfamily
  1806.10606}}].

\bibitem{Chianese:2019epo}
M.~Chianese, B.~Fu and S.F.~King, \emph{{Minimal Seesaw extension for Neutrino
  Mass and Mixing, Leptogenesis and Dark Matter: FIMPzillas through the
  Right-Handed Neutrino Portal}},
  \href{https://doi.org/10.1088/1475-7516/2020/03/030}{\emph{JCAP} {\bfseries
  03} (2020) 030} [\href{https://arxiv.org/abs/1910.12916}{{\ttfamily
  1910.12916}}].

\bibitem{Davidson:2000er}
S.~Davidson and S.~Sarkar, \emph{{Thermalization after inflation}},
  \href{https://doi.org/10.1088/1126-6708/2000/11/012}{\emph{JHEP} {\bfseries
  11} (2000) 012} [\href{https://arxiv.org/abs/hep-ph/0009078}{{\ttfamily
  hep-ph/0009078}}].

\bibitem{Mambrini:2013iaa}
Y.~Mambrini, K.A.~Olive, J.~Quevillon and B.~Zaldivar, \emph{{Gauge Coupling
  Unification and Nonequilibrium Thermal Dark Matter}},
  \href{https://doi.org/10.1103/PhysRevLett.110.241306}{\emph{Phys. Rev. Lett.}
  {\bfseries 110} (2013) 241306}
  [\href{https://arxiv.org/abs/1302.4438}{{\ttfamily 1302.4438}}].

\bibitem{Kim:2017mtc}
J.~Kim and J.~McDonald, \emph{{Clockwork Higgs portal model for freeze-in dark
  matter}}, \href{https://doi.org/10.1103/PhysRevD.98.023533}{\emph{Phys. Rev.
  D} {\bfseries 98} (2018) 023533}
  [\href{https://arxiv.org/abs/1709.04105}{{\ttfamily 1709.04105}}].

\bibitem{Kim:2018xsp}
J.~Kim and J.~Mcdonald, \emph{{Freeze-In Dark Matter from a sub-Higgs Mass
  Clockwork Sector via the Higgs Portal}},
  \href{https://doi.org/10.1103/PhysRevD.98.123503}{\emph{Phys. Rev. D}
  {\bfseries 98} (2018) 123503}
  [\href{https://arxiv.org/abs/1804.02661}{{\ttfamily 1804.02661}}].

\end{thebibliography}\endgroup

\end{document}